\documentclass[
bibnotes,amsmath,amssymb,
aps, physrev,
longbibliography
]{revtex4-2}

\usepackage[colorinlistoftodos]{todonotes}

\usepackage{xcolor}

\usepackage{amsthm,bbm,appendix,braket,nicefrac,verbatim} 
\usepackage{float}
\usepackage{subfig}
\usepackage{physics}
\usepackage{graphicx}
\usepackage{dcolumn}
\usepackage{bm}
\usepackage{hyperref}
\usepackage{tikz}
\usetikzlibrary{quantikz2}
\usepackage{mathtools}
\usepackage{booktabs}

\let\originalleft\left
\let\originalright\right
\renewcommand{\left}{\mathopen{}\mathclose\bgroup\originalleft}
\renewcommand{\right}{\aftergroup\egroup\originalright}
\def\({\mathopen{}\left(}
\def\){\right)\mathclose{}}

\newcommand*{\eqdef}{\mathrel{\vcenter{\baselineskip0.5ex \lineskiplimit0pt\hbox{.}\hbox{.}}}=}

\newcommand{\invcm}{$\mathrm{cm}^{-1}$}
\newcommand{\BE}[1]{\mathcal{B}\left[#1\right]}
\newcommand{\SE}{Schr\"{o}dinger equation}
\def\id{\mathbbm{1}}

\def\F{\mathbbm{F}}

\def\Z{\mathbbm{Z}}

\def\cD{\mathcal{D}}

\def\cN{\mathcal{N}}

\def\cR{\mathcal{R}}

\def\supp{\mathrm{supp}}

\def\WH{\mathrm{WH}}

\def\PFX{\mathrm{PFX}}
\def\QFT{\mathrm{QFT}}

\newtheorem{theorem}{Theorem}[section]

\newtheorem{corollary}[theorem]{Corollary}
\newtheorem{remark}[theorem]{Remark}

\begin{document}

\title{\textbf{Simulating high-accuracy nuclear motion Hamiltonians using discrete variable representation and Walsh–Hadamard QROM on fault-tolerant quantum computers} 
} 

\author{Michał Szczepanik}
\author{\'{A}kos Nagy}
\author{Emil Zak}
\email{Contact author: emil@beit.tech}
\affiliation{BEIT Canada Inc., Toronto, Canada}
\altaffiliation{https://www.beit.tech}
\affiliation{BEIT sp. z o.o., Mogilska 43, 31-545 Krak\'ow, Poland}
\altaffiliation{https://www.beit.tech}

\begin{abstract}
We present a quantum algorithm for simulating rovibrational Hamiltonians on fault-tolerant quantum computers. The method integrates exact curvilinear kinetic energy operators and general-form potential energy surfaces expressed in a hybrid finite-basis/discrete-variable representation. The Hamiltonian is encoded as a unitary quantum circuit using a quantum read-only memory construction based on the Walsh–Hadamard transform, enabling high-accuracy quantum phase estimation of rovibrational energy levels and dynamics simulations.
Our technique provides asymptotic reductions in both logical qubit count and T-gate complexity that are exponential in the number of atoms and at least polynomial in the total Hilbert-space size, relative to existing block-encoding techniques based on linear combinations of unitaries and variational basis representation. Compared with classical variational methods, it offers exponential memory savings and polynomial reductions in time complexity.
The quantum volume required for computing the rovibrational spectrum of water can be reduced by up to $10^5\times$ compared with other quantum methods, increasing to at least $10^6$ for a classically intractable 30-dimensional (12-atom) molecular system. For this case with a six-body coupled potential, estimating spectroscopic-accuracy energy levels would require about three months on a 1 MHz fault-tolerant quantum processor with fewer than 300 logical qubits, versus over 30 000 years on the fastest current classical supercomputer. These estimates are approximate and subject to technological uncertainties, and realizing the asymptotic advantage will require substantial quantum resources and continued algorithmic progress.
\end{abstract}

\maketitle
\newpage

\section{Introduction}
\label{sec:I-intro}
The accurate quantum-mechanical simulation of molecular systems becomes exponentially more challenging as system size increases. Each additional particle multiplies the Hilbert-space dimensionality, computational cost, and memory requirements, creating a fundamental exponential barrier to classical computation as system complexity grows. This \textit{curse of dimensionality} affects quantum-mechanical simulations across major areas of theoretical physics and chemistry, including condensed-phase systems such as molecular crystals, clusters, liquids, and bimolecular environments, as well as gas-phase systems including molecular complexes relevant to atmospheric, exoplanetary, and interstellar chemistry and spectroscopy. For example, achieving spectroscopic sub-wavenumber accuracy in energy-level calculations is essential for modeling absorption and emission spectra in molecular physics, astrophysics, and exoplanetary-atmosphere research~\cite{Gordon2017,Gordon2022,Tennyson2020,Huang2021,Jing2025}. First-principles thermodynamic and kinetic modeling, such as the computation of heat capacities, internal vibrational relaxation, or transport properties in molecular junctions~\cite{Galperin2007,Jahangiri2021}, typically tolerates energy-level uncertainties of tens of wavenumbers, yet still requires relatively high-quality and computationally demanding models, which succumb to the same curse of dimensionality.

Within molecular sciences, the calculation of rotational-vibrational (rovibrational) energy levels and associated absorption spectra exemplifies this computational bottleneck.  Rovibrational Hamiltonians are intrinsically high-dimensional, often involve strongly coupled large-amplitude motions, and require accurate representations of both kinetic and potential energy operators in curvilinear internal coordinates. Despite decades of algorithmic advances, state-of-the-art classical methods remain restricted to relatively small systems, typically fewer than ten atoms when spectroscopic accuracy is required~\cite{Bowman2008,Sarka2021,Carrington2017,Avila2019,Lauvergnat2023, Sunaga2024, Glaser2023, Felker2023,Mtyus2023, Larsson2025}. Even for such systems, classical calculations frequently rely on approximations, including reduced dimensionality models, separable or sum-of-products representations of the potential energy surface, or basis set truncations that compromise accuracy and transferability~\cite{Carrington2017}.

The limitations of classical approaches are most pronounced for floppy molecules, weakly bound clusters, and systems exhibiting dense Hamiltonian spectra. In these cases, low-frequency modes are strongly anharmonic and coupled, leading to rapid growth in basis size and the breakdown of perturbative or separable methods~\cite{Bukowski2008,Dawes2013,Wang2016,Yu2022,Wang2018,Schwan2020,Szalewicz2009,Zhu2023,Felker2024,Simk2025,Derbali2025}. While classical techniques such as multi-configuration time-dependent Hartree~\cite{Wang2003,Manthe2008,Vendrell2011,Lindoy2021,Xie2015,Wodraszka2020,Wodraszka2024} or other tensor-network methods~\cite{Baiardi2017,Baiardi2019a,Larsson2025} have achieved remarkable success, their computational scaling ultimately constrains their applicability to either modest numbers of atoms or simplified Hamiltonians appropriate for semi-rigid systems. Notably, scenarios that require accurate modeling of protein–drug binding processes, the properties of liquid water, and hydrogen bonding and networks of water molecules demand a quantum-mechanical description, which has so far remained computationally elusive~\cite{Liu1996,Morrone2008,Reddy2016,Cheng2019,Gong2022}.

Quantum computing offers a fundamentally different computational paradigm for addressing high-dimensional quantum dynamics. By encoding microscopic molecular Hamiltonian as a quantum-computational state, quantum algorithms can in principle represent and manipulate exponentially large Hilbert spaces using polynomial resources~\cite{Lloyd1996,AspuruGuzik2005,childs2012,low2017}. Early work on quantum simulation focused primarily on electronic structure problems~\cite{babbush2018,lee2021,Goings2022,Lee2023,rocca2024,loaiza2024,Deka2025}, motivated by the promising scaling of quantum phase estimation on block-encoded two-body electronic Hamiltonian, relative to exact classical diagonalization. More recently, attention has expanded to vibrational dynamics, and molecular spectroscopy, where quantum algorithms promise exponential memory savings and, in certain regimes, asymptotic reductions in time complexity.

Several quantum algorithms for vibrational structure and dynamics have been proposed~\cite{mcardle2019,Sawaya2021,Ollitrault2021,Miessen2022,Malpathak2025,Trenev2025,motlagh2024a,loaiza25}, utilizing a range of Hamiltonian representations and Hamltonian simulation techniques. These include Trotter-based approaches modeling second-quantized vibrational modes~\cite{motlagh2024a,malpathak2024,Trenev2025,loaiza25}, grid-based encodings~\cite{Ollitrault2020v,Ollitrault2021}, and block-encoding constructions using linear combinations of unitaries~\cite{Majland2025}. While these methods establish the theoretical feasibility of quantum vibrational simulations, many rely on restrictive assumptions, such as low-degree polynomial or separable potential energy surfaces, simplified kinetic energy operators, making up Hamiltonians for which quantum resources either scale unfavorably with system size or present limited model accuracy.

In this work, we present a quantum algorithm for simulating general rovibrational Hamiltonians that overcomes several of these limitations. Our approach supports exact curvilinear kinetic energy operators and general-form potential energy surfaces, without assuming a sum-of-products decomposition. 
The Hamiltonian is represented in a hybrid finite-basis/discrete-variable representation~\cite{Light2000}, enabling systematic control of accuracy while preserving favorable scaling properties. 

Central to our construction is a quantum read-only memory (QROM) based on the Walsh–Hadamard transform, which allows efficient, fault-tolerant loading of Hamiltonian matrix elements into a quantum circuit.  QROM, particularly in the context of block-encoding Born–Oppenheimer potential energy surfaces, constitutes a major bottleneck for efficient quantum simulation. By compressing the information input using the Walsh–Hadamard transform, our QROM implementation achieves an exponential reduction in quantum computational volume, defined as the product of logical qubit count and Clifford+$T$-gate complexity, relative to existing quantum approaches.
These gains in QROM oracles propagate directly to the cost of Hamiltonian block-encoding, where we load PES elements explicitly using QROM and a canonical operator method: our alternative to the method of multiplexed rotations~\cite{Gosset:2024}. 
Taken together, our algorithm for an efficient partitioning and block-encoding discrete-variable representation Hamiltonian with Walsh–Hadamard QROM (WH-QROM) achieves asymptotic reductions in quantum volume that are exponential in the number of atoms, relative to existing block-encoding techniques based on linear combinations of unitaries and variational basis representations.

Notably, the rovibrational Hamiltonians considered here constitute a particularly challenging class of Schrödinger equation instances. They involve an exact representation of strongly coupled rotational and vibrational degrees of freedom in a kinetic energy operator that captures Coriolis couplings, and a general-form PES that determines the achievable accuracy. For such systems, both Hamiltonian matrix-element evaluation and spectral resolution become exponentially more difficult as the system size increases, often referred to as \textit{double-exponential curse of dimensionality}.

To the best of our knowledge, this work constitutes the first fault-tolerant quantum computing proposal dedicated to overcoming this double-exponential curse of dimensionality in performing high-accuracy rovibrational simulations for general Hamiltonians. Earlier works~\cite{mcardle2019,Sawaya2021,Ollitrault2021,Miessen2022,Malpathak2025,Trenev2025,motlagh2024a,loaiza25} addressed Hamiltonians of lower complexity that are generally insufficient for spectroscopic accuracy rovibrational calculations, particularly for weakly bound molecular systems. 
We provide a resource analysis of the proposed algorithm, explicitly accounting for logical qubits, T-gate counts, and ancillary overheads. These estimates are benchmarked against representative classical methods and existing quantum approaches, highlighting the computational quantum advantage regimes. 
Our study includes a series of molecules with increasing complexity and computational requirements. First, we estimate the quantum resources required to compute the rovibrational spectrum of the water molecule with $n$ basis functions per mode.
Our resource estimates demonstrated a $10^{5}\times$ reduction in T-count with asymptotic reduction in quantum volume scaling as $\mathcal{O}(n^{6.5})$, due to rovibrational Hamiltonian block-encoding algorithm alone, with standard SELECT–SWAP QROM~\cite{low_trading_2024}. Application of WH-QROM further reduces resources compared to SELECT–SWAP by an order of magnitude, with advantage growing exponentially as the molecule's size increases.

Relative to block-encoding based on linear combinations of unitaries and variational basis representations, we find that for the highly rotationally excited methane (nine dimensions) the quantum volume is reduced by at least five orders of magnitude, and for a classically intractable 30-dimensional model molecule by more than six orders of magnitude.
For this latter molecule our algorithm enables sub-wavenumber calculations via quantum phase estimation with fewer than 300 qubits, compared to several thousand with other techniques. 
Because our approach requires fewer than 300 logical qubits, rather than thousands, and exhibits a modest Clifford+T gate count with polynomial advantages over standard techniques, we anticipate that it can substantially advance the practical usefulness of quantum simulation. At the current pace of hardware development, with steady growth in qubit-count capabilities, this reduction could translate into the realization of such simulations on quantum hardware on the order of a decade earlier than would otherwise be expected.

Remarkably, beyond its advantages for vibrational calculations, our method also gives significant cost reductions in modeling rotational excitation. Whereas classical algorithms  exhibit multiplicative linear scaling  $\mathcal{O}(J)$ with the rotational quantum number, our approach achieves additive square-root scaling $\mathcal{O}(\sqrt{J})$. This improvement is particularly beneficial for highly excited rotational states.

Beyond its immediate impact on molecular spectroscopy, the framework developed here establishes a general route toward quantum simulation of high-dimensional nuclear motion problems, scattering, and other systems succumbing to the curse of dimensionality. 
The ability to treat general kinetic energy operators and non-separable potentials is essential for extending quantum algorithms to larger molecules, molecular clusters, and condensed-phase environments. This capability carries direct implications for fields that require simulations of complex photochemical reaction mechanisms and proton tunneling processes in solvent environments relevant to drug discovery, among others. Finally, the prospect of first-principles calculations eventually complementing or even replacing laboratory experiments, by providing systematic error control and avoiding complex sample preparation, instrumental uncertainties, limited spectral coverage, and long acquisition times, provides a motivation for the development of quantum computation in molecular science, which we follow in this work.

The remainder of this paper is organized as follows. In Sec.~\ref{sec:method}, we introduce the rovibrational Hamiltonian and its representation in a hybrid finite-basis/discrete-variable framework.
Sec.~\ref{sec:I-classical} and \ref{sec:III-quantum-computation} respectively review classical and quantum computational methodologies and their limitations. 
Section~\ref{sec:block_encoding_reducing_costs} describes the block-encoding of the rovibrational Hamiltonian, and Sec.~\ref{sec:whqrom} presents the Walsh–Hadamard QROM construction. Sec.~\ref{sec:resource_estimation} provides quantum resource estimates and comparative analyses.

\section{Quantum-mechanical rotational-vibrational energy calculations}
\label{sec:method} 
\subsection{Rotational-vibrational Hamiltonian in curvilinear internal coordinates}
\label{sec:Hamiltonian} 
The rovibrational Hamiltonian can be written as
\begin{equation}\label{eq:nuclear_hamiltonian}
\hat{H} = \hat{K}+\hat{V},
\end{equation}
where $\hat{K}$ denotes the kinetic energy operator (KEO) for nuclear motion and $\hat{V}$ is the Born-Oppenheimer potential energy surface (PES).

\paragraph{KEO.}
We consider an $A$-atomic molecule described by a set of $D = 3A-6$ internal coordinates $\boldsymbol{q}$ and three Euler angles $\boldsymbol{\Theta}=(\theta,\phi,\chi)$ specifying the orientation of the body-fixed molecular frame (MF) with respect to the space-fixed frame. The nuclear KEO in the MF can be expressed as \cite{Gatti::2017}
\begin{equation}\label{eq:KEO_PGP}
	\hat{K} = \sum_{i,j=1}^{3A-3} \hat{P}_i^\dag G_{ij}(\boldsymbol{q})\hat{P}_j,
\end{equation}
where $G_{ij}(\boldsymbol{q})$ is a metric tensor whose explicit form depends on the choice of MF embedding and the definition of the internal coordinates $\boldsymbol{q}$. The operators $\hat{P}_i$ for $i \leq D$ are momenta conjugate to the internal coordinates $q_i$, while $\{\hat{P}_i\}_{i>D}~=~\{\hat{J}_\alpha\}_{\alpha= x,y,z}$  correspond to the components of the body-fixed angular momentum operator. The KEO in Eq.~\eqref{eq:KEO_PGP} can be decomposed into three contributions associated with pure vibrational motion, Coriolis couplings, and rotational motion, respectively \cite{Gatti::2017},
\begin{equation}
    \hat{K}(\boldsymbol{q},\boldsymbol{\Theta}) = \hat{K}^{vib}+ \hat{K}^{cor}+\hat{K}^{rot},
\end{equation}
where
\begin{equation}\label{eq:KEO_def}
    \hat{K}^{vib}(\boldsymbol{q}) = \frac{1}{2} \sum_{i,j=1}^{D}\hat{P}^\dag_i g_{ij}(\boldsymbol{q})\hat{P}_{j},
\end{equation}
\begin{equation}\label{eq:KEO_cor}
    \hat{K}^{cor}(\boldsymbol{q},\Theta) =  \sum_{\alpha={x,y,z}}\hat{J}_\alpha\sum_{j=1}^{D}\frac{1}{2}\left( \Gamma_{\alpha j}(\boldsymbol{q})\hat{P}_{j}+\hat{P}_{j}^\dag \Gamma_{\alpha j}(\boldsymbol{q})\right),
\end{equation}
\begin{equation}\label{eq:KEO_rot}
    \hat{K}^{rot}(\boldsymbol{q},\Theta) = \frac{1}{2} \sum_{\alpha, \beta={x,y,z}}\mu_{\alpha\beta}(\boldsymbol{q})\hat{J}_\alpha\hat{J}_\beta.
\end{equation}
The matrices $\mathbf{g}(\boldsymbol{q})$, $\boldsymbol{\Gamma}(\boldsymbol{q})$, and $\boldsymbol{\mu}(\boldsymbol{q})$ depend on the chosen internal coordinates and the MF embedding.
Among the various MF embedding schemes, the Eckart frame is particularly advantageous, as it minimizes the coupling between rotational and vibrational degrees of freedom \cite{Eckart1935,Wei1997,Szalay2015}. A systematic analysis of the influence of MF embedding on the convergence properties of variational rovibrational calculations can be found in Refs.~\cite{Gatti2009,Gatti::2017}. The present discussion remains agnostic to the specific choice of MF embedding and internal coordinates. However, in Sec.~\ref{sec:polyspherical-coordinates} we adopt polyspherical internal coordinates in order to provide concrete quantum resource estimates for block-encoding the rovibrational Hamiltonian in a representative embedding and coordinate choice.

We also note that eq.~\ref{eq:KEO_PGP} can also be generalized by introducing quasi-momentum operators $\mathcal{P}_i= f_i(\boldsymbol{q})P_i$, where $f_i(\boldsymbol{q})$ is an arbitrary function of internal coordinates. This leads to the modified expression
\begin{equation}
    \hat{K} = \sum_{i,j=1}^{3A-3}\hat{\mathcal{P}}_i\mathcal{G}_{ij}(\boldsymbol{q})\hat{\mathcal{P}_j}
\end{equation}
where $\mathcal{G}_{ij}(\boldsymbol{q})=\frac{G_{ij}(\boldsymbol{q})}{f_i(\boldsymbol{q})f_j(\boldsymbol{q})}$. Expressing the KEO through the quasi-momentum operators gives additional flexibility useful for quantum-computational cost reductions. 

\paragraph{PES.}
The PES is a smooth function of the $D$ internal coordinates,
\begin{equation}
\hat{V} : \mathbb{R}^D \ni \boldsymbol{q} \mapsto V(\boldsymbol{q}) \in \mathbb{R}_+ .
\end{equation}
A commonly used representation of the PES is an expansion in terms involving an increasing number of coupled coordinates, known as the $L$-mode representation (here denoted as LMR) \cite{Carter1997},
\begin{equation}
V(\mathbf{q}) = V(\mathbf{q}_0)+\sum_{l_1=1}^{L_1}V_{l_1}^{(1)}(q_{l_1})+\sum_{l_1<l_2}^{L_2}V_{l_1,l_2}^{(2)}(q_{l_1},q_{l_2})+...+\sum_{l_1<l_2<...<l_M}^{L_M}V_{l_1,l_2,...,l_M}^{(M)}(q_{l_1},q_{l_2},...,q_{l_M})
\label{eq:multimode}
\end{equation}
where the truncation order $M$ is chosen according to the available computational resources and the desired accuracy.
The PES in eq.~\eqref{eq:multimode} can be approximated by a sum-of-products (SOP) form, which offers a substantial speedup of calculations and extends the range of tractable dimensionalities from a few (typically fewer than 20) to on the order of $10^2$ degrees of freedom, at the expense of reduced accuracy. In contrast, non-SOP PES representations rely on direct functional fits with predefined model forms or on machine-learning approaches~\cite{Unke2021,Behler2021}, including neural networks (NNs)~\cite{Jiang2013,Manzhos2020,ShanavasRasheeda2022,Nandi2022}, which currently provide some of the most accurate potential energy surfaces available.

\subsection{Rotational-vibrational basis sets and Hamiltonian representation}

The Hamiltonian given in eq.~\eqref{eq:nuclear_hamiltonian} defines the Schrödinger equation
\begin{equation}\label{eq:SE}
\hat{H}|\psi\rangle = E |\psi\rangle ,
\end{equation}
which determines the rovibrational energy levels $E$ and eigenstates $|\psi\rangle$. The rovibrational Hamiltonian may be represented as a matrix in a finite basis set,
\begin{equation}
|\psi\rangle = \sum_{n=1}^{N} u_n |\phi_n\rangle ,
\label{eq:se-ansatz}
\end{equation}
where $\lbrace\ket{\phi_n}\rbrace_{n=1}^{N}$ denotes the chosen basis functions and $u_n$ are the corresponding expansion coefficients.
Substituting eq.~\eqref{eq:se-ansatz} into eq.~\eqref{eq:SE} and projecting onto the same basis set yields the generalized matrix eigenvalue problem
\begin{equation}
\mathbf{H}\mathbf{U} = \mathbf{S}\mathbf{U}\mathbf{E},
\label{eq:HVBR}
\end{equation}
where $\mathbf{H}$ is the Hamiltonian matrix with elements $H_{ij} = \langle \phi_i | \hat{H} | \phi_j \rangle$, $\mathbf{S}$ is the overlap matrix with elements $S_{ij} = \langle \phi_i | \phi_j \rangle$, $\mathbf{U}$ is the coefficient matrix whose columns contain the expansion vectors $\textbf{u}$, and $\mathbf{E}$ is the diagonal matrix of energy levels.

The rovibrational state $|\psi\rangle$ can be expanded as a sum of products of rotational and vibrational basis functions as follows:
\begin{equation}\label{eq:full_sys_wavefunction}
|\Psi_{rv}^{(J,h,\Gamma,m,i)}\rangle = \sum_{k=-J}^J
|\Phi_{k}^{(J,h,i)}\rangle |J,k,m\rangle ,
\end{equation}
where the complete rotational basis is conventionally chosen as the symmetric-top basis represented by Wigner functions,
\begin{equation}
|J,k,m\rangle = \sqrt{\tfrac{2J+1}{8\pi^2}}  (-1)^k
, \mathcal{D}^{J*}_{m,-k}(\theta,\phi,\chi),
\end{equation}
which may be straightforwardly adapted to account for parity in systems with parity symmetry\cite{Bunker1999}. Here, $J$ denotes the total angular momentum quantum number, $k$ its projection onto the body-fixed $z$-axis, $m$ its projection onto the space-fixed $Z$-axis, $h$ enumerates rovibrational eigenstates, $\Gamma$ labels the symmetry species of the state, and $i$ distinguishes different electronic states. The vibrational basis functions may be constructed as direct products of single-coordinate basis functions for each of the $D$ internal coordinates:
\begin{equation}\label{eq:vib_wavefunction}
|\Phi_{k}^{(J,h,i)}\rangle =\sum_{v_1,\dots,v_D} a^{(J,k,h,i)}_{v_1,\dots,v_D}\bigotimes_{c=1}^D |v_c\rangle ,
\end{equation}
or alternatively, as a contracted basis set~\cite{Carrington2017}. The index $v_c = 0,1,2,\dots,n_c-1$ specifies the vibrational excitation along coordinate $c$, up to the $n_c$-th energy level. Consequently, the dimension of the basis (and hence the Hamiltonian matrix) is
\begin{equation}
N=J(J+1)\prod_{c=1}^D n_c ,
\end{equation}
which reduces to $J(J+1)n^D$ in the case of an equal number $n$ of basis functions per vibrational coordinate.
With this choice of basis, the total Hilbert space decomposes into a direct sum,
\begin{equation}
\mathcal{H} = \bigoplus_{J=0}^{J_{\max}} \bigoplus_{\Gamma=0}^{n_{\Gamma}-1} \mathcal{H}_{J\Gamma},
\end{equation}
where $\Gamma$ denotes symmetry labels of the Hamiltonian other than the total angular momentum $J$ quantum number. The block-diagonal structure of the Hamiltonian matrix allows the eigenvalue problems associated with each pair $(J,\Gamma)$ to be solved independently, thereby reducing the dimension of each individual Hilbert space to $J\prod_{c=1}^{D} n_c / n_{\Gamma}$, where $n_{\Gamma}$ is the number of irreducible representations of the molecular symmetry group \cite{Bunker1999}. Still, the total basis set size scales exponentially with the number of internal coordinates $D$, although the exploitation of rotational and molecular symmetries reduces the associated prefactor.
The determination of rovibrational energy levels $E_{J,h,i,\Gamma}$ and the corresponding eigenstates $|\Psi_{rv}^{(J,h,\Gamma,m,i)}\rangle$ is the central objective of this work.

\section{Rovibrational calculations with classical computers and their limitations}
\label{sec:I-classical}

Assessing the potential advantage of a quantum computing method for rovibrational calculations requires first identifying the key components and bottlenecks of classical computational techniques.
Among those, the variational method is the most systematic and accurate. The variational method involves forming the Hamiltonian matrix in a chosen basis set and computing its eigenvalues and eigenvectors, as shown by eq.~\ref{eq:HVBR} in Appendix~\ref{sec:variational}, where further details are given.
In the variational approach, solving the rovibrational Schrödinger equation generally involves the following steps~\cite{Bowman2008, Carrington2017}:
\begin{enumerate}
\item Select a set of internal coordinates $\mathbf{q}$ (e.g., bond lengths, angles, dihedral angles) and derive the corresponding KEO within a MF embedding.
\item Compute the PES over the configuration space relevant to the interesting energy range.
\item Choose appropriate basis functions for the internal and rotational coordinates and construct the Hamiltonian matrix representation by evaluating its matrix elements. 
\item Solve the resulting eigenvalue problem.
\end{enumerate}
These steps are common across simulations of few-body quantum-mechanical systems \cite{Brocks1983,Bunker1999,Sutcliffe1986}. In practice, however, system-specific improvements can often be introduced, for example by avoiding explicit matrix-element evaluation through the use of iterative eigensolvers, or by calculating PES on-the-fly in quantum dynamics calculations.
Alternative approaches to the variational methodology, such as perturbation theory (e.g., VPT2 \cite{Franke2021}), can be applied to larger systems containing up to roughly 50 atoms \cite{Mendolicchio2024}. However, these methods typically lack sufficient accuracy and rigorous error control, particularly for systems exhibiting large-amplitude motions, where the zeroth-order approximation (usually harmonic) becomes qualitatively inadequate \cite{Sutcliffe1986}.

\subsection{Principal bottlenecks: the double-exponential curse of dimensionality}
The variational workflow faces three fundamental bottlenecks. First, the size of the basis set required for spectroscopic accuracy grows exponentially with the number of internal coordinates, leading to prohibitive memory requirements for storing wavefunctions and Hamiltonians. Second, strongly coupled systems and large-amplitude motions generate dense Hamiltonian matrices and high densities of rovibrational states, which further increase computational resources and severely limits the convergence and robustness of classical iterative eigensolvers such as Lanczos or Davidson methods~\cite{Saad2011}. Third, accurate PESs for polyatomic systems are often available only in general, non-SOP form, making the evaluation of matrix elements by quadratures and matrix-vector products exponentially costly as the number of atoms increase. 

\paragraph{Memory and floating-point operations scaling.}
To illustrate the unfavorable basis set size scaling in rovibrational calculations, consider the memory required to store a single state vector. For a triatomic molecule such as H\textsubscript{2}O, using 10 basis functions per internal coordinate, the vibrational ($J=0$) state vector contains $10^{3}$ elements - on the order of kilobytes. A four-atom molecule such as NH\textsubscript{3}, with six internal degrees of freedom, already requires megabytes of memory. For ethylene (C\textsubscript{2}H\textsubscript{4}) with 12 internal coordinates, the state vector grows to $\approx 10^{12}$ elements, corresponding to terabytes of RAM. A nine-atom system such as dimethyl sulfide (CH\textsubscript{3}–S–CH\textsubscript{3}), with 21 internal coordinates, would require storage of approximately $10^{21}$ elements, which is far beyond the capacity of any existing classical computing architecture. Computation of highly excited rotational states worsens this situation by a further factor of $\mathcal{O}(J)$. As a result, approximations or model reductions become unavoidable and a range of classical techniques has been developed to alleviate these problems.
Table~\ref{tab:rovib-classical} summarizes selected computational studies aimed at high-accuracy rovibrational energy levels calculations.
Memory is not the only limiting factor for classical computation. The number of floating-point operations (FLOP) required to converge a sufficient number of energy levels can also become prohibitive. 
Iterative eigensolvers require matrix-vector multiplications, which in the non-SOP Hamiltonian case cannot be decomposed coordinate-wise~\cite{Carrington2018}. For such calculations SOP form is useful. When matrix representation is sparse, and when not too many eigenvaules are required the iterative eigensolvers combined with SOP PES (and SOP KEO) are most efficient, scaling as $\mathcal{O}(\rho NM_{kr})$ in time-complexity and $\mathcal{O}(N)$ memory, where $\rho$ is matrix sparsity, $M_{kr}$ is the size of Krylov space (typically $\mathcal{O}(\log(1/\varepsilon))$, where $\varepsilon$ is target accuracy and $N \sim \mathcal{O}(Jn^{D})$ is the total number of basis functions. Direct diagonalization methods's time complexity scale as $\mathcal{O}(N^3)$ and  $\mathcal{O}(N^2)$ memory, limiting their use to roughly less than 1 million basis functions. Condition number and eigenvalue separation may also contribute to this scaling in a non-obvious way. Thus, the time complexity grows exponentially with the number of dimensions and at least linearly with the total number of basis functions, rendering high-accuracy rovibrational calculations infeasible for all but the smallest systems ($\approx$ 10 atoms), although approximate schemes have extended this barrier at the expense of modest reductions in accuracy~\cite{Thomas2018}, as shown in Table~\ref{tab:rovib-classical}.
Studies listed in Table~\ref{tab:rovib-classical} are typically limited to roughly 6-12 atoms, depending on the level of approximation employed~\cite{Leclerc2014,Thomas2018,Carrington2017}. Simulations of larger systems are either insufficiently accurate to meet current scientific and industrial requirements or computationally too expensive. 

A widely used strategy for reducing basis set and Hamiltonian matrix size is basis contraction, in which the full-dimensional basis is constructed through successive diagonalizations and truncations of lower-dimensional effective Hamiltonians \cite{Felker2023,Felker2024,Mant2018,Tennyson2004,Bowman2008}. While contraction yields more compact matrix representations, it typically does so at the cost of reduced sparsity. Contraction schemes are implemented in several major computational packages, including MULTIMODE \cite{Bowman2003}, TROVE \cite{Yurchenko2007}, and DVR3D \cite{Tennyson2004}. In particular, MULTIMODE employs an $L$-mode representation (LMR)~\cite{Carter1997} of the PES, expanding terms of increasing dimensionality and contracting the basis accordingly. However, this approach suffers from rapid growth in the number of PES terms for high-dimensional expansions and limited accuracy when truncated at low order \cite{Christiansen2004}. LMR is attractive and will be studied further in this work because, for many molecules, there exists a range of values of 
$L$ for which the accuracy is sufficient and the number of PES terms remains manageable, while the overall computational complexity and memory requirements are classically prohibitive.

An alternative route to reducing Hamiltonian size is basis set pruning, in either real or momentum space \cite{Brown2016,Halverson2015,Simmons2023,Wodraszka2024}. Pruned basis techniques have been successfully implemented in several codes \cite{Halverson2015,Sarka2021,Wodraszka2024}, but the construction of general, systematic pruning protocols remains challenging. For discrete-variable-representation-based methods in particular~\cite{Dawes2004}, pruning requires removing quadrature grid points in equal number to the eliminated basis functions, rendering the procedure increasingly complex and labor-intensive as dimensionality grows.

Multi-dimensional time-dependent Hartree (MCTDH) methods address basis convergence by adapting the basis through optimized single-particle functions \cite{Wang2003,Manthe2008,Vendrell2011,Lindoy2021}. These methods rely on SOP PES representations and typically scale poorly beyond 12–15 dimensions unless accuracy is sacrificed or the system is semi-rigid \cite{Vendrell2009,Xie2015}. Recent refinements of MCTDH \cite{Wodraszka2020,Wodraszka2024} offer improvements, but remain constrained by grid sizes, carrying the same structural limitations. Other notable neural-network solvers for nuclear-electron dynamics have been proposed recently~\cite{Zhang2025} offering a promising unorthodox approach.

Related techniques, including vibrational configuration interaction (VCI) and vibrational coupled cluster (VCC) methods \cite{Christiansen2004a,Christiansen2022,Schrder2022,Ziegler2019}, allow for systematic accuracy improvement but retain exponential scaling with the number of internal coordinates. Tensor network methods, such as vibrational density matrix renormalization group (vDMRG), compress both the basis and the Hamiltonian operator \cite{Baiardi2017,Rey2019,Glaser2023,Larsson2025}. However, they are effective primarily for systems with local mode couplings and struggle to capture general PES forms with high accuracy \cite{Baiardi2017,Baiardi2019,Baiardi2019a}. Consequently, vDMRG methods are generally unsuitable for highly anharmonic or floppy systems with strong global couplings, although calculations with up to 24 coupled vibrational modes have been reported \cite{Glaser2023}.

\paragraph{Hamiltonian representation.}
The Hamiltonian matrix elements can be calculated by directly integrating the KEO and PES over the selected basis set, with the resulting matrix elements stored in memory. Such an approach is mandated when direct diagonalization methods are required, especially when many eigenvalues are needed. Storing all Hamiltonian matrix elements and associated grids in memory becomes impossible for larger systems with approximately more than six atoms. 
For the PES contribution, matrix elements
\begin{equation}
\begin{split}\label{eq:FBR-PES}
V_{ij}=\langle \phi_i(\mathbf{q})|V(\mathbf{q})|\phi_j(\mathbf{q})\rangle=\int_{\mathbb{R}^{D}\times\mathbb{R}^3}d\mathbf{q}d\mathbf{\Theta}\phi_i(\mathbf{q})V(\mathbf{q})\phi_j(\mathbf{q})
\end{split}
\end{equation}
are evaluated over the full configurational space of internal coordinates $\mathbf{q} \in \mathbb{R}^{D}$ and rotational coordinates $\boldsymbol{\Theta} = (\theta,\phi,\chi) \in \mathbb{R}^{3}$. 
Analytical evaluation of these integrals is rarely possible, necessitating high-dimensional numerical quadrature schemes such as Smolyak grids~\cite{Avila2015}. 
Quadrature approximations replace the integral given in eq.~\ref{eq:FBR-PES} with a weighted sum over an exponentially large grid of size $K$:
\begin{equation}
V_{ij}\approx V^{FBR}_{ij}=\sum_{k=0}^{K-1}\frac{w_k}{\omega(q_k)}\phi_i(q_k)V(q_k)\phi_j(q_k)
\end{equation}
a procedure known as the finite-basis representation (FBR). In the absence of more efficient integration strategies, this step constitutes a dominant computational bottleneck, as the size of these grids can grow faster than the number of basis functions, giving rise to a double exponential curse of dimensionality. In contrast to electronic structure theory, even the computation of individual Hamiltonian matrix elements becomes prohibitive for nuclear motion problems at high dimensionality, especially with a general nSOP PES. Rovibrational calculations therefore suffer from two coupled exponential scalings: (i) exponential growth of the state vector and Hamiltonian matrix with the number of internal coordinates, and (ii) exponential scaling of the matrix-element evaluation cost when the Born–Oppenheimer PES lacks a compact SOP form~\cite{Avila2019}. 

\paragraph{Ab initio electronic structure calculations.}
The construction of accurate global PESs constitutes another major computational bottleneck. Developing a PES as a function of all $3A-6$ internal coordinates requires both high-accuracy ab initio energies over a large region of configuration space and an enormous number of electronic structure calculations \cite{Bunker1999}. This burden is intrinsic to Born–Oppenheimer–based approaches.
Even coarse direct-product grids rapidly become infeasible. For example, a modest four-point-per-coordinate grid for dimethyl sulfide already requires on the order of $10^{12}$ electronic structure evaluations. While individual calculations may be tractable, the aggregate workload demands extreme parallelism and vast HPC resources. Assuming a single CCSD(T) calculation costs $\sim 10^{13}$ FLOPs, generating such a PES would require between several years on a leading exascale HPC such as El Capitan (1.8 ExaFlops, October 2025).

At present, no scalable and efficient method, classical or quantum, exists for evaluating electronic energies across large nuclear configuration spaces at controllable accuracy. Machine-learned PESs~\cite{Behler2021,Unke2021}, including neural-network potentials \cite{Jiang2013,Manzhos2020,Nandi2022,Kser2023}, can significantly reduce the number of required ab initio points, but generally lack a SOP structure, which is critical for classically-efficient Hamiltonian evaluation. High-quality PES refinement therefore often relies on labor-intensive, physics-informed fitting strategies \cite{Polyansky2012,Huang2008a}. Although tensor factorization techniques such as POTFIT attempt to recover approximate SOP forms from general PESs \cite{Brommer2007}, they introduce additional approximations and incur growing computational overhead as the number of terms increases \cite{Carrington2017,Thomas2018}. Fully non-SOP PESs, in turn, require high-dimensional integrals whose cost becomes prohibitive beyond roughly 6-8 atoms \cite{Avila2015,Lauvergnat2023}.

While modern HPC facilities can generate sufficiently accurate ab initio data for many systems, strongly correlated electronic structures such as transition-metal complexes and weakly bound clusters remain beyond classical reach, due to orbital active space size. Notable examples include the FeMo cofactor in nitrogen fixation \cite{Li2019,lee2021}, cobalt phthalocyanine in CO\textsubscript{2} reduction, and the active site of cytochrome P450 \cite{Goings2022,Deka2025,Lee2023}. For these systems, achieving chemical accuracy or better for even a single geometry exceeds current classical capabilities. Recent quantum algorithms targeting electronic structure \cite{babbush2018,lee2021,rocca2024,loaiza2024,Deka2025} offer a potential pathway forward, and their coherent integration into a quantum nuclear motion framework may ultimately be required.

\begin{table}[H]
\centering
\begin{tabular}{|l|c|c|l|}
\hline
\textbf{System } & \textbf{Dim. (D)} & \textbf{Comment} & \textbf{Reference} \\
\hline
HCOOH  & 9 & Smolyak/non-SOP PES & Mart\'in Santa Dar\'a  \emph{et al.}~\cite{MartnSantaDara2022} \\
Water dimer & 12 & non-SOP PES,  $<1$~cm$^{-1}$ & Wang \emph{et al.}~\cite{Wang2018,Wang2025}, Jing \emph{et al.}~\cite{Jing2025} \\
CH$_4$F & 12 & $<0.1$~cm$^{-1}$ & Papp  \emph{et al.}~\cite{Papp2022} \\
Water trimer & 12 & non-SOP PES & Simko \emph{et al.}~\cite{Simk2025} \\
Water cluster & 12 & $<0.01$~cm$^{-1}$, 5000 states & Sarka  \emph{et al.}~\cite{Sarka2021}; Larsson~\cite{Larsson2025} \\
HF trimer & 12 & non-SOP & Felker  \emph{et al.}~\cite{Felker2023} \\
Ethylene & 12 & SOP PES/TROVE & Mant  \emph{et al.}~\cite{Mant2018} \\
Methanol & 12 & Smolyak/non-SOP PES & Sunaga \emph{et al.}~\cite{Sunaga2024,Sunaga2025-kt} \\
Ethylene oxide & 15 & SOP PES & Kallullathil \emph{et al.}~\cite{Kallullathil2023} \\
Malonaldehyde & 21 &  Smolyak/(non)-SOP PES & Lauvergnat \emph{et al.}~\cite{Lauvergnat2023} \\
Methyloxirane  & 24 & HDMR/DMRG & Glaser \emph{et al.}~\cite{Glaser2023} \\
Glycine &  24 & MULTIMODE & Qu \emph{et al.}~\cite{Qu2021} \\
Benzene  & 30 & $\sim 10^6$ energy levels, momentum-space pruning & Halverson \emph{et al.}~\cite{Halverson2015} \\
Naphthalene / Uracil & 48 &  RRBPM method, SOP PES & Thomas \emph{et al.}~\cite{Thomas2018} \\
\hline
\end{tabular}
\caption{Selected notable classical rovibrational calculations, ordered by dimensionality $D$. Smolyak means the method uses Smolyak quadratures for evaluating Hamiltonian matrix elements~\cite{Avila2015}. }
\label{tab:rovib-classical}
\end{table}

\paragraph{Coupling strength and density of states}
In systems with strong vibrational and rovibrational couplings, such as Coriolis, Fermi, and $l$-doubling interactions \cite{Bunker1999}, particularly those involving large-amplitude motions, the number of basis functions required for convergence can become prohibitively large, leading to extreme memory demands. For example, achieving high-accuracy energy levels for the water dimer, a 12-dimensional system, requires hundreds of gigabytes of RAM \cite{Wang2018}. Scaling such calculations to substantially larger systems is currently infeasible.
Prominent examples include weakly bound molecular clusters \cite{Derbali2025} relevant to atmospheric spectroscopy, such as water clusters \cite{Bukowski2008,Yu2022,Wang2018,Schwan2020,Szalewicz2009,Zhu2023,Simk2025} (e.g. vapor-continuum problem), CH\textsubscript{5}\textsuperscript{+} \cite{Wang2016}, halide-water complexes \cite{Viglaska2025}, solvated organic molecules, and other van der Waals complexes and reactive scattering systems \cite{Dawes2013,Felker2024}. In these cases, accurate treatment of strong coupling and resonance effects is essential, as rovibrational eigenvalues are highly sensitive to small wavefunction errors, rendering most classical approaches inadequate. In such strongly coupled systems, particularly when low-energy large-amplitude modes interact with high-frequency vibrations, the Hamiltonian matrix becomes dense and the eigenvalue spectrum congested, leading to memory-intensive calculations and slow convergence. In such regimes, approximate contraction schemes and classical iterative eigensolvers may fail due to the high density of states.

These difficulties are further exacerbated for highly excited states of polyatomic molecules, which often exhibit both strong couplings and a high density of states \cite{tennyson1990,Gordon2017,Ribeiro2002,Saleh2025a,Field}. When strong coupling precludes a separable representation of the PES, non-SOP forms become unavoidable and are computationally demanding for classical algorithms. Although SOP-based methods have enabled vibrational calculations in spaces of up to several hundred dimensions \cite{Xie2015,Halverson2015,Glaser2023}, they rely on approximations that restrict their general applicability. 

Consequently, quantum computational advantage is most likely to arise in regimes requiring general, non-SOP PESs \cite{Bukowski2008,Behler2021,Manzhos2020,Kser2023}, where classical methods encounter exponential barriers in both memory and runtime. For highly anharmonic or floppy systems with multiple PES minima, basis-set convergence is often slow \cite{Ribeiro2002}, and harmonic oscillator bases perform poorly \cite{Mtyus2023,Rey2023}. Although adaptive basis constructions tailored to internal coordinates have been explored \cite{Saleh2025}, systematic and transferable strategies remain elusive and pose an opportunity for quantum advantage.

\paragraph{Kinetic energy operator.}
In the aforementioned cases of strong coupling and weakly bound systems, achieving spectroscopic accuracy requires capturing Coriolis and other subtle vibrational couplings~\cite{Bunker1999}. Therefore, only approaches that employ either an exact KEO (e.g., Sutctliffe-Tennyson~\cite{Sutcliffe1986}, Polyspherical~\cite{Gatti::2017}) or a systematically converged approximate expansion (e.g., TROVE~\cite{Yurchenko2007}) or numerically exact KEO (GENIUSH~\cite{Mtyus2018}, ElVibRot~\cite{Lauvergnat2002}) are suitable for high-accuracy rovibrational calculations. A quantum computing procedure ought to adopt one of these representations as well. The selection of appropriate internal coordinates and MF embedding determines both the convergence properties of energy levels with respect to basis set size and the treatment of singularities in the Hamiltonian, making it a non-trivial, system-specific task~\cite{Gatti::2017,Sutcliffe2000,Saleh2025}. The form of the KEO can significantly affect computational convergence, i.e. memory and computational complexity, meaning that an appropriate choice of coordinates and MF embedding remains relevant to determining quantum computing advantage too. 

Once internal coordinates are chosen, the KEO can be derived through several approaches: analytical derivation, symbolic algebra software, or numerical evaluation at grid points distributed throughout the configurational space~\cite{Brocks1983,Yurchenko2007,Yachmenev2015,Gatti::2017}, either equidistant (finite-element methods~\cite{Jelovina2015}, collocation~\cite{Zak2019}) or based on quadrature grids (finite-basis and discrete variable representations~\cite{Tennyson2004,Yurchenko2007,Mtyus2018}. For larger molecular systems, the derivation of the kinetic energy operator becomes considerably more involved and often requires the use of symbolic algebra software or numerical techniques~\cite{Lauvergnat2002,Yurchenko2007,Yachmenev2015,Marsili2022}. 

Computing methodologies are commonly classified according to their choice of internal coordinates: rectilinear coordinates (including normal coordinates)~\cite{Bunker1999}, curvilinear internal coordinates~\cite{Bunker1999,Sutcliffe2000,Gatti::2017} or more recent machine-learning-optimized generalized coordinates~\cite{Saleh2025}, as well the choice of the MF embedding. In this work, we remain agnostic to the choice of MF and internal coordinates, and give example calculations for polyspherical coordinates shown in sec.~\ref{sec:polyspherical-coordinates}.

\section{Quantum computing techniques for simulating nuclear motion dynamics: state-of-the-art and this work's contribution}
\label{sec:III-quantum-computation}
Quantum computational methods are, in principle, anticipated to overcome the fundamental limitations of classical algorithms for high-complexity molecular systems in three key respects. First, quantum computation addresses the memory bottleneck by encoding exponentially large Hilbert spaces into a polynomial number of qubits.
Second, it provides the ability to calculate eigenvalues of a given Hamiltonian with arbitrary precision and accuracy, directly controlled by simulation parameters~\cite{AspuruGuzik2005,sawaya2020,Ollitrault2021}. Third and most debated is the possibility of a computational speedup over classical algorithms~\cite{Lee2023}.

Rovibrational energy levels can be calculated with the Quantum Phase Estimation (QPE) algorithm~\cite{AspuruGuzik2005}, commonly regarded as the most efficient method for eigenvalue estimation on a FTQC~\cite{babbush2018}. QPE estimates eigenvalues by accumulating the phase from the time evolution operator $e^{iHt}$, given suitable input eigenstate and precision $\varepsilon$. To implement $e^{iHt}$ efficiently, different frameworks can be used, including Trotterization~\cite{Ollitrault2021,Sawaya2021,Malpathak2025,Trenev2025}, quantum signal processing (QSP)~\cite{low2017} and qubitization~\cite{low2019}. Existing studies of few-body vibrational Hamiltonians have explored FTQC algorithms based on Trotterization~\cite{Sawaya2021,Trenev2025,Malpathak2025,motlagh2024a}, qubitization~\cite{Majland2025}, as well as several NISQ-era proposals~\cite{mcardle2019,Ollitrault2020v,Ollitrault2021,Sawaya2021,R2025}. QSP-based QPE achieve favourable asymptotic scaling of $\mathcal{O}(\zeta t + \log(1/\varepsilon))$ calls to unitary Hamiltonian oracle $\mathcal{B}[H]$ encoding a scaled Hamiltonian $H/\zeta$ matrix. The total number of calls to $\mathcal{B}[H]$ required by QPE scales as $\mathcal{O}(\zeta)$, where $\zeta$ is the block-encoding scaling constant. Thus, minimizing $\zeta$ is critical for reducing the quantum computational cost, equally important as an efficient $\mathcal{B}[H]$ implementation. While extensive techniques have been developed for reducing $\zeta$ in the context of electronic structure simulations~\cite{babbush2018,lee2021,Deka2025,rocca2024,loaiza2024}, such strategies remain underexplored in the context of rovibrational Hamiltonians, except our previous work adopting Finite Basis Representation and Linear Combination of Unitaries form for block-encoded vibronic Hamiltonians~\cite{kamakari25}. 

In this work, we address the classical computational bottlenecks identified in the previous section, namely memory and time complexity limitations, the double exponential curse of dimensionality associated with Hamiltonian matrix representations, and the challenges posed by strong mode coupling and high densities of states in floppy molecular systems. Our technique addresses the aforementioned challenges as follows:
\begin{itemize}

\item \textbf{Basis set size.}
Large-amplitude nuclear motion, as encountered in floppy molecules, requires a large number of basis functions per coordinate for convergence \cite{Viglaska2020,Rey2023,Avila2019,Felker2023,Mtyus2023}. By encoding the rovibrational wavefunction in qubit registers, our algorithm achieves an exponential reduction in memory requirements relative to classical representations. Specifically, the ratio of classical bits to quantum qubits scales as $\tilde{\mathcal{O}}\left(n^{D-\frac{7}{2}}/D^2\right)$.

\item \textbf{Coupling strength.}
Strong vibrational and rovibrational couplings invalidate perturbative or separable approximations. Our method is based on a variational formulation employing an exact KEO, with the achievable accuracy determined solely by the quality of the underlying general-form PES.

\item \textbf{Hamiltonian representation.}
No prior quantum algorithm has addressed the general class of high-accuracy rovibrational Hamiltonians combining an exact curvilinear KEO with a non–sum-of-products (non-SOP) PES. Existing approaches typically rely on simplified rectilinear KEOs, harmonic or polynomial PES expansions, or low-dimensional mode truncations \cite{mcardle2019,Sawaya2021,Malpathak2025,Trenev2025}, which limit accuracy to approximately 10–100~cm$^{-1}$, well below spectroscopic standards \cite{Brocks1983,Carter1986,Szalay2015}. Applying these algorithms directly to general rovibrational Hamiltonians leads to prohibitive qubit and gate costs or significant accuracy loss \cite{Sawaya2021,mcardle2019,motlagh2024a,kamakari25,Majland2025,Trenev2025,Ollitrault2021}.

Our approach utilizes direct-product discrete variable representations (DVRs) for the PES and selected components of the KEO, ensuring converged matrix elements even for non-SOP potentials. Combined with a novel QROM construction based on the Walsh–Hadamard transform, this strategy yields exponential reductions in gate counts and qubit requirements relative to conventional FBR/QROM implementations.

The choice of basis functions is critical for efficient block-encoding. While harmonic-oscillator bases dominate earlier quantum studies~\cite{mcardle2019,Ollitrault2020v,Ollitrault2021,Sawaya2021}, they are poorly suited for large-amplitude motion, precisely the regime where quantum advantage is expected~\cite{Richerme2023}. Our framework supports more expressive bases, including contracted functions from reduced-dimensional Schrödinger equations, Morse oscillator functions, three-dimensional harmonic-oscillator eigenfunctions (Laguerre polynomials), and DVRs constructed from these bases, which are particularly advantageous for non-SOP PESs.

Quantum approaches to pre–Born–Oppenheimer dynamics have also been proposed~\cite{ollitrault2020-nonad,su2021}, but their applicability is currently limited by the need for joint nuclear–electronic basis design, singularity management, and poor excited-state convergence~\cite{su2021,Sibaev2020,Sasmal2020,Sutcliffe2000}, restricting their use to relatively small systems~\cite{Saly2023,Mtyus2018}.

\item \textbf{Density of states.}
Highly excited or floppy systems exhibit dense rovibrational spectra, necessitating the computation of many eigenvalues. This challenge can be addressed by increasing the precision of QPE and by combining our block-encoding strategy with quantum eigenvalue landscape scanning techniques~\cite{GRM25}.

\end{itemize}

Combination of these elements yields simultaneous improvements in memory usage (qubit count versus classical bit count) and computational cost (T-gates versus classical floating-point operations). The resulting overall advantage may be summarized as
\begin{equation}
\text{memory saving} \times \text{speedup} = \mathcal{\tilde{O}}\left(n^{D-\frac{7}{2}}/D^2\right) \times \mathcal{\tilde{O}}\left(n^{D(1-\alpha)}\right), \;\;\  \alpha <1,
\label{eq:advantage}
\end{equation}
where $D$ is the number of internal coordinates and $n$ the number of basis functions per coordinate.

In the next section we introduce a new algorithm for Hamiltonian simulation tailored to high-accuracy rovibrational problems. 
When combined with QPE, it enables the computation of rovibrational energy levels, and its applicability extends naturally to time-dependent dynamics, intramolecular vibrational relaxation, and other observables derived from time-evolving rovibrational wavefunctions. Central to this framework are improved Hamiltonian partitioning strategies based on DVRs and a new QROM construction aimed at reducing the block-encoding normalization factor $\zeta$ and lowering the Clifford+T gate and qubit complexity of the block-encoding unitary $\mathcal{B}[H]$.

\section{Discrete Variable Representation of the rotational-vibrational \SE equation} 
\label{sec:DVR}
The \SE equation given in the variational basis in eq.~\ref{eq:HVBR} can be transformed into DVR by first applying quadrature approximation to the PES matrix elements and transforming the resulting FBR equation to a new basis via the FBR-DVR unitary transformation matrix $\mathbf{T}$.  Below we sketch the derivation of the block-encoded Hamiltonian, with further details given in appendix~\ref{sec:appendix-DVR}. In the next section we present a quantum algorithm that block-encodes the DVR Hamiltonian in a unitary circuit.

The FBR Hamiltonian is obtained through quadrature approximation to the PES in the variational Hamiltonian, which can be written as:
\begin{equation}
\mathbf{H}^{\mathrm{VBR}} \approx \mathbf{H}^{\mathrm{FBR}}
= \mathbf{K}^{\mathrm{VBR}} + \mathbf{T}^{\mathrm{T}} \mathbf{V}^{\mathrm{DVR}} \mathbf{T},
\label{eq:FBR1D}
\end{equation}
where $\mathbf{K}^{\mathrm{VBR}}$ is the kinetic energy matrix in the variational basis and $\mathbf{V}^{\mathrm{DVR}}$ is the diagonal PES matrix evaluated on the quadrature grid, $V^{\mathrm{DVR}}_{ll'}=V(q_l)\delta_{ll'}$.
When Gaussian quadrature is used to construct the transformation matrix $\mathbf{T}$ in eq.~\eqref{eq:dvr-definition}, the resulting representation is referred to as the Gaussian FBR and takes the form
\begin{equation}
T_{kj} = \tilde{N}_j \sqrt{w_k}, p_j(q_k),
\qquad k,j = 0,\ldots,N-1,
\label{eq:dvr-definition}
\end{equation}
where $\tilde{N}_j = |p_j|^{-1}_{L^2(\mu)}$ is a normalization constant, $w_k$ are the Gaussian quadrature weights, and $p_j(q_k)$ denotes the degree-$j$ orthogonal polynomial defining the quadrature, evaluated at the quadrature node $q_k$ \cite{Stoer2002}. For rovibrational Schrödinger equation solvers, commonly used Gaussian quadratures are associated with orthogonal polynomials corresponding to analytically solvable model systems. These include Hermite polynomials for harmonic oscillators (bond stretching), Legendre polynomials for spherically symmetric problems (bending and torsional motion), Laguerre and associated Laguerre polynomials for hydrogenic and Morse oscillator potentials, and Lobatto polynomials for problems with fixed boundary conditions \cite{AbramowitzStegun1964,Light2000}. More general potential-optimized DVRs have also been developed \cite{Bowman2008}.

Within the FBR approximation, the Schrödinger equation can be cast in the generalized matrix eigenvalue problem form
\begin{equation}
\left( \mathbf{K}^{\mathrm{VBR}} + \mathbf{T}^{\mathrm{T}} \mathbf{V}^{\mathrm{DVR}} \mathbf{T} \right) \mathbf{U}
= \mathbf{T}^{\mathrm{T}} \mathbf{T}\mathbf{U}\mathbf{E},
\label{eq:FBR1DSE}
\end{equation}
where the overlap matrix on the right-hand side is also evaluated using the quadrature approximation,
\begin{equation}
\mathbf{S} \approx \mathbf{T}^{\mathrm{T}} \mathbf{T}.
\end{equation}
Embedding the MF aims to maximally decouple the overall $SO(3)$ rotation of the molecule from its internal quantum-mechanical degrees of freedom. In practice, this is achieved by defining Euler angles (and translationally invariant internal coordinates) and integrating over the three rotational coordinates using the complete Wigner $\mathcal{D}^{J}_{m,k}(\Omega)$ function basis. The resulting effective rovibrational Hamiltonian couples vibrational modes through rotational quantum numbers ($J$ and $k$, where $J$ is the total angular momentum and $k$ represents the projection onto the MF $z$-axis):
\begin{equation}
H^{J,i,\Gamma}_{\mathbf{n},\mathbf{n}',k,k'}
= \langle \mathbf{v}^{\Gamma}| \hat{h}_{kk'} |\mathbf{v'}^{\Gamma'}\rangle
 \delta_{JJ'}  \delta_{ii'}  \delta_{\Gamma\Gamma'},
 \label{eq:matelem-ham}
\end{equation}
where $|\mathbf{v}^{\Gamma}\rangle$ denotes a vibrational product state associated with a given irreducible representation of the molecular symmetry group. The resulting Hamiltonian representation is block-diagonal with blocks labeled by the total angular momentum quantum number $J$ for a given electronic state $i$ and symmetry species $\Gamma$. In the absence of extenral fields the size of each $J,i,\Gamma$ block is given by $N=(2J+1)N_{vib}$, where $N_{vib}$ is the size of the vibrational problem.

Upon multiplication of eq.~\ref{eq:FBR1DSE} from the left by $\mathbf{T}$ and defining a new basis by $\mathbf{Z}=\mathbf{T}\mathbf{U}$ we retain the DVR form of the \SE:
\begin{equation}
\left(\mathbf{T}\mathbf{K}^{VBR}\mathbf{T}^{\dag}+\mathbf{V}^{DVR}\right)\mathbf{Z}=\mathbf{Z}\mathbf{E}
\label{eq:DVR1DSE}
\end{equation}
We adopt a mixed DVR-FBR representation for the KEO given in eq.~\ref{eq:DVR1DSE} for the purposes of quantum-computing efficiency:
\begin{equation}
	\boldsymbol{K}^{DVR} = \boldsymbol{T}\sum_{i,j=1}^{D+3}(\boldsymbol{P}_i^{FBR})^\dagger \boldsymbol{T}^\dagger \boldsymbol{G}_{ij}^{DVR} \boldsymbol{T}\boldsymbol{P}_j^{FBR}\boldsymbol{T}^\dagger,
    \label{eq:KEODVR}
\end{equation}
where $\boldsymbol{P}_j^{FBR}$ is the FBR representation of the $j$-th momentum operator.

The FBR and DVR representations extend to many-dimensions with direct-product basis sets~\cite{Dawes2004}, in which case the $D$-dimensional basis functions have $D$ indices assigned as follows: 
\begin{equation}
    \lbrace \phi_{n_1,n_2,...,n_D} := \phi_{n_1}(q_1)\phi_{n_2}(q_2)...\phi_{n_D}(q_D)\rbrace_{n_c=1,2,...,N_c; c=1,..,D}.
\end{equation} 
Evaluation of multidimensional matrix elements with such a basis requires a suitable quadrature grid. A direct-product quadrature grid is obtained by combining one-dimensional grids such that the multidimensional indices remain independent. In this case, the multidimensional $\mathbf{T}$ is a Kronecker product of one-dimensional DVR transformation matrices,
\begin{equation}
\mathbf{T} = \mathbf{^{(1)}T} \otimes \mathbf{^{(2)}T} \otimes \dots \otimes \mathbf{^{(D)}T},
\end{equation}
with elements $\mathbf{^{(c)}T}_{k_c n_c} = \tilde{N}_{k_c}\sqrt{\mathbf{^{(c)}w}_{k_c n_c}}, \mathbf{^{(c)}p}_{k_c n_c}$, and $\mathbf{^{(c)}p}_{k_c n_c}=p_{n_{c}}(q_{k_c})$. Preserving the direct-product DVR structure is advantageous for quantum computation, given an efficient $D$-dimensional $\mathbf{T}$-unitary implementation, which can be realized in the depth of a 1-dimensional $\mathbf{T}$ transformation, the cost and construction of which is discussed in appendix~\ref{sec:appendix-DVR} and in ref.~\cite{plis25}.

\section{block-encoding the rovibrational Hamiltonian}
\label{sec:block encoding}
Following the DVR Hamiltonian formulation shown in Eq.~(\ref{eq:DVR1DSE}), we construct a general quantum circuit that block-encodes the corresponding Hermitian matrix into a unitary circuit, using block-encoded representations of its component operators. In doing so we do not assume any particular form for the QROM oracle. Further in Sec.\ref{sec:resource_estimation} we demonstrate advantages of utilizing a QROM constructed with the use of the Walsh-Hadamard transform, as introduced in Sec.\ref{sec:heuristics}.

\subsection{Qubit state encoding}
To encode the vibrational wavefunction in eq.~\eqref{eq:vib_wavefunction} on a quantum computer, we introduce $D$ qubit registers, one for each vibrational mode. Each register consists of $\eta_c = \lceil \log_2 n_c \rceil$ qubits, where $n_c$ is the number of basis functions for mode $c$. We denote the state of an $\eta$-qubit register by $\ket{\psi}_\eta$. For each mode, the excitation-ordered basis states $\ket{v_c}$ are mapped to qubit states using a binary encoding,
\begin{align}\label{eq:vib_state_mapping}
    \ket{v_c}_{\eta_c} = \bigotimes_{l=1}^{\eta_c}\ket{v_c^{(l)}}_1  && v_c = \sum_{l=1}^{\eta_c}v_c^{(l)}2^{\eta_c-l}
\end{align}
where $v_c^{(l)} \in \mathbb{F}_2$. The qubits are ordered such that the first qubit corresponds to the most significant bit. Rotational states are encoded analogously. Specifically, the states $\ket{J,k,m}$ are mapped to qubit registers of size $\eta_J = \lceil \log_2(2J+1) \rceil$ according to
\begin{align}
    \ket{J,k,m}_{\eta_J} = \bigotimes_{l=1}^{\eta_J}\ket{k^{(l)}}_1 && k = \sum_{l=1}^{\eta_J}k^{(l)}2^{\eta_J-l}-J
\end{align}
with $k^{(l)}\in \mathbb{F}_2$ and  $\eta_J = \lceil\log_2(2J+1)\rceil$. The full rovibrational state defined in eq.~\eqref{eq:full_sys_wavefunction} is therefore encoded using
\begin{equation}
    \eta_S = \sum_{c=1}^D \eta_c+\eta_J
\end{equation}
qubits and can be written as
\begin{equation}
    \ket{\Psi_{rv}^{(J,h,\Gamma, m,i)}}_{\eta_S} = \sum_{k=-J}^Ja_{v_1,\dots,v_D}^{(J,k,h,i)}\bigotimes_{c=1}^D\ket{v_c}_{\eta_c}\otimes\ket{J,k,m}_{\eta_J}.
\end{equation}

\subsection{Block-encoding the KEO}
\label{sec:BE_KEO}
In the following sections, we propose block-encoding circuit for the rovibrational KEO, by combining the linear combination of unitaries (LCU) and $\rho$-sparse methods~\cite{su2021}. The proposal is evaluated in subsequent sections, with particular attention to sensitivity of quantum resources to the choice of QROM.

\subsubsection{Block-encoding the KEO in a Mixed FBR–DVR Representation}
The mixed FBR-DVR KEO matrix introduced in eq.~(\ref{eq:KEODVR}) can be expressed as~\cite{Avila2019}: 
\begin{equation}
	\boldsymbol{K}^{DVR} = \boldsymbol{T}\sum_{i,j=1}^{D+3}(\boldsymbol{P}_i^{FBR})^\dagger \boldsymbol{T}^\dagger \boldsymbol{G}_{ij}^{DVR} \boldsymbol{T}\boldsymbol{P}_j^{FBR}\boldsymbol{T}^\dagger,
\end{equation}
where $\boldsymbol{T}$ is the unitary DVR transformation, $\boldsymbol{P}_i^{FBR}$ denotes the FBR representation of the momentum operator, and $\boldsymbol{G}_{ij}^{DVR}$ is a diagonal matrix containing the $G$-matrix elements evaluated at the DVR grid points. Alternative representations, such as DVR and FBR alone are less favorable in general and are discussed in appendix~\ref{sec:appendix-BE-DVR-alone} and compared to our best method in sec.~\ref{sec:resource_estimation}.

Our approach assumes a LCU block-encoding of the Hamiltonian, where each component operator is block-encoded using the $\rho$-sparse model. The role of the DVR is to enable sparse matrix representations and thereby facilitate efficient use of the $\rho$-sparse framework. We first describe the LCU layer of the block-encoding circuit.

Consider a matrix expressed as a sum of components,
\begin{equation}
    \boldsymbol{A} = \sum_{k=0}^{K-1}\boldsymbol{A}^{(k)}.
\end{equation}
If block-encodings $\BE{\boldsymbol{A}^{(k)}}$ are available for each term, a block-encoding of $\boldsymbol{A}$ can be constructed as
\begin{equation}\label{eq:Block_enoding_sum}
    \BE{\boldsymbol{A}} = \sum_{k=0}^{K-1}\ket{k}\bra{k}_{a'} \otimes \BE{\boldsymbol{A}^{(k)}},
\end{equation}
where
\begin{equation}\label{eq:BE_sum}
    \left(\id_{a'}\otimes\bra{0}_a\otimes\id_\eta\right)\BE{\boldsymbol{A}^{(k)}}\left(\id_{a'}\otimes\ket{0}_a\otimes\id_\eta\right) = \id_{a'}\otimes\frac{\boldsymbol{A}^{(k)}}{\zeta_k}.
\end{equation}
Here $a$ denotes the ancilla register used for block-encoding $\boldsymbol{A}^{(k)}$, which acts on $\eta$ system qubits, and $\zeta_k$ is the corresponding normalization constant. The projector $\ket{k}\bra{k}_{a'}$ acts on an additional ancilla register of size $a'=\lceil\log_2 K\rceil$. If $K$ is not a power of two, the sum in Eq.~\eqref{eq:Block_enoding_sum} must be augmented with identity terms to ensure unitarity; these do not affect the effective action and are omitted for clarity.

To complete the LCU construction, the following ancilla state is prepared:
\begin{equation}\label{eq:ancila_state_G}
    \boldsymbol{G}\ket{0}_{a'} = \ket{G}_{a'} = \sum_{k=0}^{K-1}\sqrt{\frac{\zeta_k}{\zeta}}\,\ket{k}_{a'},
\end{equation}
where $\zeta = \sum_{k=0}^{K-1} \zeta_k$. The resulting block-encoding satisfies
\begin{equation}
    \left(\bra{G}_{a'}\otimes\bra{0}_a\otimes\id_{\eta}\right)\BE{\boldsymbol{A}}\left(\ket{G}_{a'}\otimes\ket{0}_a\otimes\id_{\eta}\right) = \frac{\boldsymbol{A}}{\zeta}.
\end{equation}

\begin{figure}[!hb]
    \centering
    \includegraphics[width=0.7\linewidth]{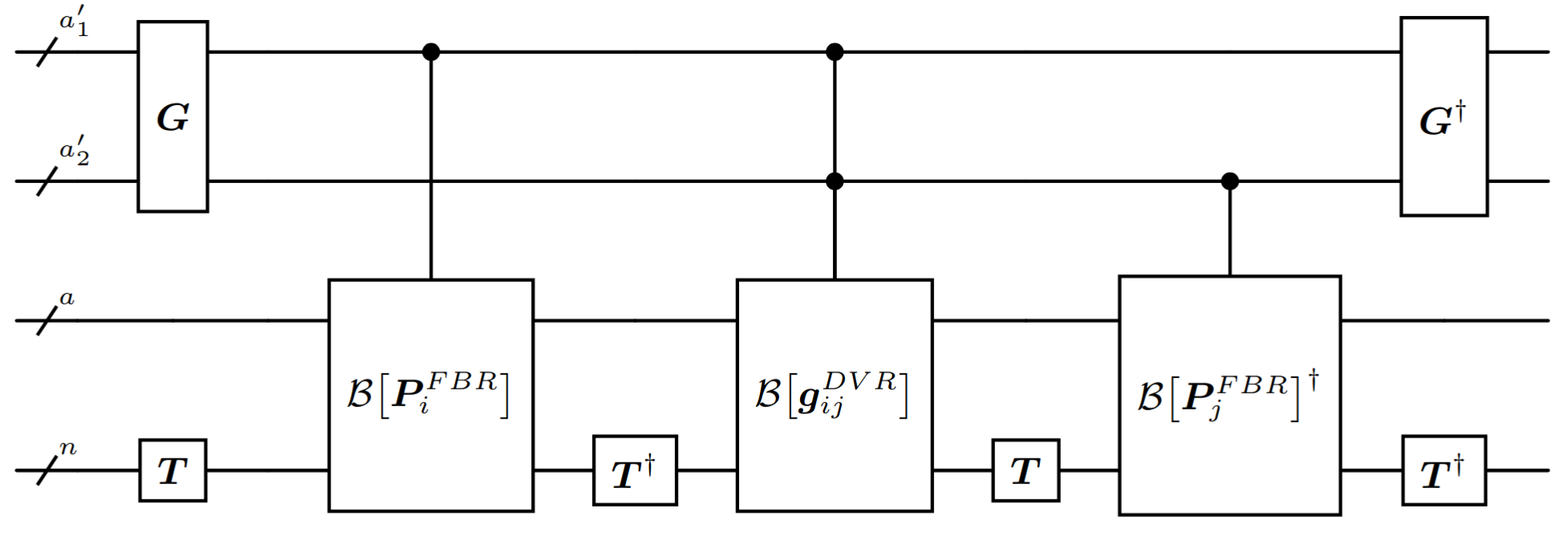}
    \caption[]{Circuit block-encoding the KEO in DVR-FBR representation as given in \ref{eq:KEODVR}. The controlled version of $ \BE{\boldsymbol{P}^{FBR}_i}$, $ \BE{\boldsymbol{G}^{DVR}_{ij}}$ and $ \BE{\boldsymbol{P}^{FBR}_j}^\dag$ means that if register $a'$ is in state $\ket{i-1}_{a'_1}\ket{j-1}_{a'_2}$ the block-encoding of $\boldsymbol{P}^{FBR}_i$, $\boldsymbol{G}_{ij}^{DVR}$ and $\left(\boldsymbol{P}_j^{FBR}\right)^\dag$ is executed. Note that here $\boldsymbol{T}$ denotes the DVR unitary and not the $T=Z^{\frac{1}{4}}$ gate.}
    \label{fig:BE_KEO_separate_citcuit}
\end{figure}

This procedure can be applied directly to the kinetic energy operator in Eq.~\eqref{eq:KEODVR}. Let $\boldsymbol{G}$ denote the unitary preparing the ancilla state in Eq.~\eqref{eq:ancila_state_G}. A block-encoding of $\boldsymbol{K}^{\mathrm{DVR}}$ is then given by
\begin{equation}\label{eq:BE_KEO_separate}
	\BE{\boldsymbol{K}^{DVR}} = \left(\boldsymbol{G}^\dagger \otimes \id_{a+\eta}\right) \boldsymbol{T}\,\boldsymbol{U}_P\,\boldsymbol{T}^\dagger \boldsymbol{U}_G \boldsymbol{T}\,\boldsymbol{U}_{P}' \boldsymbol{T}^\dagger \left(\boldsymbol{G}\otimes\id_{a+\eta}\right),
\end{equation}
with
\begin{align}\label{eq:BE_separate_U_P}
	\boldsymbol{U}_P &= \sum_{i=0}^{3A-4}\ket{i}\bra{i}_{a'_{1}}\otimes\id_{a'_{2}}\otimes \BE{\boldsymbol{P}_{i+1}^{FBR}}^\dagger, &
	\boldsymbol{U}_P' &= \sum_{j=0}^{3A-4}\id_{a'_{1}}\otimes\ket{j}\bra{j}_{a'_{2}}\otimes \BE{\boldsymbol{P}_{j+1}^{FBR}},
\end{align}
and
\begin{equation}\label{eq:BE_separate_U_G}
	\boldsymbol{U}_G = \sum_{i,j=0}^{3A-4}\ket{i}\bra{i}_{a'_{1}}\otimes\ket{j}\bra{j}_{a'_{2}}\otimes \BE{\boldsymbol{G}^{DVR}_{i+1,j+1}}.
\end{equation}
Here $\eta$ denotes the number of qubits encoding the rovibrational state in Eq.~\eqref{eq:full_sys_wavefunction}. Exploiting the symmetry of the metric tensor, Eq.~\eqref{eq:BE_separate_U_G} can be rewritten as
\begin{equation}\label{eq:BE_separate_U_G_reduced}
	\boldsymbol{U}_G = \sum_{i=0}^{3A-4}\ket{i}\bra{i}_{a'_{1}}\otimes\ket{i}\bra{i}_{a'_{2}}\otimes \BE{\boldsymbol{G}^{DVR}_{i+1,i+1}}
    + \sum_{i>j}^{3A-4}\left(\ket{i}\bra{i}_{a'_{1}}\otimes\ket{j}\bra{j}_{a'_{2}}
    + \ket{j}\bra{j}_{a'_{1}}\otimes\ket{i}\bra{i}_{a'_{2}}\right)\otimes \BE{\boldsymbol{G}^{DVR}_{i+1,j+1}},
\end{equation}
so that only the independent components of $\boldsymbol{G}^{\mathrm{DVR}}_{ij}$ require explicit block-encoding. The corresponding circuit is shown in Fig.~\ref{fig:BE_KEO_separate_citcuit}.
A block-encoding of the full Hamiltonian is obtained by extending $\boldsymbol{U}_G$ to include the PES contribution,
\begin{equation}\label{eq:Separate_method_circuit}
    \tilde{\boldsymbol{U}}_{G} = \sum_{ij=0}^{3A-4}\ket{i}\bra{i}_{a'_{1}}\otimes\ket{j}\bra{j}_{a'_{2}}\otimes  \BE{\boldsymbol{G}^{DVR}_{i+1,j+1}}+\ket{3A-3}\bra{3A-3}\otimes\ket{3A-3}\bra{3A-3}\otimes\BE{\boldsymbol{V}^{DVR}}
\end{equation}    
Details of an alternative method for block-encoding the KEO, which relies fully on the DVR or FBR representation, are given in appendix~\ref{sec:appendix-BE-DVR-alone}. Both methods are compared in Sec.~\ref{sec:resource_estimation}.

\subsubsection{Implementation of partial block-encodings}
\label{sec:block_encoding_reducing_costs}
In the general case, we construct all partial block-encodings $\BE{\boldsymbol{P_i}^{FBR}}$ and $\BE{\boldsymbol{G_{ij}}^{DVR}}$ directly using the $\rho$-sparse method~\cite{su2021}. Howver further resource reductions can be achieved when the metric tensor and the momentum operators are block-encoded with different algorithms. Accordingly, the cost of implementing $\BE{\boldsymbol{G}_{ij}^{DVR}}$ can be significantly lowered if the metric tensor is decomposed into a sum of products from 
\begin{equation}\label{eq:metric_tensorm_sop}
    \boldsymbol{G}_{ij}(\boldsymbol{q}) = \sum_{k=1}^{K_{ij}}\prod_{l=1}^{D}\boldsymbol{h}_{ij}^{(k,l)}(q_l)
\end{equation}
with $K_{ij} = \mathcal{O}(1)$. Since the matrices $\boldsymbol{h}^{(k,l)}_{ij}(q_l)$ depend only on a single coordinate, they can be efficiently block-encoded using the $\rho$-sparse method with the cost of $\mathcal{O}(\sqrt{n_i})$ T-gates and ancillas. Each term $\prod_{l=1}^{D}\boldsymbol{h}_{ij}^{(k,l)}(q_l)$ can be block-encoded as product of operators. Block-encoding for the metric tensor elements $\BE{\boldsymbol{G}_{ij}^{DVR}}$ can then be constructed using the LCU technique for block-encoding sums of operators. 

Another consequence of using DVR for the metric tensor and FBR for the momentum operator (mixed FBR–DVR), is lowered value of the block-encoding normalization constant $\zeta_{P_i} = \rho_i \lVert \boldsymbol{P}_i \rVert_{\max}$, at the cost of additional $\tilde{\mathcal{O}}(n_i^{2})$ Clifford+T gates and ancilla qubits. As discussed in Appendix.~\ref{sec:comment_about_norms}, for $\rho$-sparse encodings the block-encoding constant is minimized by the operator's eigenbasis. 
It is therefore favorable to apply a spectral decomposition of the momentum operator, $\boldsymbol{P}_i = \boldsymbol{U}_i\boldsymbol{E}_i\boldsymbol{U}_i^\dag$, where the diagonal matrix $\boldsymbol{E}_i$ can be block-encoded using the $\rho$-sparse method, achieving the optimal normalization. The unitaries $\boldsymbol{U}_i$ and $\boldsymbol{U}_i^\dagger$ can be implemented using a general unitary synthesis procedure with at most $\tilde{\mathcal{O}}(n_i^{2})$ gates.
If $\boldsymbol{P}_i$ is non-Hermitian or non-diagonalizable, an analogous construction based on singular value decomposition can be used, giving the same asymptotic resource scaling.

\subsection{$\rho$-sparse block-encoding with fused oracle}\label{sec:rho-sparse}
The reductions in logical qubit count and T-gate complexity achieved by our method originate from two complementary sources: (i) a structured partitioning of the Hamiltonian combined with a hybrid FBR–DVR representation of selected terms, and (ii) the integration of our newly introduced QROM construction. The resource and performance gains coming from the first component can be directly benchmarked against FBR-based representations, while those due to the second are compared against the SELECT–SWAP QROM technique~\cite{low2024}.

Below, we introduce a $\rho$-sparse block-encoding model that offers distinct advantages when combined with the mixed FBR–DVR representation of the kinetic energy operator introduced in Sec.~\ref{sec:BE_KEO}, particularly when combined with the specialized QROM model presented in Sec.~\ref{sec:whqrom}.

Given a matrix $\boldsymbol{A}$ acting of $n$-dimensional space let $\BE{\boldsymbol{A}}$ be a block-encoding circuit
\begin{equation}
    \left(\bra{0}_{a}\otimes \id_{\eta}\right)\BE{\boldsymbol{A}}\left(\ket{0}_{a}\otimes \id_{\eta}\right) = \frac{\boldsymbol{A}}{\zeta_A},
\end{equation}
where $\ket{0}_{a}$ denotes the state of ancilla register with $a$ qubits, $ \eta= \log_2n$ is the number of system qubits and $\zeta_A$ is the block-encoding scaling constant. The $\rho$-sparse block-encoding model assumes that the matrix representation of $\boldsymbol{A}$ has at most $\rho$ non-zero elements in each row. This approach requires implementing 2 oracles~\cite{Camps:2024}:
\begin{equation}\label{eq:dSpares_oracles1}
    O_A\ket{0}_{\eta_a}\ket{j}_{\eta_\rho}\ket{i}_\eta= \left({\frac{A_{s(i,j)i}}{||A||_{max}}}\ket{0}_{\eta_a}+\ket{\perp}_{\eta_a}\right)\ket{j}_{\eta_\rho}\ket{i}_\eta,
\end{equation}
and
\begin{align}\label{eq:dSparse_oracles}
     O_F\ket{j}_{\eta_\rho}\ket{i}_\eta = O_F\ket{j}_{\eta_\rho}\ket{s(i,j)}_\eta,
\end{align}
where $s(i,j)$ describes the index of $j$-th non-zero element of the $i$-th row of $A$, $\eta_\rho~=~\lceil\log_{2}{\rho} \rceil$, $\ket{\perp}_{\eta_a}$ is an arbitrary state orthogonal to $\ket{0}_{\eta_a}$ and $||A||_{max}=\max{A_{ij}}$. The number of ancilla qubits ${\eta_a}$ depends on the implementation method. The block-encoding circuit can then be constructed as \cite{Camps:2024}
\begin{equation}
  \BE{\boldsymbol{A}} =(\id_{\eta_a}\otimes HAD^{\otimes \eta_\rho} \otimes \id_\eta)(\id_{\eta_a}\otimes O_F)O_A(\id_{\eta_a}\otimes HAD^{\otimes \eta_\rho}\otimes \id_\eta)  
\end{equation}
where $HAD$ denotes the Hadamard gate.
 For diagonal matrices, the block-encoding circuit can be simplified to  $\BE{\boldsymbol{A}}=O_A'$ with
\begin{equation}\label{eq:dSpares_oracles_p}
    O_A'\ket{0}_{\eta_a}\ket{i}_\eta = \left(\frac{A_{ii}}{||A||_{max}}\ket{0}_{\eta_a}+\ket{\perp}_{\eta_a}\right)\ket{i}_\eta
\end{equation}
Note that by considering indices $j$ and $i$ in Eq.~(\ref{eq:dSpares_oracles1}) as a single index $k = i+2^\eta j$, oracle $O_A$ can also be thought of as a block-encoding a diagonal matrix $\mathcal{A}_{kk'} = A_{s(i,j)i}\delta_{kk'}$. 
Oracles $O_A$ and $O_F$ can be constructed with QROM. As we demonstrate in Sec.~\ref{sec:resource_estimation}, a substantial reduction in both T-count and ancillary qubits can often be achieved with WH-QROM introduced in Sec.~\ref{sec:whqrom}.
We consider and compare two methods of block-encoding diagonal matrices used for implementing $O_A$ and $O_A'$: via multiplexed rotation and by means of a position-like operator, here referred to as the \textit{canonical operator}.

\subsubsection{Multiplexed rotation method}
\label{sec:multiplexed_rotations}
The simplest approach to block-encoding a diagonal matrix $\boldsymbol{D}$ involves multiplexed rotation, which can be expressed in the general form  as
\begin{equation}\label{eq:multiplexed_rot}
\cR_f \ket{a}_1 \ket{x}_\eta
=
\left(
R_Y \left( \tfrac{2\pi}{2^d} f(x) \right)
\ket{a}_1
\right)
\ket{x}_\eta ,
\end{equation}
where $f : \mathbb{F}_2^\eta \to \mathbb{Z}$ and $R_Y(\theta) = \exp(-i \theta Y / 2)$. Let 
\begin{equation}\label{eq:multiplexed_fun_1}
    f_D(x) = 2\left[\frac{2^{d}}{2\pi}\arccos{\left({\frac{D_x}{||D||_{max}}}\right)}\right]
\end{equation} 
with $[x]$ - closest integer to $x$. Then $\cR_{f_D}$ provides an approximation of order $O(2^{-d})$ for $\BE{\boldsymbol{D}}$.
The multiplexed rotation given in Eq.~(\ref{eq:multiplexed_rot}) can be constructed as follows. Let $U_f$ denote a QROM oracle implementing
\begin{equation}\label{eq:qrom}
U_f \ket{x}_\eta \ket{y}_d \ket{z}_\eta
=
\ket{x}_\eta \ket{y + f(x)}_d \ket{z}_\eta .
\end{equation}
Diagonal unitaries can then be realized efficiently using phase kickback. In particular,
\begin{equation}
	U_f \ket{x}_\eta (\QFT_{\eta+d} \ket{-1}_{\eta+d}) = \exp \( \tfrac{2 \pi i}{2^d} f(x) \) \ket{x}_\eta (\QFT_{\eta+d}\ket{-1}_{\eta+d}).
\end{equation}
where $\ket{-1}_{\eta+d} = \bigotimes_{k=1}^{\eta+d}\ket{1}_1$. This realizes the diagonal unitary  $\cD_f \ket{x}_\eta := \exp ( \tfrac{2 \pi i}{2^d} f(x) ) \ket{x}_\eta$
with a single call to $U_f$, achieving an approximation error of order $O(2^{-d})$.
Using this construction, the multiplexed rotation can be written as
\begin{equation}
	\cR_f = \( \( S^\dagger H \) \otimes \id_{2\eta+d} \) \circ \cD_F \circ \( \( H S \) \otimes \id_{2\eta+d} \),
\end{equation}
where $F(a,x) = (-1)^a f(x)$. Defining the circuit $\widetilde{\cR}_f$ shown in Eq.~(\ref{circ:R-diag}), 
\begin{equation}
	\begin{quantikz}
		\lstick{$\ket{a}_1$}		        &	\qwbundle{1}	&	\gate{S}	&	\gate{H}	&	\ctrl{2}	&					&	\ctrl{2}	&	\gate{H}	&	\gate{S^\dagger}	&	\qw	\\
		\lstick{$\ket{x}_\eta$}		        &	\qwbundle{\eta}	&	\qw			&	\qw			&	\qw			&	\gate[2]{U_f}	&	\qw			&	\qw			&	\qw					&	\qw	\\
		\lstick{$\QFT_{\eta+d} \ket{{-1}}$}	&	\qwbundle{\eta + d}	&	\qw			&	\qw			&	\targ{}		&					&	\targ{}		&	\qw			&	\qw					&	\qw
	\end{quantikz}
    \label{circ:R-diag}
\end{equation}
one readily verifies that
\begin{equation}
	\widetilde{\cR}_f \ket{a}_1 \ket{x}_\eta (\QFT_{\eta+d} \ket{-1}_{\eta+d}) = (\cR_f \ket{a}_1 \ket{x}_\eta) (\QFT_{\eta+d} \ket{-1}_{\eta+d}).
\end{equation}
Note that, when using the phase-kickback method for diagonal unitary synthesis, only a single call to the QROM is required, in contrast to two calls in the general construction. A potential drawback of this approach, however, is the need to encode the functions Eq.~(\ref{eq:multiplexed_fun_1}) which have unbounded derivatives. For QROM implementations that perform poorly on such data structures, an alternative block-encoding strategy for diagonal matrices may be preferable; this is discussed in the next section.

\subsubsection{Canonical operator method}
\label{sec:be_without_rotations}
Standard methods for block-encoding diagonal operators via multiplexed rotations~\cite{low_trading_2024,low_halving_2021} require QROM for elements of the form $\arccos{\frac{D_x}{||\boldsymbol{D}||_{max}}}$, which is not optimal for use with certain QROMs, such as the one introduced in sec.~\ref{sec:whqrom} (WH-QROM), due to unbounded derivatives. Instead, we introduce an algorithm for implementing block-encoding diagonal matrices without the need to load values of $\arccos{x}$, adopted for specifically efficient use with our WH-QROM. To overcome the problem of divergent derivative, we directly implement QROMs defined by Eq.~(\ref{eq:qrom}) corresponding to the following functions 
\begin{equation}
    g_{D}(x) = \left[(2^d-1)\frac{D_x}{||D||_{max}} \right]
\end{equation}
Now, consider the  matrix $\boldsymbol{N}$ defined by $N_{ij} = (j-1)\delta_{ij}$ such that
\begin{equation}
    \boldsymbol{N}\ket{x}= x\ket{x}
\end{equation}
with $\ket{x} = \bigotimes_{b-1}^d\ket{x_b} $, $ x = \sum^d_{b=1}x_b2^{d-b}$ and $x_b \in \mathbb{F}_2$. By direct calculation one finds

\begin{equation}
    \bra{0}_d\bra{j}_\eta U_{g_D}^\dag \boldsymbol{N} U_{g_D}\ket{0}_d\ket{i}_\eta = \left[(2^d-1)\frac{D_{i}}{||D_{max}||}\right]\delta_{ij}
\end{equation}
However, since $\boldsymbol{N}$ is not unitary the matrix needs to be block-encoded leading to the modified expression
\begin{equation}\label{eq:canonica_block_encoding_diag}
    \bra{0}_{\eta_N}\bra{0}_{d}\bra{j}_{\eta} U_{g_D}^\dag\BE{\boldsymbol{N}}U_{g_{D}}\ket{0}_{\eta_{N}}\ket{0}_d\ket{i}_\eta = \left[(2^d-1)\frac{D_{i}}{||D||_{max}}\right]\frac{\delta_{ij}}{2^d-1}
    \end{equation}
assuming the block-encoding constant is $||\boldsymbol{N}||_{max}=2^d-1$. Here $\eta_N$ in  number of qubits required to block-encode $\boldsymbol{N}$. Eq.~\ref{eq:canonica_block_encoding_diag} shows that the circuit $U_{g_D}^\dag \BE{N}U_{g_D}$ provides an approximation of order $\mathcal{O}(2^{-d})$ of $\BE{\boldsymbol{D}}$. The canonical operator can be written as:
\begin{equation}\label{A_as_sum}
    \boldsymbol{N} = \sum_{b=1}^d 2^{d-b}\boldsymbol{\Pi}_b
\end{equation}
where $\boldsymbol{{\Pi}}_a$ are single qubit projector operators defined as $\boldsymbol{{\Pi}}_b = \bigotimes_{r=1}^{b-1}\id\otimes \ket{1}\bra{1}\bigotimes_{r=b+1}^d \id $.
The form of eq.~\ref{A_as_sum} follows from the property of single qubit projector operators:
\begin{equation}
    \boldsymbol{{\Pi}}_b\ket{x}_d = x_b\ket{x}_d 
\end{equation}
Using identity $\ket{1}\bra{1} = \frac{1}{2}(I-Z)$, eq.~\ref{A_as_sum} can be rewritten as
\begin{equation}\label{LUC_A}
    \boldsymbol{N} = \frac{1}{2}\sum_{b=1}^d 2^{d-b} (I-Z_b) = \frac{1}{2}(2^d-1)I+\sum_{b=1}^d 2^{d-b-1} (-Z_b)
\end{equation}
where $Z_b= \bigotimes_{r=1}^{b-1}I\otimes Z \bigotimes_{r=b+1}^d I$.
The right-hand side of eq.~\ref{LUC_A} is a decomposition of $\boldsymbol{N}$ into LCU. This leads to the block-encoding $B[\boldsymbol{N}]$ with $\zeta = 2^d-1$, which requires $O(d)$ multi-controlled $Z$'s controlled on $O(\log_2(d))$ ancilla qubits. Therefore, the total $T$-cost is 
\begin{equation}
    C_T = \mathcal{O}(d\log_2d)+\mathcal{O}\left(\sqrt{d\log_2\frac{1}{\epsilon}}\right)
    \label{eq:diag-cost}
\end{equation}
with, $\epsilon <<2^{-d}$. The second term in Eq.~(\ref{eq:diag-cost}) is the cost of preparing the state of ancillas.

\subsection{Walsh-Hadamard QROM}
\label{sec:whqrom}
Both block-encoding methods for diagonal operators discussed in the previous section can be used, with the appropriate choice depending on the structure of the matrix (KEO, DVR PES) and the specific QROM implementation. Below, we introduce a new QROM construction that offers substantial advantages over the SELECT–SWAP QROM approach when used in conjunction with either method.

\subsubsection{Implementation for general functions}
QROM using the Walsh-Hadamard transform (WH-QROM) can be constructed as follows. Let $f : \F_2^\eta \rightarrow \Z$ and
\begin{equation}
	\WH (f) (z) \eqdef \sum\limits_{x \in F_2^\eta} (- 1)^{x \odot z} f(x),
\end{equation}
be its Walsh-Hadamard transform. If $f$ can be represented on $d$ bits, that is, if $- 2^{d - 1} \leqslant f(x) < 2^{d - 1}$, for all $x$, then $\WH (f)$ can be represented on (at most) $b \eqdef \eta + d$ bits. Furthermore, $\WH^2 (f) = 2^\eta f$, or, equivalently,
\begin{equation}
	2^\eta f(x) \eqdef \sum\limits_{z \in F_2^\eta} (- 1)^{x \odot z} \WH (f) (z),
\end{equation}

Let $z \in \F_2^\eta$ and define the \textit{$z$-parity-controlled fanout-$X$} gate, or $\PFX_z$, as the following quantum circuit on $\eta + d$ qubits:
\begin{equation}
	\PFX_z \ket{x}_\eta \ket{y}_b = \left\{ \begin{array}{ll} \ket{x}_\eta \ket{y}_b, & \mbox{if } x \odot z = 0, \\ \ket{x}_\eta X^{\otimes d} \ket{y}_b, & \mbox{if } x \odot z = 1. \end{array} \right.
\end{equation}
These gates can be implemented using only CNOT gates. Their depth can be made as shallow as $2 \log_2 (h(z) b) + O(1)$ while using only $2 (h(z) + b) + O(1)$ gates. Note that $X^{\otimes b} \ket{y}_b = \ket{- y - 1}_b$. Let $k \in \Z$ and $A_b (k)$ be a $b$-qubit, $k$-adder oracle, that is, $A_b (k) \ket{y}_b = \ket{y + k \: \( \mathrm{mod} \: 2^b \)}_b$. Now define $W_{f, z} \eqdef \PFX_z \circ \( \id_n \otimes A_b \( \WH (f) (z) \) \) \circ \PFX_z$. Then, following \cite{nagy_novel_2024}, we get
\begin{equation}
	W_{f, z} \ket{x}_\eta \ket{y}_b = \ket{x}_\eta \ket{y + (- 1)^{x \odot z} \WH (f) (z)}_b.
\end{equation}
Thus, let
\begin{equation}
	U_f \eqdef \prod_{z \in \F_2^\eta} W_{f, z}. \label{eq:qrom_def}
\end{equation}
Then $U_f \ket{x}_\eta \ket{y}_d = \ket{x}_\eta \ket{y + 2^\eta f(x)}_b$. Note that $\left[ W_{f, z_1}, W_{f, z_2} \right] = 0$, and so the order on the left-hand side does not matter. Furthermore, $\PFX_{z_2} \circ \PFX_{z_1} = \PFX_{z_1 \oplus z_2}$, thus one can typically cancel out some amount of CNOT gates, especially when $h(z_1 \oplus z_2) \leqslant h(z_1) + h(z_2)$ is small, for example, in the case of Grey codes.

We can rewrite the action of $U_f$ as follows: for any $x, z \in \F_2^n$ and $y \in \F_2^b$, we have
\begin{equation}
	U_f \ket{x}_\eta \ket{y}_d \ket{z}_\eta = \ket{x}_\eta \ket{y + f(x)}_d \ket{z}_\eta.
\end{equation}
In other words, the first $\eta + d$ qubits implement the QROM associated with $f$, using $\eta$ ancillas that can be dirty.
We remark that there exist natural generalizations of the above design where ancillas can be traded for depth reduction. For example, if we $\supp \( WH(f) \) = S_1 \prod S_2$, where $S_1$ and $S_2$ are approximately the same size, and
	\begin{equation}
		f_i (x) \eqdef \tfrac{1}{2^\eta} \sum\limits_{z \in S_i} (- 1)^{x \odot z} \WH (f) (z),
	\end{equation}
	then the above construction can be used to create a QROM implementing
	\begin{equation}
		\ket{x}_\eta \ket{y}_b \ket{0}_b \mapsto \ket{x}_\eta \ket{y + 2^\eta f_1 (x)}_b \ket{2^\eta f_2 (x)}_b,
	\end{equation}
	with (essentially) the same amount of gates but at (essentially) half the depth. This, composed with an adder circuit (adding $2^\eta f_2 (x)$ to $y + 2^\eta f_1 (x)$), realizes the circuit at the same gate cost, up to an $O(1)$ difference.

\subsubsection{An algorithm for approximate QROMs using the Walsh--Hadamard transform}
\label{sec:heuristics}
Having discussed in the previous section a generic method for implementing QROM with the WH transform, in this section, we describe the method by which our QROMs are designed. The inputs are:

\begin{itemize}

	\item A function, $\theta : \F_2^\eta \rightarrow \left[ - 1, 1 \right)$.

	\item A binary precision, $d \in \Z_+$.

	\item An error bound, $\epsilon > 0$.

\end{itemize}

The number $d$ specifies the rounding precision for $\theta$, already on the classical level. More precisely, the function $f = f_{\theta, d} : \F_2^\eta \rightarrow \Z \cap \left[ - 2^{d - 1}, 2^{d - 1} \right)$, which we use in the QROM design, is given by $f (x) \eqdef \lfloor 2^{d - 1} \theta (x) \rfloor$. Equivalently,
\begin{equation}
	\theta (x) = \tfrac{f (x)}{2^{d - 1}} + \tfrac{\rho (x)}{2^d},
\end{equation}
where $0 \leqslant \rho (x)< 1$. Let $\theta_d (x) \eqdef \tfrac{f (x)}{2^{d - 1}}$.

The error bound, $\epsilon > 0$, deserves a bit more explanation as well. We note that our QROM is not an approximation of the exact QROM of $f$, up to some error in norm (such as operator or diamond), but rather an implementation of $U_g$ for
\begin{equation}
	g : \F_2^\eta \rightarrow \( \Z / 2^\eta \) \cap \left[ - 2^{d - 1}, 2^{d - 1} \right),
\end{equation}
so that, $\chi (x) \eqdef \tfrac{g (x)}{2^{d - 1}}$, is close to $\theta_0 (x)$, numerically. Since one of the most common usecases for QROM in this work is diagonal unitary synthesis, as discussed in sec.~\ref{sec:multiplexed_rotations}, we quantify errors via
\begin{equation}
	\epsilon \( f, g \) \eqdef \| \cD_{\pi \theta_0} - \cD_{\pi \chi} \|,
\end{equation}
where $\| \cdot \|$ is the operator norm. This expression can be simplified as
\begin{equation}
	\epsilon (f, g) = 2 \left\| \sin \( \pi \( \theta_0 - \chi \) \) \right\|_{L^\infty \( \F_2^\eta \)} = 2 \left\| \sin \( \tfrac{2 \pi}{2^d} \( f - g \) \) \right\|_{L^\infty \( \F_2^\eta \)}.
\end{equation}
For block-encoding of rovibrational FBR–DVR Hamiltonians, the precision parameter $d$ corresponds to the accuracy with which the Hamiltonian matrix elements are represented, while the error bound $\epsilon$ is set by the desired accuracy of the computed energy levels. A detailed discussion of the relationship between matrix element errors and eigenvalue accuracy in QPE-based algorithms is provided in the Supplementary Information of Ref.~\cite{kamakari25}.

With these inputs, we construct:
\begin{itemize}

	\item A function, $g : \F_2^\eta \rightarrow \( \Z / 2^\eta \) \cap \left[ - 2^{d - 1}, 2^{d - 1} \right) \cong \F_2^{\eta + d}$, such that $\epsilon \( f, g \) \leqslant \epsilon$.

	\item A QROM for $g$, that is, an oracle on $2 \eta + d$ qubits, $U_g$, such that
	\begin{equation}
		U_g \ket{x}_\eta \ket{y}_{\eta + d} = \ket{x}_\eta \ket{y + 2^\eta g(x)}_{\eta + d},
	\end{equation}
	with at most $\eta + d$ further ancillas.

\end{itemize}
Our proposal can be summarized in the following steps:
\begin{center}
    \fbox{\begin{minipage}{0.95\linewidth}
        \begin{center} \textbf{QROM construction with inputs $\theta$, $d$, and $\epsilon$} \end{center}
        \begin{enumerate}
			\item Compute $f \eqdef \lfloor 2^{d - 1} \theta \rfloor$.
			\item Compute $\hat{f} \eqdef \WH(f)$.
			\item Let $\hat{g}_k$ be equal to $\hat{f}$ at the $k$ largest components (with respect to their absolute value), otherise zero, and let $k_{f, \epsilon}$ be the smallest integer for which $\epsilon \( f, \WH^{- 1} \( \hat{g}_{k_{f, \epsilon}} \) \) < \epsilon$.
			\item Let $g \eqdef \WH^{- 1} \( \hat{g}_{k_{f, \epsilon}} \)$.
			\item Construct $U_g$ as in sec.~\ref{app:pair_cancellation} using Gidney's adder.
		\end{enumerate}
    \end{minipage}}
\end{center}

\begin{remark}
    The (time and space) complexity of Step 1. is $O \( 2^n \)$. The complexities of Steps 2--4 are all $O \( n \cdot 2^n \)$. That is because the fast Walsh--Hadamard transform can be performed via $O \( n \cdot  2^n \)$ additions and Step 3 can be performed via sorting the absolute values of $\hat{f}$, taking $O \( n \cdot 2^n \)$ time, then applying a binary search to find $k_{f, \epsilon}$. The binary search would take $O \( \log_2 \( 2^n \) \) = O (n)$ rounds and each round we perform an increasingly sparser Walsh--Hadamard transform. Since spare Walsh--Hadamard transform, with sparsity $O \( 2^k \)$, for some $k < n$, can be performed in $O \( k (n - k) 2^k \)$ time, we get that the total time of this binary search is also $O \( n 2^n \)$.
\end{remark}

Note that the above construction can immediately be applied to create (approximate) diagonal unitaries and multiplexed rotations. The idea of using (truncated) Walsh--Hadamard transforms for (diagonal) unitary synthesis is not new; cf.~\cite{welch_efficient_2014,welch,Camps2022,sun_asymptotically_2023,Henderson2023,zylberman_efficient_2025}, in particular, \cite{zylberman_efficient_2025}*{Theorem IV.1.}. However, our usage is more adapted to efficient fault tolerant, more precisely, Clifford + T decomposition, as shown later in the paper. Other methods used Clifford + $T$ decompositions of single-qubit rotations for implementing the addition of the Walsh--Hadamard components. We found that our phase-kickback method yields better results, in terms of $T$ gate counts while resulting in only a modest, $O(\eta + d)$ increase in ancilla qubits. This is further amplified if diagonal unitaries are applied multiple times in an algorithm, as the $\QFT_{n + d} \ket{- 1}_{\eta + d}$ state can be reused.

\section{Estimating eigenvalues of rotational-vibrational Hamiltonians using quantum phase estimation}
\label{sec:IV-QPE}
With block-encoded Hamiltonian, eigenvalues can be calculated using the QPE algorithm. For QPE to return eigenvalue $E_{j}$ with binary precision $2^{-\beta}$, that is $\beta$-bits, one must provide an input state $\ket{\phi_0}$ that has a non-zero overlap with the eigenstate $\ket{\psi_j} $ of the Hamiltonian, i.e. satisfies the SE: $\hat{H}\ket{\psi_j} =E_j\ket{\psi_j}$. The total T-gate cost for QPE can be written as $C_{QPE}= \mathcal{O}(\zeta C_{H}/\varepsilon)$, where $\varepsilon = 2^{-\beta}$ and $C_H$ is the cost of block-encoding Hamiltonian $\hat{H}$ with scaling constant $\zeta$. 
In QPE, the $\beta$  qubits are measured to retrieve an approximation of the eigenvalue. Each measurement returns a single estimate of a given eigenvalue. When excited states are of interest, one has the following choices: a) adapt the input state $\ket{\phi_0}$ to have a large overlap with the desired excited state; b) shift the Hamiltonian by a classically feasible estimate of the eigenvalue(s) of interest, followed by the application of an eigenstate filtering technique~\cite{Lin2020}; c) utilize an eigenvalue landscape scanning method~\cite{GRM25}.
The gate and qubit complexity for choice a) strongly depends on the specific case of constructing an appropriate ansatz state. For electronic structure calculations, this approach suffers from the \emph{orthogonality catastrophe}~\cite{tubman:hal-04942349}. For choice b), the complexity is associated with the overhead of constructing a properly localized spectral filtering function of an operator, which can be implemented using quantum signal processing (QSP). Choice c) offers up to a quadratic reduction in the number of QPE calls, is less sensitive to the input state, and is particularly advantageous when many eigenvalues are required in generalized eigenvalue problems~\cite{GRM25}.

In the following, we present resource estimates for implementing block-encoded rovibrational Hamiltonians, which can serve multiple purposes, including quantum dynamical simulations. In addition to the block-encoding costs, we also estimate the resources required specifically for QPE in energy levels calculations.

\section{Resource estimation}\label{sec:resource_estimation}
Having discussed the details of our method for combining the DVR representation of the rovibrational Hamiltonian with WH-QROM, we now turn to the discussion of quantum resources required for block-encoding rovibrational Hamiltonians. In the following section~\ref{sec:general_scaling}, we outline the main contributions to the total qubit count, Clifford+T gate complexity, and circuit depth. Section~\ref{sec:water} presents numerical resource estimates for the water molecule expressed in valence internal coordinates, which serve as a benchmark for high-accuracy rovibrational calculations \cite{Polyansky2012,Sutcliffe2000}. In these estimates, we remain QROM-agnostic or employ the standard SELECT–SWAP QROM in order to isolate and highlight the advantages coming solely from FBR–DVR Hamiltonian partitioning relative to other approaches.
Section~\ref{sec:polyspherical-coordinates} explores an alternative choice of internal coordinates, namely polyspherical coordinates. Finally, Section~\ref{sec:block_encoding_pes} quantifies the resource reductions achieved with polyspherical coordinates when block-encoding DVR PES terms for selected polyatomic molecules using the WH-QROM.

\subsection{General scaling}
\label{sec:general_scaling}
We consider $n_i$ basis functions for each coordinate $q_i$. For a given $J$, the field-free molecular system involves $2J+1$ rotational basis functions. Let $C_Q(n,d)$ denote the Clifford+$T$ cost of encoding $n$ classical data points into $d$ qubits using QROM, and let $C_D(n)$ represent the cost of block-encoding an $n$-dimensional diagonal matrix. We recall the Hamiltonian in mixed FBR-DVR:
\begin{equation}
	\boldsymbol{H} = \boldsymbol{T}\sum_{i,j=1}^{D+3}(\boldsymbol{P}_i^{FBR})^\dagger \boldsymbol{T}^\dagger \boldsymbol{G}_{ij}^{DVR} \boldsymbol{T}\boldsymbol{P}_j^{FBR}\boldsymbol{T}^\dagger+	\boldsymbol{V}^{DVR}.
    \label{eq:HDVR-FBR}
\end{equation}
The matrices $\boldsymbol{G}_{ij}^{DVR}$ are diagonal for all index combinations. We denote the corresponding matrix sizes as $n_{G_{ij}}$. If a matrix depends on a subset of internal coordinates $\{q_i\}_{i\in I}$, where $I$ is the set of relevant indices, its dimension is $n = \prod_{i\in I} n_i$. For notational clarity, we assume that $\boldsymbol{G}_{ij}^{DVR}$ always depends on $q_i$ and $q_j$. We denote the sparsity of $\boldsymbol{P}_i^{FBR}$ by $\rho_i$. For the rotational degrees of freedom, adopting a symmetric-top basis yields a diagonal representation for the $z$-component of the angular momentum, while the $x$- and $y$-components are $2$-sparse. For the purpose of implementing the $O_A$ oracle, we assume that all relevant matrix elements can be encoded using $d$ qubits.

The $T$-gate count for block-encoding a $\rho$-sparse $n \times n$ matrix using the oracles defined in Eq.~(\ref{eq:dSparse_oracles}) is given by
\begin{equation}\label{eq:block encoding_T_count_modified}
    C_T =\left\{ \begin{array}{ll}       C_{D}(\rho n) + C_F, & \mbox{if } \rho>1, \\     C_{D}(n), & \mbox{if } \rho=1. \end{array} \right. 
\end{equation} 
where $C_F$ denotes the cost of the $O_F$ oracle.
A unitary controlled on $m$ qubits can then be implemented using single-controlled unitaries and two $m$-controlled $X$ gates~\cite{barenco1995}. Since the number of Hamiltonian terms scales as $\mathcal{O}(D^2)$, this introduces an additional cost of $\mathcal{O}(D^2 \log_2 D)$ Clifford+$T$ gates. The circuit also requires preparation of the ancilla state defined in Eq.~(\ref{eq:ancila_state_G}), which can be realized with cost $\mathcal{O}\big(D \sqrt{\log_2(1/\epsilon)}\big)$ Clifford+$T$ gates using the method of Ref.\cite{Gosset:2024}. For larger molecules, or when appropriate vibrational couplings are neglected, the scaling of the number of Hamiltonian terms reduces to $\mathcal{O}(D)$.

Let us now consider the cost of implementing $O_F$ for different parts of the circuit. In DVR, the elements $\boldsymbol{G}_{ij}^{DVR}$ are diagonal. The only non-diagonal terms are angular momentum operators $\boldsymbol{J}_x$ and $\boldsymbol{J}_y$ and momentum operators $\boldsymbol{P}_i^{FBR}$. For angular momentum, the corresponding $(2J+1)$-dimensional matrices are 2-sparse, for which the $O_F$ can be expressed as
\begin{align}\label{eq:OF_angular_momentum}
    O_F\ket{0}_1\ket{i}_\eta = \left\{ \begin{array}{ll} \ket{j}_1\ket{i-1}_{\eta}, & \mbox{if } i>0, \\ \ket{j}_1\ket{i+1}_{\eta}, & \mbox{if } i=0. \end{array} \right. &&O_F\ket{1}_1\ket{j}_{\eta} = \left\{ \begin{array}{ll} \ket{j}_1\ket{i+1}_{\eta}, & \mbox{if } j<2J, \\ \ket{j}_1\ket{i-1}_\eta, & \mbox{if } j=2J. \end{array} \right.
\end{align}
The transformation defined by Eq.~(\ref{eq:OF_angular_momentum}) can be realized using only controlled adders and Clifford gates, giving the total Clifford+T cost of $\mathcal{O}(\log_2{(2J+1)})$. 
In the case of mixed FBR-DVR representation of the KEO, the $O_F$ cost depends on the structure of the momentum operator matrices. However, for popular basis set choices, such as the harmonic oscillator basis or Legendre polynomials, $O_F$ can be realized with $\mathcal{O}(\log_2n_i)$ Cliffrod+T gates.
Parameters and quantities required to express the costs of partial block-encodings and the total rovibrational Hamiltonian block-encoding are summarized in Table~\ref{tab:T_cost_partial_FBR}. 

\begin{table}[]
    \centering
    \begin{tabular}{c|c|c|c|c|c}
                    &$\rho$ &$n$&Number  \\ 
        \hline
         $\BE{\boldsymbol{P}_i}$ & $\rho_i$ &$n_i$& $2D$ \\
         $\BE{\boldsymbol{J}_z}$ &$1$  & $2J+1$& $1$ \\
         $\BE{\boldsymbol{J}_{\alpha =x,y}}$&$2$ &$2J+1$& $2$ \\
         $\BE{\boldsymbol{G}_{ij}}$&$1$ &$n_{G_{ij}}$& $\frac{(D+3)(D+4)}{2}$ \\
         $\BE{\boldsymbol{V}}$&$1$ &$\prod_{i=1}^D n_i$& $1$ \\
    \end{tabular}
    \caption{Parameters relevant to the cost of block-encoding the rovibrational Hamiltonian in the mixed FBR-DVR representation. $\rho$ denotes the maximal number of non-zero element in a row of a matrix and $n$ is the dimension of a matrix. The last column shows the number of unitaries of a given type required to block-encode the full Hamiltonian.}
    \label{tab:T_cost_partial_FBR}
\end{table}
The combined cost of all partial block-encodings, denoted as $C_{BE}$, can be calculated by inputting parameters from Tables~\ref{tab:T_cost_partial_FBR} into Table~\ref{eq:block encoding_T_count_modified}. The cost of $C_{BE}$ is also increased by the cost of implementing four DVR-FBR transformation unitaires $\boldsymbol{T}$ (cf. Figure~\ref{fig:BE_KEO_separate_citcuit}), each corresponding to the following Clifford+T count~\cite{plis25}
\begin{eqnarray}
    C^{DVR} = 2\sum_{i=1}^D \left\lfloor{\frac{\pi\sqrt{n_i}}{4}}\right\rfloor C_Q(n_i^2,d).
\end{eqnarray}
A detailed discussion of implementation of $\boldsymbol{T}$ and resource estimation is given in Appendix~\ref{sec:appendix-DVR} and further in ref.~\cite{plis25}. Here, specifically for the case of $D$-dimensional direct-product Hilbert space, assuming equal number of basis functions $n$ per internal coordinate the circuit cost for the DVR-FBR unitary is given by: $C^{DVR}=\mathcal{O}\left(D(2n^2+n(4d+1)\log(n))\right)$, hence is linear in the number of modes. $d$ denotes the precision of the $\mathbf{T}$ matrix elements.

In summary, the total Clifford+T gate cost for block-encoding the FBR-DVR Hamiltonian is given by
\begin{equation}
\label{eq:FBR-DVR-Ham-resources}
   C_H =   C_{BE} + O(D^2\log_2D) + O\left(D\sqrt{\log_2\frac{1}{\epsilon}}\right) + \mathcal{O}(\log_2{2J+1})+4    C^{DVR} .
\end{equation}
For a clearer intuitive picture of the scaling, let us consider equal number of basis functions for each coordinate $n_i = n$, for all $i$. Additionally, let us assume that all matrix elements of $\boldsymbol{G}_{ij}(\boldsymbol{q})$ are functions of at most $m$ coordinates, with $m>2$. That leads to $n_{G_{ij}} = n^{m}$. Finally, let each matrix have at most $\rho=\rho_i$ non-zero elements. The total of block-encoding Hamiltonian is given by
\begin{equation}
\label{eq:DVR-FBR-cost}
    C_{BE} =4DC_D(\rho n)+\frac{(D+3)(D+4)}{2}C_D(n^{m})+2C_D(2J+1)+8C_D(4J+2)+C_D(n^D).
\end{equation}

As shown in Table~\ref{tab:T_cost_partial_FBR}, the cost of block-encoding the potential scales most unfavorably with the size of the basis. This is because the potential is the only Hamiltonian component that depends on all internal coordinates $\boldsymbol{q}$. Moreover, if $m$ is independent of the molecular size, the potential is the only contribution that scales exponentially with the number of vibrational modes $D$, and hence with the number of atoms. For instance, in polyspherical coordinates (discussed in section~\ref{sec:polyspherical-coordinates}), the form of the KEO implies $n_{G_{ij}}\leq n^9$ for any number of atoms (cf. Appendix C in Ref.~\cite{Gatti2009}). In practice, the number of coupled modes ($m$) in the PES is smaller than the number of internal degrees of freedom, fixed for polyspherical coordinates and in the PES (denoted $L$) reaches constant value as the system size increases. 

\subsection{Example: water molecule in valence coordinates} \label{sec:water}
To illustrate the scaling of $T$-gate and qubit requirements on a concrete example, we consider a specific molecule and a choice of internal coordinates. In this section we estimate the cost of block-encoding the Hamiltonian for the  $J=0$ H$_2$O molecule using valence coordinates that are particularly useful in high accuracy calculations~\cite{Sutcliffe1986,Sutcliffe2000}. A general vibrational KEO given in Eq.~(\ref{eq:KEO_def}) for a triatomic molecule using valence coordinates can be expressed as \cite{Gatti2009}\cite{Hua1992}:
\begin{equation}
    \begin{split}
        \hat{H} & = \frac{\hat{P}_{1}^2}{2\mu_{1}}+\frac{\hat{P}_{2}^2}{2\mu_{2}}+\hat{P}_\theta^\dag\left(\frac{1}{2\mu_1 R_1^2}+\frac{1}{2\mu_2 R_2^2}-\frac{\cos\theta}{\mu_{12}R_1R_2}\right)\hat{P}_\theta\\
        &+\frac{2\cos{\theta}\hat{P}_{1}\hat{P}_{2}}{2\mu_{12}}+\left(\frac{\hat{P}_1}{2\mu_2R_2}+\frac{\hat{P}_2}{2\mu_1R_1}\right)\left(\hat{P}_\theta^\dag\sin\theta+\sin\theta P_\theta\right)+\hat{V}
    \end{split}
\end{equation}
with $\hat{P}_i = -i\hbar\frac{\partial}{\partial R_i}$ and $\hat{P}_\theta = i\hbar\frac{\partial}{\partial \theta}$. For the water molecule
$\mu_1^{-1} = \mu_2^{-1} = m_H^{-1}+m_O^{-1}$,$ \mu_{12} = m_O$, where $m_O$ and $m_H$ are the masses of the oxygen and hydrogen nuclei, respectively.

Below, we compare Clifford+T and qubit count estimates for several methods of block-encoding the vibrational Hamiltonian: a) LCU Pauli string representation; b) the whole Hamiltonian block-encoded in DVR; c) LCU with component parts block-encoded in DVR and mixed FBR-DVR (our primary method). Quantum resources for these methods are summarized in Tables~\ref{tab:HO2_T_counts} and \ref{tab:HO2_cost}.

\paragraph{LCU.} 
The most straightforward implementation writes the Hamiltonian as a sum of Pauli strings and employs the LCU method. In the worst-case scenario, the number of Pauli strings in the Hamiltonian is $N^2 = n^{4}_Rn_\theta^2$. This corresponds to $n_P = \frac{3}{4}N^2\log_2{N}$ single-qubit non-identity Pauli operators. These operators must be controlled on $\log_2(N^2)$ ancilla qubits. Following the method from ref.~\cite{babbush2018}, such controls can be implemented using $n_P$ Toffoli gates and $3n_P$ Clifford gates, with an additional $\mathcal{O}(\log(n_P))$ ancillas. In this scheme, relative-phase Toffoli gates can be used, which decompose into $4$ T-gates and $5$ Clifford gates. Therefore, the upper bound for the T-count with the LCU method is given by
\begin{equation}
    C_T^{(LCU)} = 3n_R^{4}n_\theta^2\log_2(n_R^{2}n_\theta) + \mathcal{O}\left(\sqrt{n_R^2n_\theta}\right)
\end{equation}
and for Clifford gates
\begin{equation}
    C_{Cliff}^{(LCU)} = \frac{27}{4}n_R^{4}n_\theta^2\log_2(n_R^{2}n_\theta) + \mathcal{O}\left(\sqrt{n_R^2n_\theta}\right)
\end{equation}
The second term in the sums is the cost of state preparation using the method from ref.~\cite{Gosset:2024}. For estimating the block-encoding scaling constant $\zeta_H = \sum_{i}^{N^2} \zeta_i$, we  use the bound for the $L_1$ norm given by
\begin{equation}
  \sqrt{\sum_{i}^{N^2}\zeta_i^2} \leq \sum_{i}^{N^2} \zeta_i \leq N\sqrt{\sum_{i}^{N^2}\zeta_i^2}
\end{equation}
where $H =\sum_i \zeta_i P_i$ and $P_i$ are Pauli strings. We calculate the $L_2$ norm numerically using the formula 
\begin{equation}
    \sqrt{\sum_{i}^{N^2}\zeta_i^2} = \sqrt{\frac{\Tr{H^2}}{N}}.
\end{equation}

\paragraph{DVR for the full Hamiltonian.} In the DVR basis the Hamiltonian's matrix sparsity can be exploited. This operator has exactly $\rho = n_R^2+2n_Rn_\theta -2n_R - n_\theta +1$ non-zero elements in each row. The cost of block-encoding the DVR Hamiltonian using the $\rho$-sparse method is given by Eq.~(\ref{eq:block encoding_T_count_modified}) with $n = n_R^2 n_{\theta}$. We assume the implementation of the $O_F$ oracle via QROM and therefore $C_F = C_Q\left(\rho n_R^2n_{\theta},\log_2{n_R^2n_\theta}\right)$.

\paragraph{FBR-DVR and DVR for Hamiltonian components. }The other two ways of block-encoding the Hamiltonian include the mixed DVR-FBR method and DVR method for individual components layer within LCU, as discussed in Sec.~\ref{sec:block encoding}. Following discussion given in Sec.~\ref{sec:block_encoding_reducing_costs}, due to the symmetry of the water molecule the block-encoding of the KEO can be realized by the following circuit
\begin{equation}
    B[\boldsymbol{H}] = CSWAP_{12}B[\boldsymbol{H}_{eff}]CSWAP_{12}
\end{equation}
where an effective Hamiltonian is defined as
\begin{equation}
    \begin{split}
        \hat{H}_{eff} &= \frac{\hat{P}^2_1}{2\mu_1}+\hat{P}_\theta^\dag\left(\frac{1}{2\mu_1 R_1^2}-\frac{\cos{\theta}}{2\mu_{12}R_1R_2}\right)\hat{P}_\theta\\
    &+\frac{\cos\theta\hat{P_1}\hat{P_2}}{2\mu_{12}}+\frac{\hat{P}_1}{2\mu_1 R_2}\left(\hat{P}_\theta^\dag\sin\theta +\sin\theta\hat{P}_\theta\right)   +\frac{1}{2}V
    \end{split}
\end{equation}
and $CSWAP_{12}$ is a gate exchanging the state of registers corresponding to $R_1$ and $R_2$ controlled on a single ancilla.
First, we consider block-encoding via the circuit given by Eq.~(\ref{eq:BE_KEO_separate}). 
The corresponding unitaries given by Eqs.~(\ref{eq:BE_separate_U_P}-\ref{eq:BE_separate_U_G}) are written as
\begin{equation}
    \boldsymbol{U}_P = \left(\ket{0}\bra{0}+\ket{2}\bra{2}+\ket{4}\bra{4}\right)\otimes\BE{\boldsymbol{P}_1}+\left(\ket{1}\bra{1}+\ket{3}\bra{3}\right)\otimes\BE{\boldsymbol{P}_\theta}+\ket{5}\bra{5}\otimes\id
\end{equation}
\begin{equation}
    \boldsymbol{U}_P' = \left(\ket{0}\bra{0}+\ket{2}\bra{2}+\ket{3}\bra{3}\right)\otimes\BE{\boldsymbol{P}_1}+(\ket{1}\bra{1}+\ket{4}\bra{4})\otimes\BE{\boldsymbol{P}_\theta}+\ket{5}\bra{5}\otimes\id
\end{equation}
\begin{equation}
    \begin{split}
        \boldsymbol{U}_g &= \ket{0}\bra{0}\otimes\id+\ket{1}\bra{1}\otimes\BE{\boldsymbol{g}_{uu}}+\ket{2}\bra{2}\otimes\BE{\cos\theta}\\
    &+\left(\ket{3}\bra{3}+\ket{4}\bra{4}\right)\otimes\BE{\frac{1}{R}_2}\BE{\sin\theta} +\ket{5}\bra{5}\otimes \BE{\boldsymbol{V}}
    \end{split}
\end{equation}
with $g_{uu} = \frac{1}{2\mu_1 R_1^2}-\frac{\cos{\theta}}{2\mu_{12}R_1R_2}$. Note, that we have used the symmetry between $R_1$ and $R_2$ to express the block-encoding of $\boldsymbol{P}_2$ as $\BE{\boldsymbol{P}_2} = SWAP_{12}\BE{\boldsymbol{P}_1}SWAP_{12}$, which introduces an additional cost of controlled SWAP gates.
The cost is therefore
\begin{equation}
    \begin{split}
        C_{BE} &= 4C_D(2n_R)+4C_D(n_\theta^2/2)+3C_D(n_\theta)+C_D(n_R)+C_D(n_R^2n_\theta)\\
        &+2C^{DVR}+\mathcal{O}(\log_2n_R^2n_{\theta})
    \end{split}
\end{equation}
The last term in the sum describes the cost of controlled SWAPS and the implementation of $O_F$ oracles.
For the second method of full-DVR Hamiltonian (cf. eq.~\ref{eq:BE_second_method_circuit} in Appendix) the cost is given by
\begin{equation}
    C_{BE} = 2C_D(n_R^2n_\theta^2)+C_D(n_R^2n_\theta)+6C_D(n_R^2)+ 2C_D(n_\theta^2)+ C_D(n_\theta)+C_D(n_R) 
\end{equation}
The asymptotic T-counts are summarized in Table~\ref{tab:HO2_T_counts}. Table~\ref{tab:HO2_cost} shows the cost for a an example choice of basis set size with $n_R=2^5$ and $n_\theta=2^6$. 

\begin{table}[]
    \centering
    \begin{tabular}{c|c|c|c}
                    &T-count&Ancillas&Cliffords  \\ 
        \hline
        LCU &$\mathcal{O}\left(n^6\right)\log_2{n}$ &$\mathcal{O}(\sqrt{n)}$ &  $\mathcal{O}\left(n^6\right)\log_2{n}$    \\
        Full DVR &$\mathcal{O}\left(n^{\frac{5}{2}}\right)$ &$\mathcal{O}\left(n^{\frac{5}{2}}\right)$ &$\mathcal{O}\left(n^{5}\right)$  \\
        Separate DVR & $\mathcal{O}\left(n^2\right)$&$\mathcal{O}\left(n^2\right)$ &  $\mathcal{O}\left(n^4\right)$     \\
        DVR+FBR & $\mathcal{O}\left(n^\frac{3}{2}\right)$& $\mathcal{O}\left(n^\frac{3}{2}\right)$&  $\mathcal{O}\left(n^3\right)$
    \end{tabular}
    \caption{Asymptotic costs of     block-encoding the vibrational Hamiltonian of $H_2O$ molecule for different method, assuming $n_R = n_\theta = n$ and $C_D(n) = \mathcal{O}(\sqrt{n)} T+ \mathcal{O}(n) \text{ Cliffords}$}
    \label{tab:HO2_T_counts}
\end{table}

\begin{table}[h]
    \centering
    \begin{tabular}{c|c|c|c|c|c}
                    &T-count &Ancillas&Cliffords&  Norm[cm$^{-1}]$  &Fixed Ancilla T-count\\ 
        \hline
        LCU FBR&$2.1\times 10^{11}$ &$0.81\times 10^3$ &$4.6\times10^{11}$&$1.3\times 10^6-8.5\times10^9$&$2.1\times 10^{11}$\\
        Full DVR &$4.4\times 10^7$ &$6.4\times 10^5$&$2.9\times 10^{12}$&$1.2\times 10^{9}$&$1.7\times10^{10}$\\
        Separate DVR & $1.7\times10^{6}$&$2.4\times10^4$&$4.7\times10^{9}$&$6.6\times10^7$&$2.7\times10^7$\\
        DVR-FBR & $4.5\times10^5$&$4\times10^{3}$&$1.1\times10^{8}$&$7.5\times10^{6}$&$7.7\times10^5$\\
    \end{tabular}
    \caption{Summary of costs for different methods for $n_{\theta} = 2^6$ and $n_R = 2^5$. The first four columns assume SELECT-SWAP implementation with optimal T-count. For a better comparison with the LCU method, the last column also shows T-count assuming there are 810 ancillas available.}
    \label{tab:HO2_cost}
\end{table}
For rotationally excited states ($J>0$), the rovibrational energy levels of water are accurately described by the method of ref.~\cite{Tennyson2004}, implemented in the well-established \textsc{DVR3D} code. For completeness, in supplementary information we discuss a variant of our quantum algorithm adapted to the \textsc{DVR3D} framework, which has had significant impact in astrophysics, exoplanetary studies, and atmospheric spectroscopy. 

The advantages of utilizing an appropriate combination of DVR and FBR representations are already evident for the water molecule, even when using a standard QROM. Table~\ref{tab:HO2_T_counts} demonstrates an $\mathcal{O}(n^{7/2})$ improvement in quantum volume (the product of T-gate and qubit count), as well as a $\tilde{\mathcal{O}}(n^{3})$ reduction in the Clifford-gate count, for the mixed DVR–FBR approach compared to the LCU method. These asymptotic improvements are reflected in the numerical estimates reported in Table~\ref{tab:HO2_cost}. For a representative basis size of approximately $6.4\times10^{4}$ functions, which is typically sufficient to achieve spectroscopic accuracy in water rovibrational calculations, we observe an improvement of roughly six orders of magnitude in T-count relative to LCU, along with reductions of several orders of magnitude compared to alternative representations, such as a full DVR or LCU combined with a purely DVR-based Hamiltonian representation.
We further improve this scaling by incorporating WH-QROM. In the following section, we present resource estimates for the water molecule and for a larger polyatomic system obtained using our mixed FBR–DVR block-encoding method and WH-QROM.

\subsection{Polyatomic molecules in polyspherical coordinates}
\label{sec:polyspherical-coordinates}
For a more systematic resource estimation for our method we consider the polyspherical coordinates~\cite{Gatti2009}, in which the Hamiltonian is expressed in terms of $A-1$ relative position vectors $\boldsymbol{R}_i$, where $A$ is the number of atoms. Each vector is parameterized using spherical coordinates $(R_i,\theta_i,\phi_i)$. Since there are only $D=3A-6$ internal degrees of freedom, not all coordinates are independent. By an appropriate definition of the Euler angles $\boldsymbol{\Theta} = (\alpha,\beta,\gamma)$, one can set $\theta_{A-1} = \phi_{A-1} = \phi_{A-2} = 0$~\cite{Gatti2009}. The resulting set of internal coordinates therefore consists of $A-1$ radial coordinates $R_i$, $A-2$ polar coordinates $\theta_i$, and $A-3$ azimuthal coordinates $\phi_i$. For convenience, one may also introduce $u_i=\cos\theta_i$ for the polar degrees of freedom. In what follows, we focus on the case where the relative position vectors $\boldsymbol{R}_i$ are orthogonal. We denote by $n_R$, $n_\theta$, and $n_\phi$ the basis sizes for the coordinates $R_i$, $\theta_i$, and $\phi_i$, respectively.

\begin{table}[]
    \centering
    \begin{tabular}{c|c|c|c|c|c}
                    &$\rho$ &$n$&Number &Clifford+T ($n=n_i$)&Reduced Clifford+T  \\ 
        \hline
        $ \BE{\boldsymbol{P}_{R_i}}$    &$2$ &$n_R$ &$A-1$  &$2C_D(2n)$&$2C_D(2n)$\\
        $ \BE{\boldsymbol{P}_{u_i}}$    &$\frac{1}{2}n_\theta$ &$n_\theta$ &$A-2$  &$2C_D(\frac{1}{2}n^2)$&$2C_D(\frac{1}{2}n^2)$\\
        $ \BE{\boldsymbol{P}_{\phi_i}}$    &$\frac{1}{2}n_\phi$ &$n_\phi$ &$A-3$  &$2C_D(\frac{1}{2}n^2)$&$2C_D(\frac{1}{2}n^2)$\\
        \hline
         $\BE{\boldsymbol{g}({u_i,u_i})}$&$1$ &$n_R^2n_\theta$& $A-2$& $C_D(n^3)$&$6C_D(n)$\\
         $\BE{\boldsymbol{g}({\phi_i,\phi_i})}$&$1$ &$n_R^3n_\theta^2 n_\phi$& $A-3$& $C_D(n^6)$&$36C_D(n)$\\
         $\BE{\boldsymbol{g}({u_i,u_j})}$&$1$ &$n_Rn_\theta^2n_\phi^2$& $\frac{(A-2)(A-3)}{2}$& $C_D(n^5)$&$10C_D(n)$\\
         $\BE{\boldsymbol{g}({\phi_i,\phi_j})}$&$1$ &$n_R^2n_\theta^3n_\phi^2 $& $\frac{(A-3)(A-4)}{2}$& $C_D(n^7)$&$42C_D(n)$\\
         $\BE{\boldsymbol{g}({u_i,\phi_i})}$&$1$ &$n_Rn_\theta^2n_\phi$& $A-3$& $C_D(n^4)$&$4C_D(n)$\\
        $\BE{\boldsymbol{g}(({u_i,\phi_j})}$&$1$ &$n_Rn_\theta^3 n_\phi^2$& $(A-3)^2$& $C_D(n^6)$&$18C_D(n)$\\
        \hline
        $\BE{\boldsymbol{\Gamma}(\alpha,u_j)}_{\alpha=x,y}$&$1$ &$n_R n_\theta n_\phi$& $2(A-2)$& $C_D(n^3)$&$3C_D(n)$\\
        $\BE{\boldsymbol{\Gamma}(z,u_j)}$&$1$ &$n_R n_\theta^2 n_\phi$& $A-2$& $C_D(n^4)$&$4C_D(n)$\\
        $\BE{\boldsymbol{\Gamma}(x,\phi_j)}$&$1$ &$n_R n_\theta^2 n_\phi$& $A-3$& $C_D(n^4)$&$8C_D(n)$\\
        $\BE{\boldsymbol{\Gamma}(y,\phi_j)}$&$1$ &$n_R n_\theta n_\phi$& $A-3$& $C_D(n^3)$&$3C_D(n)$\\
        $\BE{\boldsymbol{\Gamma}(z,\phi_j)}$&$1$ &$n_R^2 n_\theta^2 n_\phi$& $A-3$& $C_D(n^5)$&$15C_D(n)$\\
        \hline
        $\BE{\boldsymbol{\mu}(z,z)}$&$1$ &$n_R^2n_\theta$& $1$ & $C_D(n^3)$&$6C_D(n)$\\
        $\BE{\boldsymbol{\mu}(\alpha,\alpha)}_{\alpha=x,y}$&$1$ &$n_R$& 2 & $C_D(n)$&$C_D(n)$\\
        $\BE{\boldsymbol{\mu}(x,z)}$&$1$ &$n_Rn_\theta$& 1 & $C_D(n^2)$&$2C_D(n)$\\
        \hline
        $\BE{\boldsymbol{J}_z}$&$1$& $2J+1$&$2$& $C_D(2J+1)$&$C_D(2J+1)$\\
        $\BE{\boldsymbol{J}_{\alpha= x,y}}$&$2$& $2J+1$&$4$&$2C_D(4J+2)$&$2C_D(4J+2)$\\
        \hline
         $\BE{\boldsymbol{V}}$&$1$&$n_R^{A-1}n_\theta^{A-2}n_\phi^{A-3}$&$1$&$C_D(n^{(3A-6)})$&$C_D(n^{(3A-6)})$
    \end{tabular}
    \caption{Summary of parameters for calculating the cost of block-encoding various parts of the rovibrational Hamiltonian and the T-count for the polyspherical internal coordinates using mixed FBR-DVR method. $\rho$ denotes the maximal number of non-zero element in a row of a matrix, while $n$ denotes the dimension of a matrix. The rightmost column shows the T-count assuming $n_R=n_\theta=n_\phi = n$. The last column shows the improved scaling when utilizing the sum of products form of $\boldsymbol{G}_{ij}$, as described in sec.~\ref{sec:block_encoding_reducing_costs}. For clarity of presentation, the subscript indices were replaced by arguments in matrices, e.g. $\boldsymbol{g}_{ab}\rightarrow \boldsymbol{g}(a,b)$.}.
    \label{tab:T_cost_partial_spherical_mixed}
\end{table}

\begin{table}[]
    \centering
    \begin{tabular}{c|c|c|c|c|c}
        Operator      &$\boldsymbol{U}_P$&$\boldsymbol{U}_G$& $\boldsymbol{V}_{(SEL-SWAP)}$& $\boldsymbol{V}_{(WH)}$ \\ 
        \hline
        T-cost  &$\mathcal{O}\left(An\right)+\mathcal{O}\left(\sqrt{2J+1}\right)$&$\mathcal{O}\left(A^2n^{\frac{7}{2}}\right)$& $\mathcal{O}\left(n^{\frac{1}{2}(3A-6)}\right)$  & $\mathcal{O}\left(n^{\alpha(3A-6)}\right)$ \\
        T-cost(Reduced)  &$\mathcal{O}\left(An\right)+\mathcal{O}\left(\sqrt{2J+1}\right)$&$\mathcal{O}\left(A^2\sqrt{n}\right)$& $\mathcal{O}\left(n^{\frac{1}{2}(3A-6)}\right)$  & $\mathcal{O}\left(n^{\alpha(3A-6)}\right)$ \\
    \end{tabular}
    \caption{Asymptotic T-count scaling for different parts of the Hamiltonian expressed in polyspherical coordinates using the mixed DVR-FBR method. The first 4 columns assume $C_D(n) = \mathcal{O}(\sqrt{n})$, achieved by SELECT-SWAP QROM for diagonal unitary synthesis~\cite{Gosset:2024,low_trading_2024}, while the last column assume WH-QROM. Here, $\alpha$ is a potential dependent constant, discussed in the following section. The second row shows the improved scaling when utilizing the sum-of-products form of $\boldsymbol{G}_{ij}$, as described in sec.~\ref{sec:block_encoding_reducing_costs}.}
    \label{tab:poly_spher_scaling-dvr-fbr}
\end{table}

 The quantities required to evaluate the total block-encoding cost, along with a breakdown of the T-count for individual contributions to the Hamiltonian, are summarized in Tables~\ref{tab:T_cost_partial_spherical_mixed},~\ref{tab:poly_spher_scaling-dvr-fbr} for the mixed FBR-DVR approach. The corresponding tables for the full DVR approach can be found in Appendix~\ref{sec:appendix-BE-DVR-alone}. Since elements of the metric tensor in polyspherical coordinates have a simple sum-of-products form given by Eq.~(\ref{eq:metric_tensorm_sop}), we use the method described in sec.~\ref{sec:block_encoding_reducing_costs} for constructing the block-encoding. Assuming diagonal unitaries are implemented using the method of ref.~\cite{Gosset:2024}, the total cost scales as $C_D = O\left(\sqrt{n\log_2{\frac{1}{\epsilon}}}\right)$. The resulting asymptotic T-count scaling is presented in Table~\ref{tab:poly_spher_scaling-dvr-fbr}.

\subsection{Block-encoding the PES.}\label{sec:block_encoding_pes}
In polyatomic molecules, the major source of computational complexity, both in classical and quantum computing is encoding and processing the PES. While the DVR representation requires only evaluations of the PES at grid points, the number of matrix elements still scales exponentially with the number of internal coordinates, rendering PES the dominant complexity contribution. SELECT-SWAP QROM still requires exponentially (with dimensionality, cf. Tab~\ref{tab:T_cost_Ham}, eq.~\ref{tab:poly_spher_scaling-dvr-fbr}) many operations and qubits to encode the PES. To alleviate this unfavorable scaling, we utilize the WH-QROM method.

In what follows, we present numerical resource estimates for several polyatomic PESs, comparing the standard SELECT-SWAP QROM with our WH-QROM approach. We investigated molecules of varying dimensionality and flexibility of bonds. Table~\ref{tab:PES-polyaotmic} reports results for  water (H$_2$O), phosphine (PH$_3$), methane (CH$_4$), dimethylsulfide (DMS), ethylthiol (ETSH), and the penta-2,4-dieniminium cation (PSB3). The selected systems are representative for a range of applications, from high-resolution spectroscopy of small molecules (water (3D), phosphine (6D), methane (9D), dimethylsulfide (21D)) relevant for example in modeling exoplanet spectra, to the biologically motivated PSB3, widely used as a model for the protonated Schiff base of retinal (rPSB), the chromophore of light-sensitive rhodopsin proteins~\cite{Hutcheson2021}, and the floppy ethylthiol (ETSH).

For each molecule, we employed $d = 33$-bit precision of matrix elements and considered five different qubit counts $\eta$ in the range $14 \leqslant \eta \leqslant 27$, which corresponds to different DVR grid sizes. In the error range $10 \lesssim \log_2 \left(\nicefrac{1}{\epsilon}\right) \lesssim 25$, we computed the corresponding Toffoli gate counts $\tau$ at 100 sample points. Finally, linear regression was used to test the hypothesis:
\begin{equation}\label{hypothesis:hypthesis}
	\log_2 \( \tau \) = c_1 \tau + c_2 \log_2 \( \log_2 \( \nicefrac{1}{\epsilon} \) \) + c_3. 
\end{equation}
with predictions and their $R^2$ scores shown in Table ~\ref{tab:PES-polyaotmic}.

\renewcommand{\arraystretch}{1.5}
\begin{table}[h!t]
	\begin{tabular}{|c|c|c|}
		\hline
				& Toffoli count																			& $R^2$ score	\\
		\hline
		CH4	\cite{rey2014}	& $2^{0.4905 \cdot \eta - 8.442} \( \log_2 \( \nicefrac{1}{\epsilon} \) \)^{4.1484}$		& 0.9898	\\
		DMS		\cite{Senent2014}& $2^{0.3763 \cdot \eta - 7.837} \( \log_2 \( \nicefrac{1}{\epsilon} \) \)^{4.9158}$		& 0.9852	\\
		ETSH	\cite{Senent2014}& $2^{0.2548 \cdot \eta + 0.7694} \( \log_2 \( \nicefrac{1}{\epsilon} \) \)^{3.9581}$		& 0.9796	\\
		H2O		\cite{Huang2008}& $2^{0.7292 \cdot \eta - 3.2896} \( \log_2 \( \nicefrac{1}{\epsilon} \) \)^{2.4458}$		& 0.9893		\\
		PH3		\cite{Ovsyannikov2008}& $2^{0.6155 \cdot \eta - 3.2295} \( \log_2 \( \nicefrac{1}{\epsilon} \) \)^{2.3755}$		& 0.9932	\\
		PSB3	\cite{VillasecoArribas2024}& $2^{0.7181 \cdot \eta + 0.3773} \( \log_2 \( \nicefrac{1}{\epsilon} \) \)^{1.9622}$		& 0.9932	\\
		\hline
	\end{tabular}
	\caption{Fitted Toffoli count formulae for the studied PES data with WH-QROM. $\eta$ is the number of qubits representing PES values (DVR grid size). The CNOT gate counts in all constructions are proportional to the Toffoli gate counts.}
    \label{tab:PES-polyaotmic}
\end{table}

Table ~\ref{tab:PES-polyaotmic} demonstrates that, for example, methane, with its 9 internal coordinates, a molecule whose simulation can exceed the capabilities of classical computing (e.g. when more than 10 basis functions per dimension are used in direct-product basis), can be simulated with exponentially fewer quantum resources than with the techniques based on SELECT-SWAP QROM. With standard SELECT-SWAP QROM and without resorting to contraction schemes or non-direct-product basis sets (which complicate the computational workflow and are not easily generalizable to other molecular systems), the computation quickly becomes prohibitive in both memory and time complexity, both in classical and quantum computing. Even with state-of-the-art contraction schemes~\cite{Felker2024,Yurchenko2007}, obtaining the high-accuracy Hamiltonian spectrum of methane with a complicated PES remains computationally challenging. Quantum computation, however, can mitigate several of these limitations, as we discuss below, by overcoming the QROM bottleneck. Indeed, for the methane molecule (and other systems shown in Table~\ref{tab:PES-polyaotmic}), we find that the leading contribution to block-encoding cost, namely, inputting the PES via WH-QROM requires exponentially fewer quantum Toffoli gates (and thus T-gates) than traditional QROM implementations based on the SELECT-SWAP technique~\cite{low_trading_2024}, as the system size increases (number of dimensions). 

The total qubit count for our WH-QROM is always $3 \eta + 2 d = 3 \eta + 66$, which is linear in $\eta$. Taking the above predictions at face value for a moment, we note the following. In three cases, for CH$_4$, DMS, and ETSH, the predicted scaling with $\eta$ is $O (2^{\alpha \eta})$, with $\alpha < \tfrac{1}{2}$. This is already better than the scaling presented in  ref.~\cite{low_trading_2024}, which is optimal in the general case. In fact, this is achieved without exponentially many ancillas. In the other cases, $\tfrac{1}{2} < \alpha < 1$, a comparable Toffoli gate count is achieved with the construction of ref.~\cite{low_trading_2024} only when $\mathcal{O}\left( 2^{(1 - \alpha) \eta} \right)$ ancillas are used. Thus quantum volume for our method is always exponentially reduced compared to other QROMs.

\smallskip

Allowing once more to extend the validity of the hypothesis given in \ref{hypothesis:hypthesis}, we can estimate the costs as a function of errors alone, if we use the results of~\ref{corollary:epsilon_approx}, assuming that $\eta = D m = O \( \log_2 \( 2 \pi \sqrt{D} \| \nabla V \|_\infty\tfrac{}{\epsilon} \) \)$. We then get the following
\begin{equation}
	\tau = O \( \( 2 \pi \sqrt{D} \tfrac{\| \nabla V \|_\infty}{\epsilon} \)^{c_1 D} \( \log_2 \( \nicefrac{1}{\epsilon} \) \)^{c_2} \).
    \label{eq:wh-qrom-cmplexity}
\end{equation}
For a fixed function ($V(\mathbf{q)}$ and $D$), we get
\begin{equation}
	\tau = O \( \tfrac{\log_2 \( \nicefrac{1}{\epsilon} \)^{c_2}}{\epsilon^{c_1 D}} \).
\end{equation}
Recall that the same complexity of the most efficient SELECT-SWAP QROM design of ref.~\cite{low_trading_2024} is $O ( \sqrt{\tfrac{\log_2 ( \frac{1}{\epsilon}  )}{\epsilon^D}} )$. Thus when $c_1 < \tfrac{1}{2}$ (or $c_1 = \tfrac{1}{2}$, but $c_2 < \tfrac{1}{2}$), then our design achieves a better scaling. When $\tfrac{1}{2} \leqslant c_2 < 1$, then SELECT-SWAP requires about $\lambda \sim \epsilon^{(1 - c_2) D}$ times more qubits to achieve the same Toffoli gate count. Note that either way, both $\epsilon \rightarrow 0^+$ and the $D \rightarrow \infty$ limits are favorable for our design. For clarity of presentation, the $c_1$ parameter is denoted as $\alpha$ in Table~\ref{tab:poly_spher_scaling-dvr-fbr} and elsewhere in the manuscript. 

\smallskip

For a quantitative comparison of the advantages of our method over the SELECT-SWAP method, we evaluate the complexities of the WH-QROM gate and qubit at $\epsilon = 2^{- 10}$ precision and at the largest input qubit number $\eta$, and compare them to those of SELECT-SWAP. For the WH-QROM, we compute the exact numbers of qubits, CNOT gates, and the Toffoli depth. In addition, we reduce the number of digits from $d = 33$ to $d = 15$, consistent with the target precision of $2^{- 10}$. Details of resource calculation procedure used for SELECT-SWAP QROM are given in Appendix~\ref{sec:appendix-SELECT-SWAP}.
\renewcommand{\arraystretch}{1}
\begin{table}[!ht]
	\begin{tabular}{|c|c|c|c|c|c|c|}

		\hline
				& qubit ratio  & Toffoli count ratio    & Toffoli depth ratio	& Toffoli volume ratio	& CNOT count ratio	& weighted cost ratio		\\
		\hline
		CH$_4$		& 286          & 1.2078		            & 0.6048		& 336				& 1505			& 158	\\
		DMS		& 207	       & 1.5452		            & 0.7759		& 210				& 1123			& 143	\\
		ETSH	& 208	       & 0.162                  & 0.0811		& 32				& 150			& 15	\\
		PH$_3$		& 286	       & 0.1128		& 0.0565		& 34				& 116			& 14	\\
		H2O		& 151	       & 0.0683		& 0.0343		& 9.8				& 29			& 3.5	\\
		PSB3	& 207	       & 0.0094		& 0.0047		& 1.9				& 6.7			& 0.759	\\
		\hline
	\end{tabular}
	\caption{Toffoli gate complexity ratios for encoding PESs with SELECT-SWAP over WH-QROMs for several molecules considered (cf. Table~\ref{tab:PES-polyaotmic}). Weighted costs was calculated as $\textnormal{Toffoli cost} = 50 \times \textnormal{CNOT cost}$~\cite{Gidney2024}. Toffoli volume is the product of qubit count and Toffoli Count.}
    \label{table:WH_vs_SelectSWAP_QROM}
\end{table}

In Table~\ref{table:WH_vs_SelectSWAP_QROM}, we report, for each PES considered, the ratios of SELECT-SWAP over WH-QROM costs. Ratios greater than one indicate an advantage in favor of the WH-QROM.

\smallskip

When block-encoding the PES via multiplexed rotations, one encodes angles of the form $\tfrac{\mathrm{PES}(x)}{\cN_{\mathrm{PES}}}$, where $\cN_{\mathrm{PES}}$ is a normalization constant. We consider $\cN_{\mathrm{PES}}$  proportional to the $L^\infty$ norm of the PES, consistent with the block-encoding requirements. Once again, we benchmark our approach against the most efficient implementations of SELECT–SWAP.
Choosing $\cN_{\textnormal{PES}}$ to be exactly the $L^\infty$ norm is disadvantageous for the WH-QROM, since its error typically scales with the $L^\infty$ norm of the derivative, and the derivative of $\arccos(x)$ diverges at $x=\pm 1$. For this reason, we use $\cN_{\textnormal{PES}} = 2 | \textnormal{PES} |_{L^\infty}$. We note that the larger the prefactor of $| \textnormal{PES} |_{L^\infty}$ in the normalization, the better the results. This behavior is expected, because away from $1$, the derivative of $\arccos$ is bounded and nearly linear. We compute the complexities of both WH-QROM and SELECT-SWAP QROM at $2^{-10}$ error, using $d = 13\text{--}14$ digits for WH-QROM. Results are shown in Table~\ref{table:WHvsSelectSWAPinfty}. The number of digits $d$ is another tunable hyperparameter that can further optimize performance, trading off accuracy.

We note that in the SELECT-SWAP approach, constructing a multiplexed rotation requires \textit{two QROM calls }plus additional gates (scaling with $\log_2(\nicefrac{1}{\epsilon})$) to implement the rotations. In contrast, the WH-QROM approach requires only a \textit{single QROM call} and a single reusable copy of $\QFT_{\eta + d} \ket{-1}_{\eta + d}$ when the phase-kickback idea of sec.~\ref{sec:multiplexed_rotations} is applied. The tables below report the ratios considering only the QROM calls.
\begin{table}[ht]
	\begin{tabular}{|c|c|c|c|c|c|c|}
		\hline
				& qubit ratio	& Toffoli count ratio	& Toffoli depth ratio	& Toffoli volume ratio	& CNOT count ratio	& weighted cost ratio		\\
		\hline
		CH4		& 229	& 0.2588		& 0.1298		& 29.7				& 240			& 24.6	\\
		DMS		& 166	& 3.2179		& 1.6169		& 268				& 1770			& 213	\\
		ETSH	& 166	& 0.4047		& 0.2025		& 33.77				& 263			& 27.2	\\
		H2O		& 119	& 0.015			& 0.0075		& 0.9				& 3.63			& 0.465	\\
		PH3		& 224	& 0.038			& 0.0191		& 4.2882			& 27.24			& 3.285	\\
		PSB3	& 166	& 0.016			& 0.0081		& 1.3403			& 7.547			& 0.848	\\
		\hline

	\end{tabular}
	\caption{Toffoli gate complexity ratios for encoding PESs with SELECT-SWAP over WH-QROMs for several molecules considered (cf. Table~\ref{tab:PES-polyaotmic}). Comparison with $L^\infty$ normalization. The complexities correspond to $\epsilon =2^{- 10}$ error tolerance.}
    \label{table:WHvsSelectSWAPinfty}
\end{table}

We conclude that the relative advantage of using the multiplexed rotations method or the canonical operator method with WH-QROM for block-encoding PESs is molecule dependent, showing a substantial overall reduction in quantum volume compared to SELECT-SWAP.

\subsection{Rotationally excited methane ($D=9$)}
To put our method of mixed FBR-DVR Hamiltonian in polyspherical coordinates with WH-QROM to the test, we estimate the cost of block-encoding the Hamiltonian for the CH$_4$ molecule using orthogonal polyspherical coordinates for two rotational excitation levels, $J=20$ and $J=400$. We use the DVR-FBR method, which provides the best asymptotic scaling, as shown in Table~\ref{tab:poly_spher_scaling-dvr-fbr}. The Hermite grid used for radial coordinates is defined be two parameters: equilibrium bond length $R_0$ and characteristic length $l=\sqrt{\frac{\hbar}{\mu\omega}}$, such that $R_i = R_0+x_il$ where $x_i$ is $i$-th zero of the appropriate Hermite polynomial. Similarly, for Legendre grid used for angular coordinates we have $\theta_i = \theta_0-\frac{\theta_{max}}{\pi}\arccos{y_i}$, where $y_i$ corresponds to $i$-th zero of a appropriate Legendre polynomial. The KEO is implemented using the $\rho$-sparse block-encoding method and SELECT-SWAP QROM, while the PES is block-encoded with the method described in sec.~\ref{sec:be_without_rotations} using WH-QROM.

\begin{table}[h]
    \centering
    \begin{tabular}{c|c|c|c|c|}
        Molecule& T-count & Ancillas & Norm [cm${-1}$]  &Quantum volume\\
        \hline
              $ \BE{H^{DVR}}_{\rho\text{-sparse}} $&- & -& -&-\\
              $ \BE{H^{DVR}}_{LCU} $&- & -& -&-\\
              $\BE{K}^{(SEL-SWAP)}_{(J=20)}$  &$1.0\times 10^{6}$& $871(6.7\times10^3)$& $3.5\times10^{11}$& $2\times 10^{9}$\\
              $\BE{K}^{(SEL-SWAP)}_{(J=400)}$  &$1.6\times 10^{6}$& $871(6.7\times10^3)$& $3.6\times10^{11}$& $2\times 10^{9}$\\
              $QROM_V^{(SEL-SWAP)}$  &$2\times 10^{11}$& $1.2\times 10^{10}$& $-$ & $1.4\times 10^{21}$\\
              $QROM_V^{(WH)}$  &$1.6\times 10^{11}$& $255$& $-$& $2.6\times10^{13}$\\
              $\BE{\hat{N}}$  &$1.3\times 10^{4}$& $548$& $6.5\times10^4$ &$7\times10^{6}$\\
    \end{tabular}
    \caption{Costs of block-encoding the CH$_4$ rovibrational Hamiltonian   in polyspherical coordinates with parameters: $\omega=5\times10^3 cm^{-1}$, $\mu=1 Da$, $R_0=1.65 $\AA, $n_\theta=n_\phi=2^{7}$, $n_R = 2^{7}$, $\theta_{max} = 0.5\frac{\pi}{2}$. $\BE{\hat{N}}$  is block-encoding of the number operator in the canonical operator method used for resource estimation. Values in the third column assume minimal number of ancillas achievable by serial implementation of partial block-encoding, while the number in the brackets gives estimates for parallelized implementation}
    \label{tab:norm_estimates_ch4_2}
\end{table}
Table \ref{tab:norm_estimates_ch4_2} demonstrates a more than six orders of magnitude reduction in quantum volume for block-encoding the rovibrational Hamiltonian for methane, compared to previous methods. Note that the dependence of T-count and quantum volume on the rotational excitation level $J$ is weak, in contrast to classical calculations. 

\subsection{Rotationally excited molecule with $D=30$ coupled vibrational modes}
To illustrate how quantum resources scale with dimensionality in our approach, we consider a hypothetical polyatomic molecule described by an exact KEO combined with a non-SOP PES that couples all degrees of freedom at arbitrary strength. Such a system would be computationally prohibitive for present day classical methods. We assume that the WH-QROM scaling behavior reported in Table~\ref{tab:PES-polyaotmic} remains valid and adopt a representative value of $\alpha=0.3$. In addition, we aim to demonstrate the scaling behavior with increasing rotational excitation. Table~\ref{tab:M30} summarizes the resulting resource estimates for this system.
\begin{table}[h]
    \centering
    \begin{tabular}{c|c|c|c|c|}
        Molecule& T-count & Ancillas & Norm [cm${-1}$]  &Quantum volume\\
        \hline
              $\BE{K}^{(SEL-SWAP)}_{(J=0)}$  &$2.7\times 10^{6}$& $295(8.9\times10^3)$& $6.8\times10^{5}$& $4.6\times 10^{8}$\\
              $\BE{K}^{(SEL-SWAP)}_{(J=100)}$  &$2.8\times 10^{6}$& $295(8.9\times10^3)$& $8.7\times10^{5}$& $4.8\times 10^{8}$\\
              $\BE{K}^{(SEL-SWAP)}_{(J=200)}$  &$2.8\times 10^{6}$& $420(1.2\times10^4)$& $1.2\times10^{6}$& $4.8\times 10^{8}$\\
              $\BE{K}^{(SEL-SWAP)}_{(J=400)}$  &$2.9\times 10^{6}$& $435(1.3\times10^4)$& $2.7\times10^{6}$& $4.8\times 10^{8}$\\
              $QROM_V^{(SEL-SWAP)}(\text{crude})$  &$9.8\times 10^{24}$& $6.1\times 10^{23}$& $-$ & $6.0\times 10^{48}$\\
                            $QROM_V^{(SEL-SWAP)}(\text{reduced})$  &$2.5\times 10^{16}$& $>75$& $-$ & $ 1.8\times 10^{18}$\\
              $QROM_V^{(WH)}(\text{crude})$  &$2.6\times 10^{21}$& $393$& $-$& $1.0\times10^{24}$\\
            $QROM_V^{(WH)}(\text{reduced})$  &$1.2\times 10^{10}$& $393$& $-$& $4.7\times10^{12}$\\
              $\BE{\hat{N}}$  &$1.3\times 10^{4}$& $548$& $1.6\times10^4$ &$7\times10^{6}$\\
    \end{tabular}
    \caption{Costs of block-encoding Hamiltonian for generic molecule with $30$ vibrational modes. The parameters used for the calculation are: $\omega=5\times10^3 cm^{-1}$, $\mu=7 Da$, $R_0=3$ \AA, $N_\theta=2^{4}$, $N_R = 2^{4}$, $\theta_{max} = 0.7\frac{\pi}{2}$. \textit{Reduced} denotes LMR representation of the PES with $L=6$ and $\binom{30}{6}$ terms~\cite{Bowman2003}, while \textit{crude} denotes fully coupled PES. $\BE{\hat{N}}$  is block-encoding of the number operator in the canonical operator method used for resource estimation. }
    \label{tab:M30}
\end{table}

We observe that the SELECT–SWAP QROM implementation introduces a substantial gate and qubit overhead. Although  WH-QROM reduces this cost by more than 25 orders of magnitude, the resulting total T-count remains prohibitively large. The significant QROM cost reported in Table~\ref{tab:M30} for the PES primarily originates from the total number of grid points at which the surface would nominally be evaluated.
However, the vast majority of these points correspond to configurations associated with very high energies. For this reason, the number of unique PES values that must actually be loaded via QROM is expected to be several orders of magnitude smaller. Moreover, any intrinsic structure or smoothness present in the PES could be further exploited to reduce the effective QROM cost.

For this reason, in addition to \textit{crude} model, where all modes are coupled simultaneously, we present in Table~\ref{tab:M30} WH-QROM resource estimates under a more physically realistic scenario (denoted as \textit{reduced}). Further reductions are however needed and possible.
Most molecular systems can be accurately described with coupling lengths lower than the total number of modes~\cite{Schrder2024}.
A LMR PES that couples up to four modes is most commonly adopted in classical computation. Extending beyond this coupling length quickly becomes intractable.
We consider a molecular system comprising 30 vibrational modes, with couplings that include up to $L = 6$ modes simultaneously, which should suffice even for a highly anharmonic or floppy molecule. No SOP form of the PES is assumed. We show resource estimation for QPE for the case of $R=1000$ terms in the $L=6$ non-SOP coupled PESs in Table~\ref{tab:1cm}.

In practice, such calculations are typically preceded by a series of low-dimensional computations solving one-dimensional Schrödinger equations along the vibrational coordinates.
The resulting potential-optimized DVR (PODVR) basis functions then provide a high-quality basis set.
Assuming $n = 10$ PODVR basis functions per mode, and that $R=1000$ $L$-body coupling configurations contribute to the PES, we estimate the total T-count for WH-QROM as $2^{\alpha \cdot\log_2(Rn^L)}\log^{\xi}(1/\varepsilon) =  2.3\times 10^{14}$ T-gates, where $\alpha = 0.3$, $\xi = 4$, and $\varepsilon = 1$cm$^{-1}$ were used in the estimation. By comparison, the SELECT–SWAP approach would require approximately $10^{16}$ $T$-gates for a similar number of ancilla qubits. Further estimates and more detailed analysis of T-gate count dependence on basis set parameters can be found in Appendix~\ref{app:polyspherical}.

The scaling behavior of the KEO block-encoding and its associated normalization constant is also favorable in our method, resulting in only a moderate increase in T-count, compared to 9-dimensional methane (cf. Table~\ref{tab:norm_estimates}). Furthermore, the overall quantum resource requirements for the 30-mode molecule do not increase substantially as the total rotational quantum number increases from $J = 0$ to $J = 400$.
Finally, we note that the high $T$-gate requirements shown in Table~\ref{tab:M30} reveal a fundamental limitation of the Born–Oppenheimer framework, the need for evaluating the PES at exponentially many grid points. A quantum oracle capable of constructing the PES by directly estimating the eigenvalues of the electronic Hamiltonian could mitigate this limitation, insensitive of the specific approximations employed.

\subsection{Resource requirements for QPE}
In Table~\ref{tab:1cm} we compare resource estimates for QPE for molecular systems with increasing dimensionality in the range $D=9-51$, along with various truncations to the LMR form of the PES. 

\begin{table}[h]
    \centering
    \begin{tabular}{c|c|c|c|c}
         Molecule&$\BE{H}$&Norm&QPE T cost & QPE ancillas  \\
         \hline
         $CH4_{L=5}$& $1.1\times10^{6}$ &$1.4\times10^6$&$1.8\times10^{13}$&$230$\\
         $D=21,L=5$& $5.9\times10^6$ &$7.0\times10^5$&$7.1\times10^{12}$&$240$\\
         $D=30,L=5$& $1.2\times10^7$ &$1.2\times10^6$&$2.4\times10^{13}$&$240$\\
         $D=51,L=5$& $2.7\times10^7$ &$2.1\times10^6$&$9.8\times10^{13}$&$250$\\
         $CH4_{L=6}$& $1.4\times10^{6}$ &$1.4\times10^6$&$5.2\times10^{11}$&$240$\\
         $D=21,L=6$& $1.1\times10^7$ &$7.0\times10^5$&$1.4\times10^{13}$&$250$\\
         $D=30,L=6$& $2.2\times10^7$ &$1.2\times10^6$&$4.7\times10^{13}$&$250$\\
         $D=51,L=6$& $5.0\times10^7$ &$2.1\times10^6$&$2\times10^{14}$&$260$\\
         $CH4_{L=9}$& $1.0\times10^6$ &$1.4\times10^6$&$2.3\times10^{12}$&$280$\\
    \end{tabular}
    \caption{Resource estimation for molecules with $n=10$,  $\epsilon_{QPE}=1$~cm$^{-1}$. For general molecules $D=21,30,51$, the parameters used are $\theta_{max}=0.6\frac{\pi}{2}$, $r_0=3.5$, $\omega=5000$~cm$^{-1}$, and $R=1000$ PES terms. For $CH_4$ we used: $\theta_{max}=0.5\frac{\pi}{2}$, $r_0=1.8$, $\omega=5000$~cm$^{-1}$.}
    \label{tab:1cm}
\end{table}

\begin{figure}[H] 
    \centering
    \includegraphics[width=0.9\linewidth]{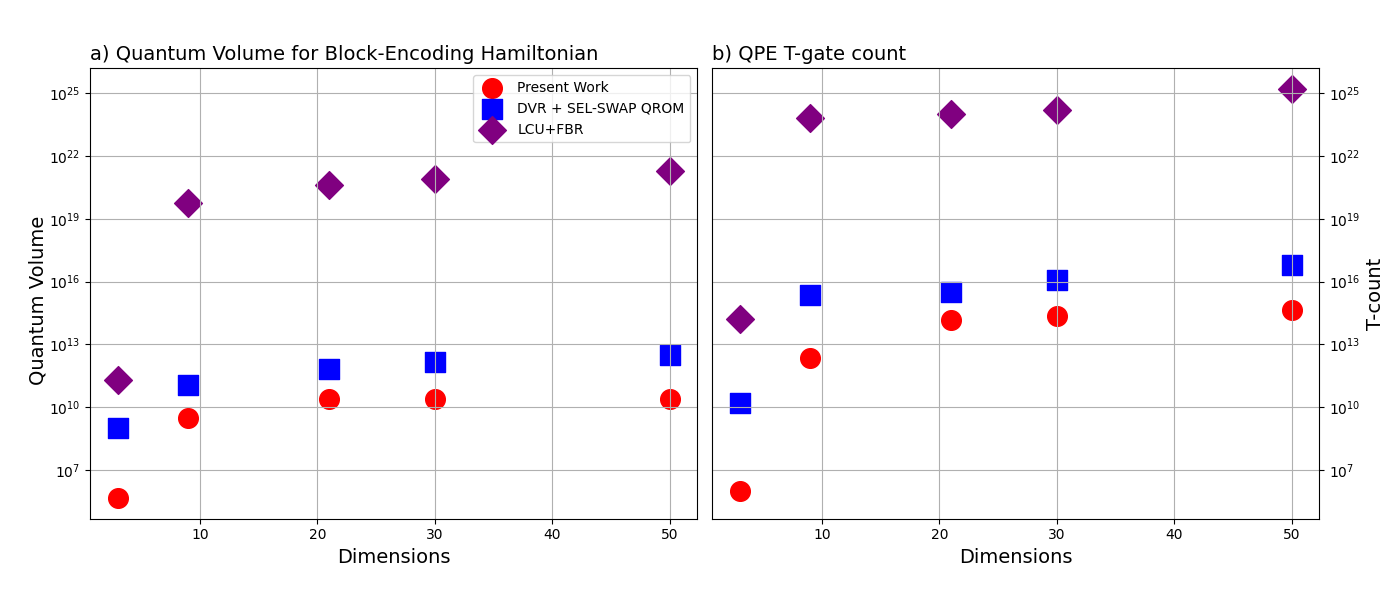}
    \caption{a) Comparison of quantum volume (T-gate times qubit count) for block-encoding rovibrational Hamiltonians at $J=0$ for water (3D), methane (9D) and model molecules ($21D$,$30D$,$51D$). Shown are: our present work using the FBR-DVR form of the Hamiltonian with WH-QROM (red filled circles), present work using LCU + FBR (purple squares), present work using SELECT-SWAP QROM for DVR-FBR Hamiltonian (blue squares). b) Respective T-gate count for QPE. $L=6$ PES is used for all cases except methane where $L=D=9$ and water where $L=D=3$.}
    \label{fig:scaling}
\end{figure}

Results of block-encoding and QPE costs are summarized in Figure~\ref{fig:scaling}. We compare quantum complexities (quantum volume and T-count) for block-encoding rovibrational Hamiltonians at $J=0$ for water (3$D$), methane (9$D$) and model molecules (21$D$,30$D$,50$D$). We observe that for all molecules considered  our method gives several orders of magnitude lower resource estimates than other techniques and this scaling becomes more favourable as the dimensionality increases. 

\section{Discussion}
\label{sec:discussion}
\begin{figure}[!h] 
    \centering
    \includegraphics[width=0.8\linewidth]{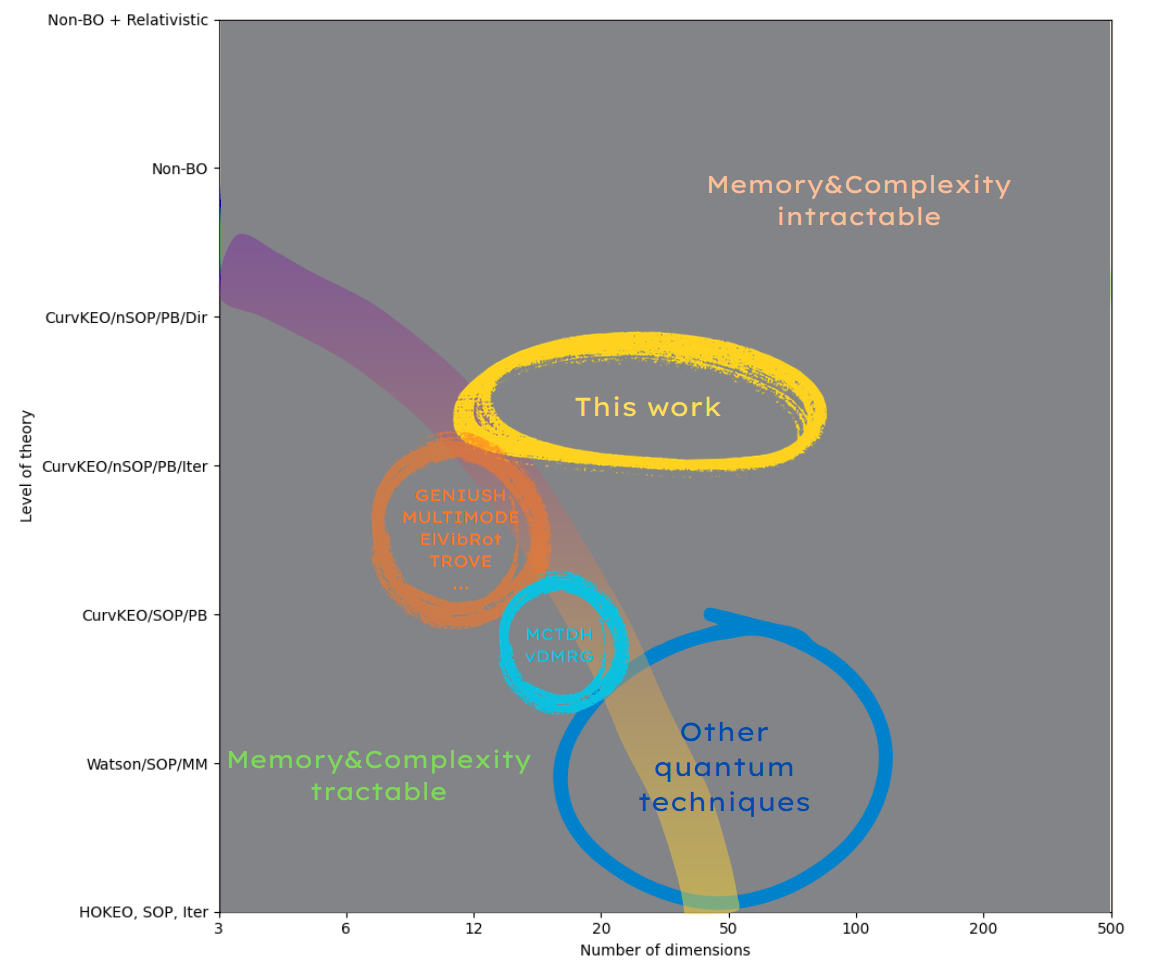}
    \caption{Sketch of computational complexity regimes in rovibrational calculations. The horizontal axis represents the number of dimensions (i.e., internal coordinates), while the vertical axis denotes increasing levels of theoretical accuracy. The memory-intractable regime corresponds to calculations requiring classical memory beyond the typical capacity of current HPC architectures, whereas the complexity-intractable regime refers to computational costs, measured in floating-point operations (FLOPs), that exceed realistic execution times on present-day CPUs and GPUs. The boundaries shown are approximate and intended for illustrative purposes only. Selected classical computational methods are indicated, along with a schematic region where existing quantum simulation techniques are expected to apply. \textit{HOKEO} denotes the harmonic-oscillator KEO, \textit{Iter} an iterative eigensolver (few eigenvalues), \textit{Dir} a direct eigensolver (many eigenvalues), \textit{Watson} the rectilinear Watson Hamiltonian~\cite{Watson1968}, \textit{PB} pruned or contracted basis methods, and \textit{non-BO} is full non–Born–Oppenheimer theory. }
    \label{fig:complexity}
\end{figure}

Comparing our quantum algorithm for rovibrational Hamiltonian simulation with other quantum simulation methods requires accounting for both the relative accuracy of the underlying model used to calculate energy levels and the associated quantum resources, namely the qubit and T-gate counts. Comparison with classical computing methods is more challenging, due to the disparity in resource handling, the costs of these resources, and the differing dynamics of their improvement. To capture this, we divide the computational problem space into computational regimes depicted in Figure~\ref{fig:complexity}. We compare quantum and classical computational methods within the context of these regimes, with the goal of giving a bird's eye view of computational landscape, highlight the state of the art, and position our method within.

Our contribution in this work can be summarized as follows. We introduced a method that:
\begin{enumerate}
    \item  uses a general rovibrational Hamiltonian with an exact KEO in internal coordinates and a non-SOP PES;
    \item  introduces a fault-tolerant quantum algorithm for simulating such Hamiltonians, via block-encoding and Quantum Phase Estimation, using a new QROM based on Walsh-Hadamard transform;
    \item utilizes discrete variable representations for representing Hamiltonians to reduce quantum resources.
\end{enumerate}
We analyzed several representations of the rovibrational Hamiltonian and assessed their relative utility for quantum computation, with particular attention to qubit count and Clifford+T gate budgets. Among the models considered, a mixed finite-basis-representation/discrete-variable-representation (FBR–DVR) form of the Hamiltonian proved to be the most advantageous, as shown in eq.~\ref{eq:DVR-FBR-cost}, Tables~\ref{tab:HO2_cost}–\ref{tab:HO2_T_counts} for the water molecule benchmark and in Fig.~\ref{fig:scaling} for other polyatomic molecules.

For water in valence coordinates, the efficiency gain comes from a reduced prefactor in block-encoding cost, which is up to two orders of magnitude lower than that of either FBR or DVR alone. Moreover, FBR–DVR block-encoding requires approximately $4000\times$ fewer Clifford gates, a resource increasingly recognized as non-negligible in fault-tolerant quantum computation~\cite{Gidney2024}. In terms of T-gates, for a fixed number of ancilla qubits, the cost for water is reduced from $10^{11}$ (FBR) to $10^{5}$ (FBR–DVR), a nearly six-order-of-magnitude improvement.
These gains stem directly from the choice of Hamiltonian representation, with the standard SELECT-SWAP QROM. This advantage arises in part because Gaussian DVRs represent local operators, such as the PES and the $G$-matrix in the KEO, as diagonal matrices, at the expense of implementing the DVR–FBR transformation unitary. This transformation however remains computationally inexpensive due to the direct-product structure of the Hilbert space and the low cost of each one-dimensional DVR–FBR unitary. Overall, the quantum resource complexity reduction compared to other methods scales at least as $\mathcal{O}(n^3)$ in both the T-gate-count times qubit-count metric and the Clifford gate count, where $n$ is the number of basis functions per internal coordinate.

For general polyatomic molecules, the computational cost depends on the choice of internal coordinates. For polyspherical coordinates, the asymptotic T-count scaling of block-encoding the FBR–DVR Hamiltonian is
\begin{equation}
    \mathcal{O}\left(D^2n^{\frac{9}{2}}+Dn^{3}+\sqrt{2J+1}+2n^{\frac{3}{2}}+n^{\alpha D}\right)
\end{equation}
which is dominated by the PES contribution. When SELECT-SWAP QROM is used throughout, $\alpha=1/2$ with exponentially many ancilla qubits ($\mathcal{O}(N^{\frac{1}{2}})$), and $\alpha=1$ with $\mathcal{O}(d)$ qubits, where $d$ is precision of matrix elements and $N=n^{D}$. The overall complexity depends on the locality of the PES, which here is assumed to be fully coupled, i.e., $D=3A-6$. For high-accuracy PESs, e.g. those obtained with machine learning, all vibrational modes can be coupled, a situation not explicitly addressed in previous works.
Despite the diagonal form of the PES in DVR, which significantly reduces block-encoding costs, the QROM input/output overhead remains substantial. To address this, we devised a WH-QROM, which mitigates the cost associated with the large volume of PES datapoints. With WH-QROM, $\alpha$ lies in the range $[0.25,0.8]$ while requiring only $\mathcal{O}(\eta+d)$ ancilla qubits with $\eta=\log_2N$. WH-QROM can be conveniently used with a block-encoding algorithm we propose that eliminates the need for multiplexed rotations~\cite{low2024,Gosset:2024}, instead using two calls to WH-QROM plus an implementation of a position-like operator (cf. sec.~\ref{sec:be_without_rotations}). This leads to a tangible reduction in quantum volume compared with alternative approaches. 

Indeed, our method of choice, which combines the FBR–DVR Hamiltonian with WH-QROM reduces quantum resource requirements relative to previously proposed approaches (FBR-based or SELECT-SWAP-based). For example, for the water molecule, loading the high-accuracy PES~\cite{Huang2008} into quantum memory scales as $2^{0.7292 \cdot \eta - 3.2896} ( \log_2  (\frac{1}{\epsilon} ) )^{2.4458} $ in contrast to the $2^\eta$ scaling of SELECT-SWAP QROM, with even more favorable behavior observed for other molecules. As shown in Table~\ref{table:WHvsSelectSWAPinfty}, for larger molecular systems the reduction in Toffoli (T-gate) and Clifford counts associated with implementing the multiplexed rotations in block-encoding of the DVR PES reaches factors of up to $1770\times$, while simultaneously requiring exponentially fewer qubits as the DVR basis size increases. For some PES types though block-encoding through multiplexed rotations is not favourable with WH-QROM. For this reason, an alternative algorithm has been proposed in sec.~\ref{sec:be_without_rotations} (canonical operator method) with results shown in Tables~\ref{table:WH_vs_SelectSWAP_QROM},\ref{tab:M30} and in supplementary information.

Taken together, these results indicate that among the approaches studied, the combination of WH-QROM with the FBR–DVR Hamiltonian provides the most resource-efficient strategy for quantum simulation of rovibrational Hamiltonians, both when block-encoding is implemented with multiplexed-rotations for representing Hamiltonian matrix elements or through the canonical operator method. 

If block-encoding with multiplexed rotations is used, then diagonal unitary synthesis, where QROM is the main cost driver, becomes more efficient. State-of-the-art algorithms for diagonal unitary synthesis based on SELECT-SWAP scale as $\mathcal{O}\left(N^{1/2}\log^{0.5}_2 (1/\epsilon)\right)$ at the expense of $\mathcal{O}(N^{1/2})$ ancilla qubits, or $\mathcal{O}(N\log_2 (1/\epsilon))$ with $\mathcal{O}(d)$ ancilla qubits. By contrast, our diagonal unitary synthesis with WH-QROM scales as $ O\left(N^{\alpha}\log^{\xi}_2(1/\epsilon)\right)$ with $O(\log_2N+d)$ ancilla qubits, where $\xi\approx 2-5$ and $\alpha$ in range 0.25-0.8 for the studied molecules. The general scaling of PES synthesis cost with WH-QROM compared to SELECT-SWAP was demonstrated in Table~\ref{tab:PES-polyaotmic}, where the gains in Toffoli count are exponential with the number of qubits. 

In summary, the use of WH-QROM significantly improves the scaling of diagonal unitary synthesis from $C_D^{\text{SEL-SWAP}} = \mathcal{O}(N \log(1/\varepsilon))$ to $C_D^{\text{QROM}} = \mathcal{O}(N^{\alpha} \log^\xi(1/\varepsilon))$ at constant ancilla overhead, where $\alpha \approx 0.25$–$0.8$ and $\xi$ is a small constant.
We emphasize that these results stem from applying WH-QROM directly to the full PES. Its effectiveness depends on properties such as the PES gradient (cf. Eq.~\ref{eq:wh-qrom-cmplexity}) and can be further improved through hybrid schemes, where WH-QROM is applied to well-behaved regions of the PES and SELECT-SWAP elsewhere. 
When finite coupling length in PES is sufficient ($L=\mathcal{O}(1)$), our method achieves scaling $\mathcal{O}(Rn^{L\alpha} + \mathrm{poly}(n)_{KEO})= \mathcal{O}(\mathrm{poly}(D)\mathrm{poly}(n))$ in Clifford+T gates and $\mathcal{O}(\log(N))$ in qubits, where we assumed $R=\mathcal{O}(poly(D))$.

\paragraph{Comparison with other methods.}
Our approach exhibits favorable scaling and is expected to become increasingly advantageous with molecular size and complexity. A particularly challenging class of systems long recognized as intractable for classical computation are floppy molecules and weakly bound clusters, characterized by high state densities~\cite{Derbali2025}, strongly coupled low-frequency vibrational modes, and significant rovibrational couplings. For these systems, where high accuracy is essential, previous quantum algorithms~\cite{mcardle2019,Kallullathil2023,Malpathak2025,motlagh2024a,Trenev2025,Majland2025} offer limited applicability. To our knowledge, the present work provides the first explicit quantum algorithm tailored to this regime, precisely where quantum computation offers the greatest potential benefits.

Compared with earlier proposals relying on simplified KEOs or SOP PESs in FBR~\cite{mcardle2019,Kallullathil2023,Malpathak2025,motlagh2024a,Trenev2025}, our method improves both accuracy and resource efficiency through a combined DVR representation and WH-QROM. SOP PESs, including polynomial forms used in prior work~\cite{ollitrault2020-nonad,mchale2017}, remain efficiently compatible with WH-QROM, while more complex non-SOP PESs, such as neural-network potentials, do not restrict its applicability.
The methods of ref.~\cite{Majland2025}, ref.~\cite{Malpathak2025}, and ref.~\cite{Trenev2025} adopt multimode-type Hamiltonians of higher quality than most other approaches summarized in supplementary materials. Their general common scaling is $\mathcal{O}(D^L n^{2L})$, compared to our $\mathcal{O}(D^2n^{\alpha L})$, with the key distinction that our Hamiltonian uses the exact KEO and a general (non-SOP) PES, enabling higher accuracy. Even in the fully coupled case ($L=D$), our method retains more favorable scaling while maintaining higher accuracy in general. For low coupling-lengths other methods can be advantageous.  Except for Ref.~\cite{Majland2025}, existing approaches generally do not provide explicit block-encodings of the rovibrational Hamiltonian, further limiting direct comparison.

\paragraph{Classical processing overheads.}
We should also be mindful of classical computer pre- and post-computation required by quantum algorithms, raising an additional overhead. The Walsh-Hadamard transform requires $\mathcal{O}(Ln^L\log n)$ FLOPS associated with classical precomputation, where $L\leq D$ is coupling length in the PES. One possibility to mitigate this overhead is to fit the WH form of PES from the beginning. 

Next, the construction of the PES requires a number of quantum chemistry calculations, size of which can be significant. Additionally, simulation of Born-Oppenheimer rovibrational dynamics requires careful handling of the PES, cost of which can quickly become prohibitive, even for quantum computers and our algorithm. Our algorithms comes with a significant constant prefactor due to block-encoding scaling constant and the volume of PES datapoints. 
One way of mitigating the large number of datapoints that must be loaded for a general PES is to use a pre-Born-Oppenheimer model~\cite{su2021,Mtyus2018}, which comes with other inconveniences. Some new ideas are here needed, and combining coherently a quantum algorithm solving electronic structure ~\cite{babbush2018,burg2021,lee2021,rocca2024,loaiza2024,Deka2025} with block-encoding of the DVR PES could be one possible way to proceed. 

Coupling coherently quantum computation for the electronic structure and nuclear motion calculation within the Born-Oppenheimer framework is beyond the scope of this work, but in particular could be achieved by qubit-encoding appropriate multidimensional grids of internal coordinates. In doing so, one would devise a quantum circuit that encodes the nuclear configuration grid into a coherent superposition and applies QPE to the electronic Hamiltonian, without measuring the estimation register. This Hamiltonian could be generated by appropriately shifting a reference-geometry electronic Hamiltonian using the translation operator $T(\vec{a}_k)=\exp(-\vec{a}_k\nabla_q)$, encoded as a quantum circuit and controlled on the nulcear geometry grid point index. The required matrix elements of the second-quantized electronic Hamiltonian are then constructed via quantum arithmetic and analytic relations. As a result, the output of the electronic QPE yields a coherent superposition of the form $\sum_{k=1}^N \ket{\mathbf{q}_k}\ket{V(\mathbf{q}_k)}$, which can be combined with our canonical operator method for block-encoding the PES.
This procedure requires circuit-level implementations of analytic formulas for wavefunction propagation in nuclear coordinate space in order to obtain nuclear-dependent matrix elements, such as $g_{pqrs}(\mathbf{q}_k)$.
While this approach may incur polynomial overheads in both logical qubits and T-gate counts, it could, in principle, alleviate the burden of explicitly evaluating the Born–Oppenheimer PES on a DVR grid. The DVR grid points $\mathbf{q}_k$ can be efficiently generated from a small number of one-dimensional quadratures, owing to the direct-product structure of the grid. Studies presented, for example, in Ref.~\cite{Sasmal2020}, which treat nonadiabatic dynamics without directly resorting to potential energy surfaces, motivate further work aimed at mitigating the reliance on Born–Oppenheimer PESs.

\subsection{Quantum vs. classical complexity scaling and resource comparison}
The usefulness of quantum algorithms must be benchmarked against established classical methods, which requires careful alignment of model accuracy with the corresponding quantum and classical computational resources. Such comparisons are challenging due to fundamental differences in how resources are defined, costed, and improved in quantum versus classical computing. As a result, quantifying quantum advantage remains nontrivial and somewhat subjective, depending on rapidly evolving factors such as quantum error-correction overheads, gate fidelities and speeds, quantum–classical control latencies, and advances in classical memory and processor architectures. Accordingly, the comparisons presented here should be viewed as primarily qualitative. Below, we summarize the relevant metrics and asymptotic complexities of representative classical and quantum algorithms.

\subsubsection{Classical Computing Scaling}
Classical computation of rovibrational energy levels uses two primary techniques: direct diagonalization and iterative methods. Direct diagonalization algorithms scale as $\mathcal{O}(N^3)$ FLOPs with memory requirements of $\mathcal{O}(N^2)$, where $N$ is the total dimension of the basis.
Iterative procedures, particularly Krylov subspace methods, exhibit superior scaling of $\mathcal{O}(N\rho M_{\text{Kr}})$ FLOPs, where $\rho$ represents the matrix sparsity and $M_{\text{Kr}}$ denotes the Krylov subspace dimension. The Krylov dimension scales logarithmically with the convergence precision, $M_{\text{Kr}} = \mathcal{O}(\log(1/\varepsilon))$, and linearly with the number of requested eigenvalues $N_{\text{eval}}$. For computing all eigenvalues, though not the optimal application of iterative methods, the scaling approaches $\mathcal{O}(N^2 \rho)$. When matrices exhibit logarithmic sparsity, $d = \mathcal{O}(\log N)$, iterative eigensolvers provide computational advantages with memory requirements of only $\mathcal{O}(N)$. 

\subsubsection{Quantum Computing Scaling}
For quantum algorithms applied to $A$-atomic molecular rovibrational Hamiltonians, the computation of a single energy level  scales as $\widetilde{\mathcal{O}}(C_H \zeta /\varepsilon)$ in Clifford+T gate count. $\zeta$ denotes the total block-encoding normalization constant, as illustrated in Table~\ref{tab:HO2_cost} and is determined by the largest eigenvalue of the Hamiltonian and its sparsity. For Hamiltonians without singularities in the KEO, the scaling is dominated by $V_{\max} d$, where $V_{\max}$ represents the maximum value of the PES, corresponding to the energy range over which the computation is valid. This parameter scales linearly with the energy and sublinearly with $N_{\text{eval}}$. For DVR rovibrational Hamiltonians, the matrix sparsity typically satisfies $d = \mathcal{O}(\log N) = \mathcal{O}(D \log n)$, influencing both quantum and classical computational scaling. For exact DVR KEOs however, the dominant contribution to $\zeta$ may arise from singular terms of the KEO sampled at quadrature grid points. Careful selection of quadrature schemes and basis set parameters is therefore essential to mitigate this issue.

\subsubsection{Resource comparison for different computational complexity regimes}
\paragraph{Classical memory and time-complexity intractable regime.} In this regime we consider iterative eigensolver for computing a single energy level and compare it with the respective quantum computing cost. Here we assume direct-product basis sets, such that the total number of basis functions is $N=n^D$. 
Our method's Clifford+T complexity scales as $\mathcal{\widetilde O}\left(\zeta (n^{\alpha D}+\sqrt{2J+1})/\varepsilon\right)$ in rotational excitation and $\mathcal{O}\left(D^2n^{\frac{7}{2}}\right)$ in qubits, compared to $\mathcal{O}(Jn^{D}\kappa^{\frac{1}{2}}\rho \log (1/\epsilon))$ FLOPs and $\mathcal{O}(n^D)$ memory in the classically most optimistic case. The $\frac{1}{\varepsilon}$ scaling with precision is inherent to Heisenberg-limited quantum algorithms. $\zeta$ scales at least as  $\mathcal{O}(n^2)$. Upon dividing the quantum computational memory complexity by classical memory requirement and the respective time-complexities, we note that the advantage in memory is 
\begin{equation}
Saving = \mathcal{\tilde{O}}(n^{D-\frac{7}{2}}/D^2)
    \label{eq:advantage-qubit}
\end{equation}
and in time-complexity
\begin{equation}
Speedup = \mathcal{\tilde{O}}(n^{D(1-\alpha)})+ \mathcal{O}(\sqrt{J})
    \label{eq:advantage-time}
\end{equation}
When the PES is given as LMR with  $L\leq D$ maximum coupling length, at the expense of $R\in \mathcal{O}(poly(D))$ terms composing the PES, the speedup can be quantified as $Speedup = \mathcal{\tilde{O}}(n^{(D-L\alpha)}/poly(D))+ \mathcal{O}(\sqrt{J})$.
By directly comparing the quantum and classical requirements, we observe that the memory advantage is exponential: the quantum memory requirement scales polynomially with $D$ and $n$, while the classical requirement scales as $\mathcal{O}(n^D)$. The time-complexity speedup is more subtle. The classical-to-quantum complexity ratio illustrates an advantage of the quantum algorithm in the limit of large $D$ and $n$, with the quantum cost growing as $\mathcal{O}(n^{\alpha D})$ (with $\alpha < 1$ in practice) versus the classical $\mathcal{O}(n^D)$. We note however, that the prefactor in the quantum scaling can be large, limiting the utility of the present method for smaller systems. Similarly, the block-encoding scaling constant can take up large values, yet its scaling with system size is mild ($\mathcal{O}(D\log N)$). Also WH-QROM requires classical data processing which scales as $\mathcal{O}(Ln^L)$. These issues need to be addressed in future studies. Estimating multiple energy levels can be achieved efficiently through quantum landscape scanning methods based on QPE~\cite{GRM25}. We discuss this method in Appendix.~\ref{sec:appendix-landscape}.

Two more specific cases can be considered. 

\paragraph{Classical memory tractable and time-complexity intractable regime.} The first is the classical memory-tractable and computational complexity intractable region. It corresponds to a scenario in which the Hilbert space basis is small enough to fit in RAM, say, less than 1PB of data, but the computation of a large number of requested energy levels becomes prohibitively expensive. In this regime, classical memory-tractable Hilbert space dimensions present only a limited window for quantum advantage. Here, in principle, one could prepare the full state vector and directly load precomputed Hamiltonian matrix elements into a quantum device.
The prefactor in scaling of iterative algorithms may become significant when many eigenvalues are required. As a result, even when storing a single vector in classical memory is feasible, and one does not need to store the Hamiltonian explicitly for MVPs, the required number of FLOPs may easily exceed the capabilities of current classical computers. This scenario locates near the low-dimensionality and high-accuracy region in Figure~\ref{fig:complexity} (top-left), where quantum computation may offer an advantage catalyzed by classical memory tractability.

An example of this scenario may include solving the rovibrational Schrödinger equation using a basis set contraction scheme, as implemented in the MULTIMODE method~\cite{Bowman2003}. In such procedures, the classical computer constructs contracted basis sets up to a given dimensionality (number of coupled coordinates) and performs a series of diagonalizations and truncations. However, for a critical dimensionality $L$, further classical truncation to $l+1$ becomes computationally infeasible. The resulting compressed, yet still large, representation of the Hamiltonian can then be fed to a quantum algorithm, such as the one presented in this work. This approach is particularly useful when many highly excited state energies are of interest, as direct diagonalization becomes necessary due to the density of the Hamiltonian matrix.
In these cases, successive basis set truncations must be less restrictive to accurately capture highly excited states, which can result in matrices that are too large for direct diagonalization, even in the compressed MULTIMODE representation. A quantum computer can block-encode the Hamiltonian in the $L$-coupled representation and extend the solution to the $>L$-coupled problem. Nonetheless, identifying a clear quantum advantage in this scenario remains subtle, as the classical computational cost is not strictly prohibitive. 

As an illustrative example, consider the 12-dimensional water dimer problem: while its memory requirements (approximately 500 GB) remain manageable with modern classical hardware, converging millions of eigenvalues through iterative matrix–vector operations could demand years of large-scale parallel computation. On the other hand, efficient iterative solvers with spectral focusing could be used to mitigate the cost, leaving the verdict case-dependent.

\paragraph{Classical memory intractable and time-complexity tractable regime.} The other specific scenario
of computational regime where classical memory requirements exceed available resources while the time complexity remains theoretically tractable. In this regime, the Hilbert space dimension is too large to store even a single state vector in random access memory, yet the necessary matrix-vector products could theoretically be performed within reasonable time scales to extract essential spectroscopic information (i.e., few energy levels of large, strongly coupled systems).

Classical approximate methods such as the rank-reduction block power method (RRBPM)~\cite{Leclerc2014,Thomas2018} provide viable solutions in this regime. Quantum advantage emerges primarily from favorable memory scaling: while classical methods require $\mathcal{O}(N)= \mathcal{O}(n^D)$ memory for state storage, quantum algorithms can exploit quantum superposition to represent exponentially large state spaces using only $\mathcal{O}(\log N)=\mathcal{O}(D n)$ qubits.

Secondly, the total execution time for quantum computation carries substantial uncertainty, even when asymptotic scaling is established. For systems approaching the memory-intractable regime, quantum algorithms may achieve practical advantages through their superior memory efficiency, even before reaching any asymptotic time complexity improvements. For a rovibrational Hamiltonian of an $A$-atomic molecule, the computation of $N_{\text{eval}}$ energy levels scales as $\mathcal{\widetilde O}(NN_{eval}(C_P+\left[n^{\alpha D} + \sqrt{2J+1}\right]\zeta/\varepsilon ))$ in Clifford+T gate count compared to $\mathcal{O}(N\rho N_{eval}\log(\frac{1}{\varepsilon}))$ FLOPS for classical computation. Here, $C_P$ is trial state preparation cost for QPE, which in general case scales as $\mathcal{\widetilde O}(N)$. $\zeta$ scales favorably with the number of modes $D$, especially when sampling of singular DVR terms in the KEO is the main contribution, in which case $\zeta$ is largely independent of $D$; i.e. $\zeta \in \mathcal{O}(1)$, albeit with a relatively large prefactor. In the classical scaling we neglected the cost of Hamiltonian matrix elements calculations, which can scale as bad as $\mathcal{O}(N^3)$.
WH-QROM improves the scaling of block-encoding the PES from $\mathcal{O}(2^{D\log_2 N})$ to $\mathcal{O}(2^{\alpha D \log N})$, with $\alpha \approx 0.25$–$0.8$, which is exponentially more favorable than the classical scaling $\mathcal{O}(n^D \rho)$. In many DVR-based rovibrational Hamiltonians, $\rho = \mathcal{O}(\log N) = \mathcal{O}(D \log n)$.

\paragraph{Back-of-the-envelope estimate of simulation times.}
To provide a sense of the relative computational resources required by our method compared with classical techniques, we present a back-of-the-envelope estimate of memory usage and runtime for representative cases. We assume a parallelized architecture for logical T-gate generation based on magic-state distillation and surface-code error correction~\cite{Gidney2024}, capable of producing up to $10^{6}$ T-gates per second. While adopting such a model as a reference clock rate introduces significant uncertainty, our intention is to offer an illustrative comparison rather than a quantitative performance prediction.

For example, simulating a single rovibrational energy level at $J=400$ for methane (six-mode coupled PES, $\approx$ 1\invcm accuracy), $10$ basis functions per coordinate and a direct-product basis in FBR, would require roughly one month on half the total capacity of the Frontier supercomputer ($\approx$ 300,000 CPUs,  $\approx$ 670 PFLOPS). The same calculation would take about 3 days on a quantum computer operating at the T-gate rate considered above.  A comparable calculation for a 21-mode molecule (e.g. dimethylsulfide)  with a basis size of $10^{21}$ and 6-body couplings (see Table~\ref{tab:M30},~\ref{tab:1cm}, and in supplementary materials) would demand more than 1000 PB of RAM and an estimated $10^{13}$ years on Frontier. In contrast, using our algorithm (WH-QROM, Table~\ref{tab:1cm}), the same computation would take less than a year and require fewer than 300 logical qubits. This disproportion grows with dimensionality, such that for a 51-dimensional molecule with 6-mode coupled PES the classical computing time for energy level down to  $\approx$ 1\invcm\ is estimated to be longer than $10^{30}$ years on Frontier with $10^{30}$ PB of RAM, whereas quantum computation would take approximately 15 months requiring fewer than 300 logical qubits. 
In our estimates, we assume generic classical iterative algorithms and variational methodologies, while noting that further optimizations on the classical side are possible. In particular, system-specific approximations to the basis-set structure and Hamiltonian representation, such as those employed in MCTDH and tensor-network methods~\cite{Wang2003,Manthe2008,Vendrell2011,Lindoy2021,Wodraszka2020,Wodraszka2024}, can significantly reduce classical runtimes. Nevertheless, when high accuracy is required, these approaches typically do not overcome the unfavorable asymptotic scaling, and the general scaling advantage of the quantum method remains difficult to match with classical techniques.

\subsubsection{Motivation for First-Principles Quantum Simulation.}
We conclude our discussion by relating it to the experimental challenges that motivate the adopted paradigm for quantum simulation. Although experimental rovibrational spectra provide essential benchmark data, they are not always available, reproducible, or cost-effective. For instance, dimethyl sulfide is of considerable astrobiological interest as a candidate biosignature gas \cite{Madhusudhan2025}, yet its high-resolution rovibrational spectrum remains poorly characterized. Laboratory spectroscopy of such systems is often hindered by complex sample preparation, systematic and instrumental uncertainties, limited spectral coverage, and long acquisition times. In these cases, high-accuracy first-principles simulations play a critical role by providing reliable theoretical reference data. Their applications span spectroscopy, remote sensing, atmospheric modeling, materials science, and biochemical modeling.

The paradigm shift from semi-empirically-based modelling to first-principles is arguably one of the leading driving forces for quantum simulation. Our framework is designed to follow this paradigm shift and enable systematic investigations of complex phenomena, including weakly bound clusters, condensed-phase environments, and chemically relevant nuclear motion processes. Potential applications range from atmospheric and astrochemical spectroscopy to biochemical reactions, enzymatic catalysis, artificial photosynthesis, and the design of covalently binding small-molecule drugs. 

The central computational bottleneck is the lack of an effective strategy for describing large-amplitude, strongly coupled nuclear motion using compact basis representations. While increasing basis-set size and optimizing internal coordinates can, in principle, improve accuracy, existing approaches are severely constrained by memory requirements and the computational cost of PES evaluation. As demonstrated in this work, quantum computing architectures offer a promising pathway to address these limitations. By encoding Hamiltonian representations directly in qubit space, quantum devices may enable near-variational simulations of rovibrational dynamics that are currently infeasible on classical hardware. To our knowledge, this work presents the first quantum simulation formalism capable of treating rovibrational Hamiltonians with an exact curvilinear KEO and a general-form PES.

\section{Conclusions}
We have proposed a quantum algorithm for simulating rovibrational Hamiltonians, utilizing the DVR representation for parts of the Hamiltonian. DVR provides the advantage that local operators, including the PES, are diagonal regardless of their form or complexity. We combined DVR with a hybrid LCU / $\rho$-sparse Hamiltonian block-encoding model, proposing a method for block-encoding PES without multiplexed arbitrary angle rotations~\cite{Gosset:2024}. 
For example, for the water molecule, our approach yields favorable complexity compared to standard LCU and variational basis approaches, achieving an improvement of $\mathcal{O}(n^{6.5})$  in quantum volume and five-orders of magnitude reduction in T-count for realistic simulation parameters, solely due to the choice of representation. This improvement is enabled by the implementation of the DVR–FBR transformation unitary~\cite{plis25}. In addition to the improved Hamiltonian representation and block-encoding, we introduce a new QROM based on the Walsh–Hadamard transform. Combining the new WH-QROM with DVR and an efficient block-encoding algorithm demonstrated further gains in qubit count and T-gate complexity compared to other techniques.  With WH-QROM, block-encoding of the DVR Hamiltonian requires exponentially fewer quantum resources (qubit and T-count) with increasing dimensionality, relative to the SELECT-SWAP QROM~\cite{low2024}. The gain in the total runtime compared to other classical and quantum methods is polynomial in the total problem size, yet exponential with system's dimensionality. We note that our model describes the rotational-vibrational dynamics with an exact KEO and accuracy determined mostly by the quality of the PES. Other techniques~\cite{motlagh2024a,Majland2025,Trenev2025,Malpathak2025} assume sum-of-product representation of the PES and KEO in the form of series expansion, which may limit their accuracy.

Comparing performance with respect to today's classical computing techniques is challenging and should be considered with caution. Our resource estimation indicates that the both memory saving (qubit count vs. bit count), and time-complexity speedup shows improvement relative to classical computing methods, summarized by: 
\begin{equation}
\text{memory saving} \times \text{speedup} = \mathcal{\tilde{O}}\left(n^{D-\frac{7}{2}}/D^2\right) \times \mathcal{\tilde{O}}\left(n^{D(1-\alpha)}\right), \;\;\  \alpha <1,
\label{eq:advantage}
\end{equation}
where $D$ denotes the dimensionality of the problem and $n$ the number of basis functions per dimension. These findings should nonetheless be interpreted with care. Although the asymptotic scaling indicates an exponential in dimensionality advantage in quantum volume and polynomial advantage in the total problem size, the prefactors remain large and may constrain practical applicability unless further algorithmic and hardware advances are achieved. 
We also emphasize that such comparisons do not account for prospective developments or technological bottlenecks in either quantum or classical computing paradigms.

The main computational bottleneck within the Born–Oppenheimer framework arises from the need to evaluate the PES at a vast number of grid points. This limitation could be alleviated by reducing the PES coupling length at the expense of accuracy loss, coupling quantum electronic-structure algorithms with the present nuclear-motion method or by employing pre–Born–Oppenheimer formulations.  

Finally, the present technique can be integrated with our recently proposed quantum landscape scanning method~\cite{GRM25}, which enables the extraction of multiple eigenvalues (spectra) from block-encoded Hamiltonians at reduced computational cost. Although the exact usefulness of the class of algorithms we discuss in this work remains to be resolved, we motivate our study by potential applications in high-accuracy quantum dynamics and spectroscopy, atmospheric and exoplanetary research, proton-transfer processes in biology, molecular junctions, rovibrational dynamics of complex polyatomic systems, and the physics of weakly bound clusters.

\section{Acknowledgments}
We thank Joel Bowman for helpful remarks.
This work was funded by the European Innovation Council accelerator grant COMFTQUA, no. 190183782. 
\appendix

\section{Encoding multivariable functions}
For $x \in \F_2^n$, let $\overline{x} \eqdef - x_0 + \sum\limits_{a = 1}^{n - 1} \tfrac{x_a}{2^a}$. Let $\Theta : [- 1, 1]^D \rightarrow [- 1, 1]$ be a function. For each $\underline{n} \eqdef \( n_1, n_2, \ldots, n_D \) \in \Z_+^D$ with $n \eqdef \sum\limits_{a = 1}^D n_a$, we define the $n$-qubit diagonal unitary, $U_{\Theta, n_1, n_2, \ldots, n_D}$, as
\begin{equation}
	U_{\Theta, \underline{n}} \ket{x_1} \ket{x_2} \ldots \ket{x_D} = e^{i \pi \Theta \( \overline{x}_1, \overline{x}_2, \ldots, \overline{x}_D \)} \ket{x_1} \ket{x_2} \ldots \ket{x_D}.
\end{equation}
When $D = 1$, in \cite{zylberman_efficient_2025}*{Theorem IV.1.} the authors prove that if $\Theta$ is continuously differentiable, then for each $\epsilon > 0$, there is $n_\epsilon$, so that if $n \geqslant n_\epsilon$, then $U_{\Theta, (n)}$ and $U_{\Theta, (n_\epsilon)} \otimes \id_{n - n_\epsilon}$ are $\epsilon$ close. In \cite{zylberman_efficient_2024}*{Section 6.}, they also prove versions of this theorem for the $D > 1$ case; cf. equation (B58) in the reference. We state and prove an independent generalization below.

\begin{theorem}
	Assume that $\Theta$ is Lipschitz continuous with Lipschitz constant, $K_\Theta$. Let $\underline{n} \eqdef \( n_1, n_2, \ldots, n_D \) \in \Z_+^D$ and
	\begin{equation}
		\epsilon = \epsilon \( \underline{n} \) \eqdef 2 \pi K_\Theta \sqrt{\sum\limits_{a = 1}^D \tfrac{1}{4^{n_a}}}. \label{eq:epsilon}
	\end{equation}
	Then, for all $\underline{n}^\prime \eqdef \( n_1^\prime, n_2^\prime, \ldots, n_D^\prime \) \in \Z_+^D$, with $n_a \leqslant n_a^\prime$, we have
	\begin{equation}
		\| U_{\Theta, \underline{n}^\prime} - U_{\Theta, \underline{n}} \otimes \id_k \| < \epsilon,
	\end{equation}
    where $k \eqdef \sum\limits_{a = 1}^D \( n_a^\prime - n_a \)$.
\end{theorem}

\begin{proof}
	Let $(n_1, n_2, \ldots, n_D), (n_1^\prime, n_2^\prime, \ldots, n_D^\prime) \in \Z_+^D$, $k \in \Z_+$, and $\epsilon > 0$, as above. Since $U_{\Theta, (n_1^\prime, n_2^\prime, \ldots, n_D^\prime)}$ and $U_{\Theta, (n_1, n_2, \ldots, n_D)} \otimes \id_k$ are both diagonal unitaries, it is enough to to show the bound for the differences of each pairs of eigenvalues. Let $\( x_1^\prime, x_2^\prime, \ldots x_D^\prime \) \in \F_2^{n_1^\prime} \times \F_2^{n_2^\prime} \times \cdots \times \F_2^{n_D^\prime}$ be arbitrary and let $\( x_1, x_2, \ldots x_D \) \in \F_2^{n_1} \times \F_2^{n_2} \times \cdots \times \F_2^{n_D}$ be the element such that the first $n_a$ bits of $x_a^\prime$ equals to $x_a$. Then we have
	\begin{equation}
		\( U_{\Theta, (n_1^\prime, n_2^\prime, \ldots, n_D^\prime)} - U_{\Theta, (n_1, n_2, \ldots, n_D)} \otimes \id_k \) \ket{x_1^\prime} \ket{x_2^\prime} \ldots \ket{x_D^\prime} = \( e^{i \pi \Theta \( \overline{x}_1^\prime, \overline{x}_2^\prime, \ldots, \overline{x}_D^\prime \)} - e^{i \pi \Theta \( \overline{x}_1, \overline{x}_2, \ldots, \overline{x}_D \)} \) \ket{x_1^\prime} \ket{x_2^\prime} \ldots \ket{x_D^\prime}.
	\end{equation}
	The modulus of this eigenvalue is
	\begin{align}
		\left| e^{i \pi \Theta \( \overline{x}_1^\prime, \overline{x}_2^\prime, \ldots, \overline{x}_D^\prime \)} - e^{i \pi \Theta \( \overline{x}_1, \overline{x}_2, \ldots, \overline{x}_D \)} \right|	&= 2 \left| \sin \( \tfrac{\pi}{2} \( \Theta \( \overline{x}_1^\prime, \overline{x}_2^\prime, \ldots, \overline{x}_D^\prime \) - \Theta \( \overline{x}_1, \overline{x}_2, \ldots, \overline{x}_D \) \) \) \right| \\
			&\leqslant 2 \tfrac{\pi}{2} \left| \Theta \( \overline{x}_1^\prime, \overline{x}_2^\prime, \ldots, \overline{x}_D^\prime \) - \Theta \( \overline{x}_1, \overline{x}_2, \ldots, \overline{x}_D \) \right| \\
			&\leqslant \pi K_\Theta \left| \( \overline{x}_1^\prime - \overline{x}_1, \overline{x}_2^\prime - \overline{x}_2, \ldots, \overline{x}_D^\prime - \overline{x}_D \) \right| \\
			&= \pi K_\Theta \sqrt{\sum\limits_{a = 1}^D \( \overline{x}_a^\prime - \overline{x}_a \)^2}.
	\end{align}
	Now note that $0 \leqslant \overline{x}_a^\prime - \overline{x}_a = \sum\limits_{b = n_a}^{n_a^\prime - 1} \tfrac{\( x_a^\prime \)_b}{2^b} \leqslant 2 \( \tfrac{1}{2^{n_a}} - \tfrac{1}{2^{n_a^\prime}} \) < \tfrac{2}{2^{n_a}}$. Thus,
	\begin{equation}
		\left| e^{i \pi \Theta \( \overline{x}_1^\prime, \overline{x}_2^\prime, \ldots, \overline{x}_D^\prime \)} - e^{i \pi \Theta \( \overline{x}_1, \overline{x}_2, \ldots, \overline{x}_D \)} \right| \leqslant 2 \pi K_\Theta \sqrt{\sum\limits_{a = 1}^D \frac{1}{4^{n_a}}} = \epsilon.
	\end{equation}
\end{proof}

Note that, by the Mean Value Theorem, if $\Theta$ is continuously differentiable, then $K_\Theta$ can be replaced by $\| \nabla \Theta \|_\infty$.

\smallskip

Fix $m \in \Z_+$. Using a slight abuse of notation, let $U_{\Theta, m} \eqdef U_{\Theta, (m, m, \ldots, m)}$.

\begin{corollary}
	\label{corollary:epsilon_approx}
	Let $F$ be continuously differentiable, $\epsilon > 0$, and define
	\begin{equation}
		m_\epsilon \eqdef \left\lceil \log_2 \( 2 \pi \sqrt{D} \tfrac{\| \nabla \Theta \|_\infty}{\epsilon} \) \right\rceil. \label{eq:m_epsilon}
	\end{equation}
	Then, for all $m \geqslant m_\epsilon$, we have that
	\begin{equation}
		\| U_{\Theta, m} - U_{\Theta, m_\epsilon} \otimes \id_{D (m - m_\epsilon)} \| < \epsilon.
	\end{equation}
\end{corollary}

Note that when $D = 1$, then Corollary \ref{corollary:epsilon_approx} recovers \cite{zylberman_efficient_2025}*{Theorem IV.1.}.

\smallskip

The above formulae allow us to estimate the resources needed to implement quantum circuits for diagonal unitaries of the form $U_{\Theta, \underline{n}}$. In \cite{gosset_quantum_2024}*{Theorem 1.2} a $T$ gate optimal method was introduced for unstructured data. In order to implement an $n$-qubit diagonal unitary within $\epsilon$ error, their method uses $O \( \sqrt{\log_2 \( \nicefrac{1}{\epsilon} \) 2^n} + \log_2 \( \nicefrac{1}{\epsilon} \) \)$ many $T$ gates and ancillas. Assuming that the data are $D$-dimensionan, the unitary is of the form $U_{\Theta, m}$, and setting the approximation error and $\epsilon$ in \ref{eq:m_epsilon} are the same, we can use \ref{eq:m_epsilon} to eliminate $n = n_\epsilon \eqdef D m_\epsilon$, and get the $T$-gate and ancilla counts of the $\epsilon$-approximation, $U_{\Theta, m}$, as a function of $\epsilon$ only:
\begin{equation}
	\begin{array}{ll} \mbox{$T$-gate and ancilla counts of $U_{\Theta, m}^\epsilon$} \\ \mbox{via \cite{gosset_quantum_2024}*{Theorem 1.2}} \end{array} =  O \( \sqrt{\log_2 \( \nicefrac{1}{\epsilon} \) 2 \pi \sqrt{D} \( \tfrac{\| \nabla \Theta \|_\infty}{\epsilon} \)^D} + \log_2 \( \nicefrac{1}{\epsilon} \) \).
\end{equation}
For a fixed $\Theta$ (and thus, $D$) we get that the complexity is $O \( \sqrt{\tfrac{\log_2 \( \nicefrac{1}{\epsilon}  \)}{\epsilon^D}} \)$. Note that this complexity gets (exponentially) worse as the number of variables, $D$, increases. Although explicitly mentioned in \cite{gosset_quantum_2024}, it is clear from their proof that their construction can be generalized to a 1-parameter family, where the parameter can trade ancillas for $T$ gates. More concretely, for each $1 \leqslant \lambda \leqslant 2^n$ corresponds an implementation with $O \( \tfrac{2^n}{\lambda} + \log_2 \( \nicefrac{1}{\epsilon} \) \( \lambda + 1 \) \)$ many $T$ gates and $O \( \log_2 \( \nicefrac{1}{\epsilon} \) \lambda \)$ ancillas. With this in mind, we have the following
\begin{align}
	\begin{array}{ll} \mbox{$T$ gate count of $U_{\Theta, m}^{\epsilon, \lambda}$} \\ \mbox{via \cite{gosset_quantum_2024}*{Theorem 1.2}} \end{array}	&= O \( \( 2 \pi \sqrt{D} \tfrac{\| \nabla \Theta \|_\infty}{\epsilon} \)^D \lambda^{- 1} + \log_2 \( \nicefrac{1}{\epsilon} \) \lambda \), \\
	\begin{array}{ll} \mbox{ancilla count of $U_{\Theta, m}^{\epsilon, \lambda}$} \\ \mbox{via \cite{gosset_quantum_2024}*{Theorem 1.2}} \end{array}	&= O \( \log_2 \( \nicefrac{1}{\epsilon} \) \lambda \).
\end{align}
When $\lambda = O \( \tfrac{2^n}{\log_2 \( \nicefrac{1}{\epsilon} \)} \)$, we recover \cite{gosset_quantum_2024}*{Theorem 1.2}. Also not mentioned in \cite{gosset_quantum_2024} is the Clifford cost, which is dominated by the CNOT contribution of the QROM of \cite{low_trading_2024}.

\subsection{Pair cancellation in $U_f$:}
\label{app:pair_cancellation}

Let
\begin{equation}
    \begin{quantikz}
        & \qwbundle{b} & \gate{k} & \qw
    \end{quantikz}
\end{equation}

denote $b$-qubit, $k$-adder oracle on a circuit diagram, that is, the oracle of $A_b (k)$ in eq.~\ref{eq:qrom_def}. The following identities are straightforward to verify:
\begin{equation}
	\begin{array}{ccccc}
		\begin{quantikz}
			&	\qwbundle{n}	&	\qw			&	\ctrl[wire style={"z"}]{1}	&	\qw			& \\
			&	\qwbundle{b}	&	\gate{k} 	&	\targ{}						&	\gate{l}	&
		\end{quantikz}
		& = &
		\begin{quantikz}
			&	\qwbundle{n}	&	\qw				&	\ctrl[wire style={"z"}]{1}	&	\ctrl[wire style={"z"}]{1}	& \\
			&	\qwbundle{b}	&	\gate{k + l} 	&	\gate{- 2 l}				&	\targ{}						&
		\end{quantikz}
		& = &
		\begin{quantikz}
			&	\qwbundle{n}	&	\qw				&	\octrl[wire style={"z"}]{1}	&	\ctrl[wire style={"z"}]{1}	& \\
			&	\qwbundle{b}	&	\gate{k - l}	&	\gate{2 k}					&	\targ{}						&
		\end{quantikz}
	\end{array}
\end{equation}

In order to quantify the costs, we need to specify the construction for the adder oracles, $A_b (k)$. In \cite{gidney_halving_2018}, Gidney constructed a quantum $+$ quantum adder circuit with $4(b - 1)$ $T$ gates and $b - 1$ ancillas. For our case one of the numbers ($k$) is classically given, which simplifies the cost to be $4 \( b - 2 - \mathrm{lsb}_k \)$ $T$ gates and $b - 2 - \mathrm{lsb}_k$ ancillas, where $\mathrm{lsb}_k$ is the index of the least significant bit of $k$. That is, $\mathrm{lsb}_k = l$, exactly when $2^l | k$, but $2^{l + 1} \not | k$. We chose this adder as it uses a low number of $T$ gates and ancillas. In this case, a controlled version of $A_b (k)$ can be implemented via $4 \( b - 1 - \mathrm{lsb}_k \)$ $T$ gates and $b - 1 - \mathrm{lsb}_k$ ancillas.

\begin{remark}
	There many other designs with different trade-offs; cf. \cite{wang_comprehensive_2025}. In particular, $A_b (k)$ can be implemented with a logarithmically shallow circuit, while keeping the $T$ gate and ancillas costs linear; cf. \cite{draper_logarithmic_2004}.
\end{remark}

Since the cost and depth of $A_b (m)$ and $CA_b (\pm 2 m)$ are the same (up to, at most, $O(1)$ Clifford gates), we get that if $k = \pm l$, then one of the second two circuits is cheaper, by the cost of $A_b (\pm k)$. In particular, \ref{eq:qrom_def} can potentially be simplified, if there are pairs, $z_1, z_2 \in \F_2^n$, such that $\WH (f) (z_1) = \pm \WH (f) (z_2) \neq 0$. Note that, at best, the cost can be halved, when all nonzero components of the Walsh--Hadamard transform can be paired up.

When there are components that cannot be paired up as above, one can still use the above identites to achieve further cost reductions, if there are pairs $z_1, z_2 \in \F_2^n$, such that $\WH (f) (z_1)$ and $\WH (f) (z_2)$ have the same least significant bit. In this case, the savings is
\begin{equation}
	4 \( \mathrm{lsb}_{\WH (f) (z_1)} - \max \( \mathrm{lsb}_{\WH (f) (z_1) + \WH (f) (z_2)}, \mathrm{lsb}_{\WH (f) (z_1) - \WH (f) (z_2)} \) \) \geqslant 4,
\end{equation}
many $T$ gates.

\section{Variational basis set and discrete variable representations of the \SE}
\label{sec:variational}
A choice of the rotational-vibrational basis set introduces a representation to the time-independent \SE\ in the form of an algebraic matrix equation $\mathbf{H}\mathbf{U}=\mathbf{S}\mathbf{UE}$, where $\mathbf{H}$ is the \textit{Hamiltonian matrix},  $\mathbf{S}$ is the \textit{overlap matrix}, $\mathbf{U}$ is the matrix of eigenvectors and $\mathbf{E}$ is the diagonal matrix of energies.
Once the basis set has been chosen, the calculation can
proceed with building (exactly) and diagonalizing the Hamiltonian matrix $\mathbf{H}$. This way of solving the \SE\ is called the variational basis representation (VBR) approach, which is by far the most popular computational method in nuclear motion theory.

The time-independent \SE\ can be written as
\begin{equation}
\hat{H}|\psi \rangle = E |\psi \rangle
\label{eq:hpsiepsi}
\end{equation}
where typically the wavefunction is represented as a linear combination of basis functions
\begin{equation}
|\psi \rangle = \sum_{n=1}^N u_n |\phi_n\rangle
\label{eq:linexp}
\end{equation}
with constant coefficients $u_n$ which need to be determined. $|\phi_n\rangle$ is the $n$-th basis function used to model the wavefunction $|\psi \rangle $. $N$ denotes the size of the variational basis set. Eq. \ref{eq:linexp} can be inserted into eq. \ref{eq:hpsiepsi} to give 
\begin{equation}
\sum_{n=1}^N u_n \hat{H}|\phi_n\rangle = E \sum_{m=1}^N u_m |\phi_m\rangle
\label{eq:linexp2}
\end{equation}
At this point, one possibility for proceeding with the representation of the \SE\  is to multiply eq. \ref{eq:linexp2} from the \textit{left} by a basis function $\langle \phi_i |$, which results in the following matrix equation:
\begin{equation}
\mathbf{H}^{VBR}\mathbf{U}=\mathbf{S}\mathbf{UE}
\label{eq:galerkin}
\end{equation}
where $H^{VBR}_{ij}=\langle \phi_i | \hat{H} | \phi_j \rangle$ is the matrix element of the Hamiltonian operator, $\mathbf{U}$ is the vector of coefficients $u_j$, where $j=1,...,N$ and $S_{ij}=\langle \phi_i | \phi_j \rangle$ is the overlap matrix. $VBR$ stands for \textit{variational basis representation}. This method of solving the \SE\ is called \textit{variational} or \textit{Galerkin} method. The second order differential \SE\ is replaced by a set of $N$ algebraic equations. The task is to solve the generalized eigenvalue problem given in eq. \ref{eq:galerkin} to determine $\mathbf{E}$ and $\mathbf{U}$.

In the Born-Oppenheimer approximation, when an exact kinetic energy operator is used, such as the one given in eq.~\ref{eq:nuclear_hamiltonian}, the only source of errors in the variational calculations are the error in the PES and the error associated with the finite variational basis set size. The former error is associated with imperfect representation to the electronic wavefunction of the molecule whereas the basis set size error indicates that the variational basis is not complete. Whenever a complete variational basis set is used, the solutions to the \SE\ can be represented exactly. Such a complete representation is given by the wavefunctions of the symmetric top model, for which analytic solutions to the \SE\ are known. Symmetric top eigenfunctions serve as a complete, finite-size basis set for the rotational motion of any molecule. 

The VBR assumes that all matrix elements of the Hamiltonian $H^{VBR}_{ij}=\langle \phi_i | \hat{H} | \phi_j \rangle$ are calculated exactly. This is often very difficult if not impossible to achieve, especially when several internal coordinates of the molecule are coupled in the PES, meaning that a multidimensional integral has to be calculated. For this reason, it is often necessary to employ approximate methods of evaluating matrix elements of the Hamiltonian.

Hamiltonian defined in eq.~\ref{eq:nuclear_hamiltonian} can be expressed in a finite basis set: 
\begin{equation}
|\psi \rangle = \sum_{n=1}^N u_n |\phi_n\rangle ,
\label{eq:vbr-ansatz}
\end{equation}
where $\lbrace\ket{\phi_n}\rbrace_{n=1}^N$ denotes the chosen basis functions and $u_n$ are the corresponding expansion coefficients.
Substituting eq.~\ref{eq:vbr-ansatz} into the \SE ~$\hat{H}|\psi \rangle = E |\psi \rangle$ and projecting onto the same basis yields the generalized eigenvalue problem:
\begin{equation}
\mathbf{H}^{\mathrm{VBR}}\mathbf{U}^{\mathrm{VBR}}=\mathbf{S}^{\mathrm{VBR}}\mathbf{U}^{\mathrm{VBR}}\mathbf{E}^{\mathrm{VBR}},
\label{eq:HVBR}
\end{equation}
where $\mathbf{H}^{\mathrm{VBR}}$ is the Hamiltonian matrix with elements $H^{\mathrm{VBR}}_{ij}=\langle \phi_i | \hat{H} | \phi_j \rangle$, $\mathbf{S}^{\mathrm{VBR}}$ is the overlap matrix with elements $S^{\mathrm{VBR}}_{ij}=\langle \phi_i | \phi_j \rangle$, $\mathbf{U}^{\mathrm{VBR}}$ is the coefficient matrix whose columns contain the expansion vectors ${u_n}$, and $\mathbf{E}^{\mathrm{VBR}}$ is the diagonal matrix of eigenvalues. The superscript VBR denotes the variational basis representation. Finding eigenvalues $\mathbf{E}^{\mathrm{VBR}}$ and eigenvectors $\mathbf{U}^{\mathrm{VBR}}$ requires calculating matrix elements $H^{\mathrm{VBR}}_{ij}$ and $S^{\mathrm{VBR}}_{ij}$, which 
if done with quadrature, sets a new finite basis representation (FBR). Matrix elements in FBR are computed with quadrature integration rule, which can be approximate. The PES matrix elements in FBR are given by the following expression: 
\begin{equation}
\begin{split}\label{eq:FBR-PES2}
V^{VBR}_{ij}=\langle \phi_i(\mathbf{q})|V(\mathbf{q})|\phi_j(\mathbf{q})\rangle=\int_{\mathbb{R}^{M}\times\mathbb{R}^3}d\mathbf{q}d\mathbf{\Theta}\phi_i(\mathbf{q})V(\mathbf{q})\phi_j(\mathbf{q})\approx \\
\approx V^{FBR}_{ij}=\sum_{k=0}^{N-1}\frac{w_k}{\omega(q_k)}\phi_i(q_k)V(q_k)\phi_j(q_k)=
\sum_{k=0}^{N-1}\mathbf{T}_{ik}^{T}\mathbf{V}^{DVR}_{kk}\mathbf{T}_{kj}
\end{split}
\end{equation}
where the integration is carried out over the configurational space of internal coordinates $\mathbf{q}\in \mathbb{R}^{M}$ and rotational coordinates $\mathbf{\Theta} \equiv (\theta,\phi,\chi) \in \mathbb{R}^{3}$ and where
 \begin{equation}
      T_{kj}=\tilde{N}_j\sqrt{w_k}p_j(q_k), \hbox{ for } k,j=0,\ldots,N-1. 
      \label{eq:dvr-definition}
 \end{equation}
$\mathbf{V}^{DVR}_{kk}$ are elements of the discrete variable representation of the PES given by the values of the PES function $V(q)$ at grid points $q_k$.  Then $\mathbf{T}$ refers to \textit{FBR-to-DVR} transformation matrix. In DVR all local operators, including the PES are diagonal. 
 When a Gaussian quadrature rule is used in construction of the $\mathbf{T}$ matrix given in eq.~\ref{eq:dvr-definition} we refer to Gaussian-DVR, for which the normalization constant is given by $\tilde{N}_j=\|p_j\|_{L^2(\mu)}^{-1}$, $w_k$ are the Gaussian quadrature weights, $p_j(q_k)$ is the value of the degree-$j$ orthogonal polynomial defining the Gaussian quadrature evaluated at the Gaussian quadrature node $q_k$~\cite{Stoer2002}. For solving the rovibrational SE, popular choices for Gaussian quadratures for FBR rely on orthogonal polynomials representing sotluions to the Schrödinger equation for specific  model systems, including Hermite polynomials for the harmonic oscillator (bond stretching), Legendre polynomials for spherically symmetric problems (bending and torsional motions), Laguerre polynomials for the hydrogen atom, associated Laguerre polynomials for the Morse oscillator (bond stretching), Lobatto polynomials for problems with fixed boundary conditions~\cite{AbramowitzStegun1964,Light2000} or general potential-optimized DVRs~\cite{Bowman2008}.

\section{Quantum circuit and resource estimation for the FBR-DVR transformation}
\begin{figure}[h!]
\includegraphics[width=1\linewidth]{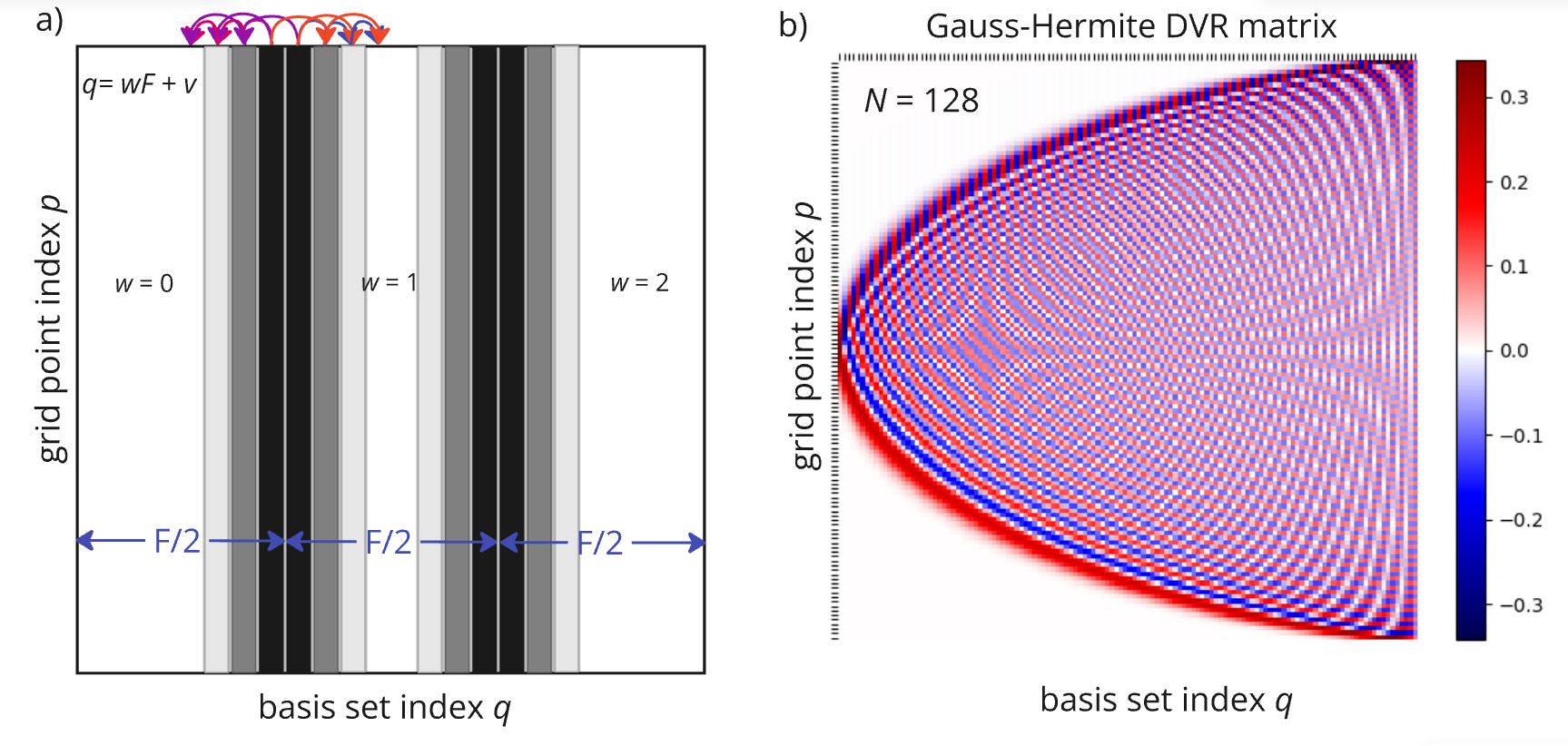}
    \caption{a) Schematic representation for the segment construction of the DVR oracle matrix. Basis set (column) index is represented as $q=wF+v$, where $w=0,1,2,...,\frac{N}{F}-1$ and $v=0,1,2,...,F-1$ for the purpose of splitting the column space in the DVR matrix into $\frac{N}{F}$ segments. Columns in each segment are constructed simultaneously in the ascending and descending horizontal direction following the three-step recursion given in eq.~\ref{eq:recursion}; b) Color map encoding values of the DVR transformation matrix for the Gauss-Hermite quadrature with $N=128$ elements. Figure adapted from ref.~\cite{plis25}.}
    \label{fig:schematic}
\end{figure}

The FBR-DVR transformation unitary can be implemented as quantum circuit using the technique presented in ref.~\cite{plis25}. Here we sketch the key idea and give resource estimation.
Our construction gives the following transformation:
\begin{equation}
    \hat{D}=\ket{0}\bra{1}\hat{T}+\ket{1}\bra{0}\hat{T}^\dagger
    \label{eq:DVRunitary-reflection}
\end{equation}
The construction assumes providing an oracle returning the $k$-th column of the DVR unitary, $\ket{u_k}= \hat{T}\ket{k}$. Let $U_k$ prepare the following state:
\begin{equation}
	\ket{w_k}=\ket{0}_a\ket{k}-\ket{1}_a\ket{u_k}
	\label{eq:reflections_state}
\end{equation}
where $\ket{k}$ represents index-state labeling columns of the DVR matrix. Then the product of $N$ reflections:
\begin{equation}
    \hat{R}_k=I-2\ket{w_k}\bra{w_k}
    \label{eq:reflections}
\end{equation}
gives $\hat{D}=\Pi_{k=1}^N \hat{R}_k = \ket{0}\bra{1}\hat{T}+\ket{1}\bra{0}\hat{T}^\dagger$. Having execution of $\hat{T}$ and $\hat{T}^{\dag}$ conditioned on the state of ancilla qubit can be useful in certain scenarios, as discussed further. 
The total cost for the $\hat{D}$ unitary is $2n^2+n(4m+1)\log(n)$, where $m$ is the precision of matrix elements. This cost consists of the state preparation cost and the reflections cost. In the following paragraphs we give a more detailed resource estimation for the procedure sketched above for constructing $\hat{D}$.
The construction of the DVR oracle, which loads DVR matrix elements uses the recursive properties of orthogonal polynomials and is schematically shown in Fig.~\ref{fig:schematic}.

\subsubsection{Quantum DVR Oracle}
\label{sec:appendix-DVR}
To implement the FBR-DVR unitary circuit, we first construct a unitary operator, referred to as the \textit{DVR oracle}. We outline the technique presented in our other recent work given in ref.~\cite{plis25}. The DVR oracle is defined as a quantum transformation that encodes the elements of the DVR matrix $T$ into qubit registers, expressed as
\begin{equation}
    \mathcal{T}\ket{p}\ket{q}\ket{0} = \ket{p}\ket{q}\ket{T_{pq}},
    \label{eq:DVRoracle-eq}
\end{equation}
where $T_{pq}$ denotes the DVR matrix elements, $\ket{q}$ is the state representing the column index (basis-state index), and $\ket{p}$ corresponds to the row index (grid-point index).

For clarity, we define the column index as $q = wF + v$, where $w = 0, 1, \ldots, \frac{N}{F}-1$ and $v = 0, 1, \ldots, F-1$. This representation partitions the column space of the DVR matrix into $\frac{N}{F}$ segments. The integer $F$, referred to as the segmentation parameter, specifies the size of each segment. The construction of the overall circuit then proceeds through several stages, as outlined below.

\paragraph{Segment initialization.}
First we construct an initialization unitary that loads $2\frac{N}{F}$ columns of the DVR matrix via QROM, defined as
\begin{equation}
\hat{U}^{(\mathrm{init})}\ket{p}\ket{w}\ket{0}\ket{z}\ket{z}
= \ket{p}\ket{w}\ket{x_p}\ket{z \oplus T_{p\tilde{q}-1}}\ket{z \oplus T_{p\tilde{q}}},
\label{eq:uinit}
\end{equation}
where $\tilde{q} = wF + \frac{F}{2}$ denotes the midpoint index within each of the $\frac{N}{F}$ matrix segments, and $\ket{x_p}$ stores the nodes of the $N$th orthogonal polynomial that defines the DVR basis.

This step involves loading $\frac{N^2}{F}$ pairs of values $(T_{p\tilde{q}-1}, T_{p\tilde{q}})$, each represented with $m$ bits. The associated $T$-gate complexity depends on the specific QROM implementation. When the SELECT–SWAP algorithm~\cite{low_trading_2024} is used, the cost scales as
\begin{equation}
C(\hat{U}{\mathrm{init}}) = 2\frac{N\sqrt{m}}{\sqrt{F}} + \sqrt{Nm},
\label{eq:init_SELSWAP}
\end{equation}
whereas for the SELECT–QROM approach it scales as
\begin{equation}
C(\hat{U}{\mathrm{init}}) = \frac{N^2}{F} + N.
\label{eq:init_SEL}
\end{equation}

\begin{figure}[h!]
    \centering
                \includegraphics[width=0.6\linewidth]{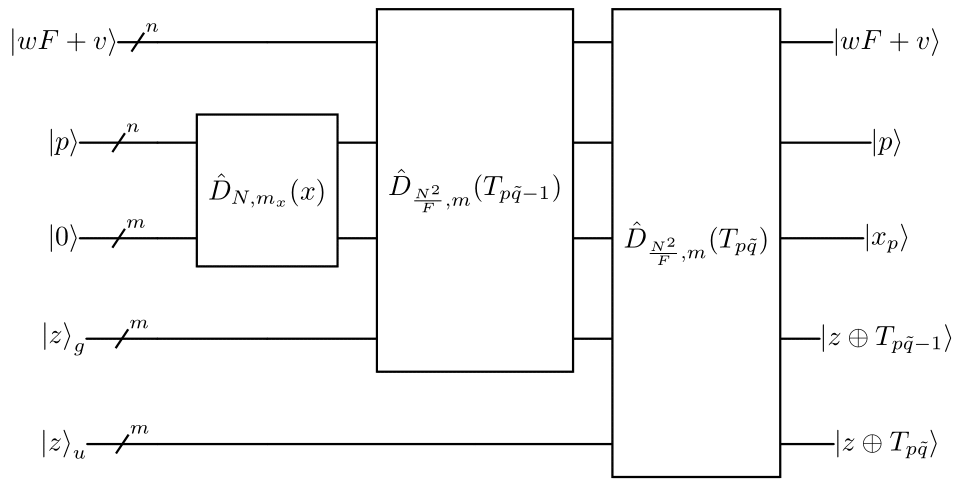}

    \caption{Quantum circuit representing initialization unitary for the DVR oracle defined in eq.~\ref{eq:uinit}. $\hat{D}_{N,m}(A)$ denotes QROM encoding data represented by function $A$ for $N$ arguments and output stored in $m$ qubits. Here $T_{p\tilde{q}}$ denotes the DVR matrix element for the $\tilde{q}$'th column, where $\tilde{q}=wF+\frac{F}{2}$. The column index $q$ is represented as $q=wF+v$. Adapted from ref.~\cite{plis25}.}
    \label{fig:DVR_initialization}
\end{figure}

\begin{figure}[h!]
    \includegraphics[width=\linewidth]{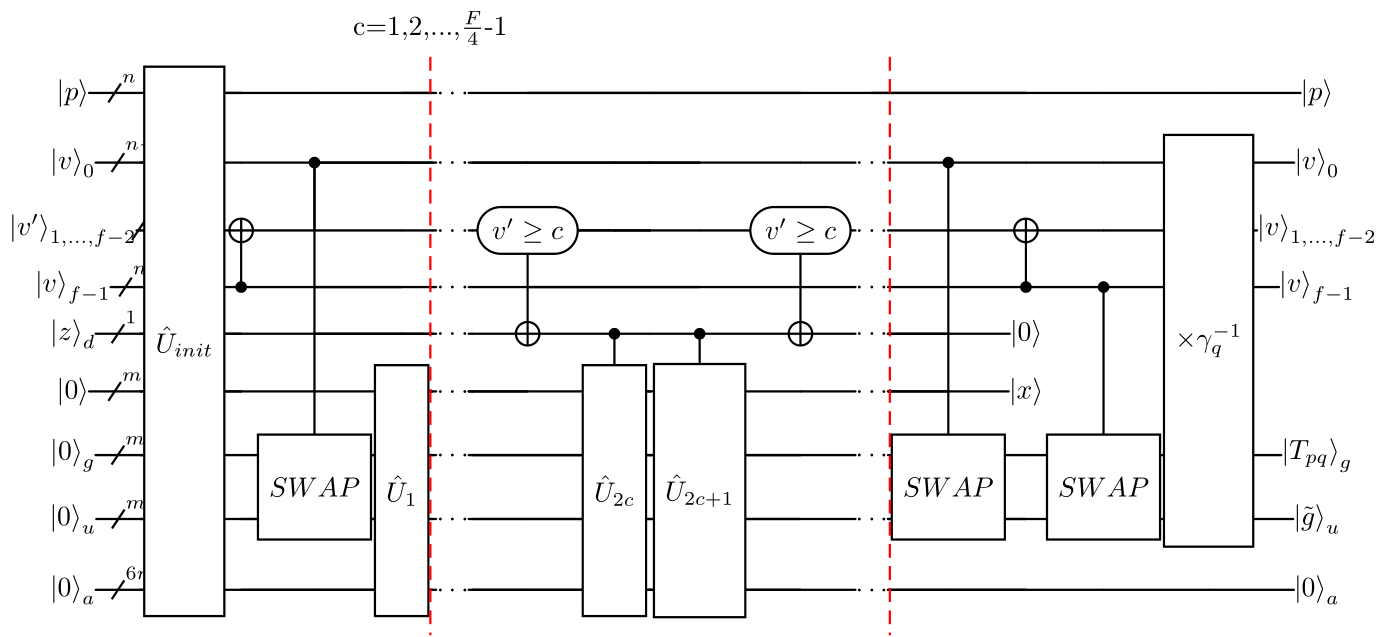}
    \caption{Quantum circuit representing the DVR oracle defined in eq.~\ref{eq:DVRoracle-eq}. $\hat{U}_{init}$ represents state initialization oracle shown in Fig.~\ref{fig:DVR_initialization}.  $\hat{U}_c$ are iteration unitaries shown in Fig.~\ref{fig:arithmetic}. $\gamma_q^{-1}$ is a gate multiplying the result by the appropriate scaling factor defined in eq.~\ref{eq:gamma}. The result is returned in $m$-qubit register $\ket{T_{pq}}_g$. Adapted from ref.~\cite{plis25}. } 
    \label{fig:DVRoracle}
\end{figure}

\paragraph{Recursive construction.}
After the initialization step, the algorithm applies a sequence of unitary operations, denoted $\hat{U}_{2c}$ and $\hat{U}_{2c+1}$, controlled by the running index $c = 1, 2, \ldots, \frac{F}{4} - 1$. These operations iterate over the range determined by the segmentation parameter $F$, as illustrated in Fig.~\ref{fig:DVRoracle}.

The recursive section of the circuit is enclosed by SWAP operations acting on two $m$-qubit output registers. Each SWAP gate is controlled by the most significant bit $v_0$ of the binary representation of
\begin{equation}
v = v_0 2^{f-1} + v_1 2^{f-2} + \ldots + v_{f-1} 2^0,
\end{equation}
where $v = 0, 1, 2, \ldots, F - 1$, and the column index is expressed as $q = wF + v$. Here, $F = 2^f$ defines the segmentation parameter in powers of two.

The two output registers hold the intermediate results of the recursive construction, while the controlled-SWAP operations ensure that the final output is directed to the $\ket{\cdot}_g$ register, independent of the parity of the queried column index. The gate cost of a single controlled-SWAP operation is $m$ Toffoli gates.

\begin{figure}[h!]
    \centering
        \includegraphics[width=0.5\linewidth]{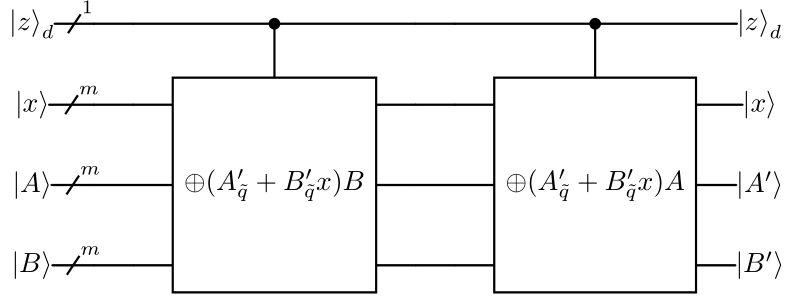}
        \caption{Quantum circuit representing  $\hat{U}_c$ iteration unitary in the the DVR oracle. $\ket{z}_d$ is the \textit{dump} qubit register shown also in Figure~\ref{fig:DVRoracle}.}
    \label{fig:arithmetic}
\end{figure}

The recursive block completes after $\frac{F}{4}-1$ steps (marked by red vertical dashed lines in Fig.~\ref{fig:DVRoracle}), with each step constructing two columns per segment. The columns are generated simultaneously in descending and ascending order of $q$, beginning from the midpoint columns indexed by $\tilde{q} = wF + \frac{F}{2}$. A schematic illustration of this procedure is provided in Fig.~\ref{fig:schematic}a. In constructing $\hat{U}_{2c}$ and $\hat{U}_{2c+1}$, we use the three-term recurrence relation satisfied by the DVR matrix elements,
\begin{equation}
T_{p;q+2} = (A_q + B_q x_p) T_{p;q+1} + C_q T_{pq},
\label{eq:recursion}
\end{equation}
where the coefficients $A_q$, $B_q$, and $C_q$ depend on the form of the underlying orthogonal polynomial.

The number of independent parameters in eq.~\ref{eq:recursion} can be reduced by introducing scaled matrix elements $T'_{p;q+2} = \gamma_q T_{p;q+2}$, which yield the following recurrence relations for the ascending and descending directions, respectively:
\begin{align}
T'_{pq} &= (A'_q + B'_q x_p) T'_{p;q-1} + T'_{p;q-2}, \qquad \hbox{for} \quad q = wF + \frac{F}{2} + k, \
T'_{pq} &= (A'_q + B'_q x_p) T'_{p;q+1} + T'_{p;q+2}, \qquad \hbox{for} \quad q = wF + \frac{F}{2} - 1 - k,
\label{eq:recursion'}
\end{align}
where $k = 1, \ldots, \frac{F}{2} - 1$. For the two initial columns in each segment, we set $\gamma_q = 1$ for $q = wF + \frac{F}{2} - 1$ and $q = wF + \frac{F}{2}$.
For the ascending direction $q = wF + \frac{F}{2} + k$, the scaling constant is
\begin{equation}
\gamma_q = \gamma_{q-2} C_{q-2}^{-1} = (C_{q-2} C_{q-4} \ldots C_i)^{-1},
\label{eq:gamma}
\end{equation}
where $i \in { wF + \frac{F}{2} - 1, wF + \frac{F}{2} }$ has the same parity as $q$. The rescaled coefficients are then given by $(A'_q, B'_q) = \frac{\gamma_q}{\gamma{q-1}} (A_{q-2}, B_{q-2})$.
For the descending direction $q = wF + \frac{F}{2} - 1 - k$, the scaling constant is $\gamma_q = \gamma_{q+2} C_q = C_q C_{q+2} \ldots C_i$, where again $i \in { wF + \frac{F}{2} - 1, wF + \frac{F}{2} }$ has the same parity as $q$. In this case, $(A'_q, B'_q) = -\frac{\gamma{q+2}}{\gamma{q+1}} (A_q, B_q)$.

Each unitary iteration $\hat{U}_{2c}$ and $\hat{U}_{2c+1}$ comprises two arithmetic subblocks, each performing two multiplications and one addition, as shown in Fig.~\ref{fig:arithmetic}. The corresponding qubit-state transformations are
\begin{align}
U_{2c} \ket q \ket x \ket A \ket B &=
\ket q \ket x \ket A \ket{B \oplus (A'_{\hat q} + B'_{\hat q} x) A},
\
U_{2c+1} \ket q \ket x \ket A \ket B &=
\ket q \ket x \ket{A \oplus (A'_{\check q} + B'_{\check q} x) B} \ket B,
\label{eq:uc-unitaries1}
\end{align}
for $|v - \frac{F - 1}{2}| > 2c$. The recursion indices appearing in eqs.~\ref{eq:uc-unitaries1} are defined as
\begin{equation}
\hat q = \hat q(v_0, w, c) =
\begin{cases}
wF + F/2 + 2c, & v_0 = 1, \\
wF + F/2 - 1 - 2c, & v_0 = 0,
\end{cases}
\end{equation}
and
\begin{equation}
\check q = \check q(v_0, w, c) =
\begin{cases}
wF + F/2 + 2c + 1, & v_0 = 1, \\
wF + F/2 - 2 - 2c, & v_0 = 0.
\end{cases}
\end{equation}

\section{Block-encoding the KEO using FBR or DVR alone}
\label{sec:appendix-BE-DVR-alone}
In this section we propose an alternative block-encoding strategy. The PES is kept in DVR, but the KEO is block-encoded either entirely in DVR or FBR basis. Below, we describe partial block-encodings for the individual KEO components defined in eqs.~\eqref{eq:KEO_def}–\eqref{eq:KEO_rot}. Encoding these terms separately is advantageous, as each component typically acts on a restricted subset of the qubit registers, leading to improved gate-count scaling with respect to the basis size.

Analogously to the mixed FBR–DVR representation, the present approach employs the LCU framework to block-encode the component terms of the KEO in the DVR representation.  
Here, however, it is advantageous to further decompose the KEO into its internal energy contributions, given by eqs.~\ref{eq:KEO_def}–\ref{eq:KEO_rot}.  
Each contribution is block-encoded using the $\rho$-sparse method.  

The purely vibrational part of the KEO can be written as
\begin{equation}\label{eq:KEO_def_expanded}
    \boldsymbol{K}^{vib}(\boldsymbol{q}) 
    = \frac{1}{2}\sum_{i=1}^{D}\boldsymbol{P}_i^\dagger \boldsymbol{g}_{ii}(\boldsymbol{q})\boldsymbol{P}_i 
    + \frac{1}{2}\sum_{i<j}^{D}\left(\boldsymbol{P}_i^\dagger \boldsymbol{g}_{ij}(\boldsymbol{q})\boldsymbol{P}_j+\boldsymbol{P}_j^\dagger \boldsymbol{g}_{ij}(\boldsymbol{q})\boldsymbol{P}_i\right).
\end{equation}
Since the operators $\boldsymbol{P}_i^\dagger \boldsymbol{g}_{ij}(\boldsymbol{q})\boldsymbol{P}_j$ and $\boldsymbol{P}_j^\dagger \boldsymbol{g}_{ij}(\boldsymbol{q})\boldsymbol{P}_i$ act on the same subspace, there is no advantage in block-encoding them separately.  
Therefore, the form in Eq.~\ref{eq:KEO_def_expanded} yields only $\tfrac{D(D+1)}{2}$ distinct terms, as opposed to the full $D^2$.

For the Coriolis contribution $\boldsymbol{K}^{cor}$ in eq.~\ref{eq:KEO_cor}, it is beneficial to block-encode the rotational and internal parts independently.  
Equation~\ref{eq:KEO_cor} can be rewritten as
\begin{equation}\label{eq:KEO_cor_decomposed}
    \boldsymbol{K}^{cor} = \sum_{\alpha=x,y,z}\boldsymbol{J}_\alpha \otimes \boldsymbol{k}_\alpha^{cor},
\end{equation}
with
\begin{equation}
   \boldsymbol{k}_\alpha^{cor}
   = \sum_{j=1}^{D}\frac{1}{2}\left(\boldsymbol{\Gamma}_{\alpha j}(\boldsymbol{q})\boldsymbol{P}_{j} + \boldsymbol{P}_{j}^\dagger \boldsymbol{\Gamma}_{\alpha j}(\boldsymbol{q})\right) 
   = \sum_{j=1}^D \boldsymbol{k}_{\alpha j}^{vib}.
\end{equation}
For each $\alpha$, the matrices $\boldsymbol{J}_\alpha$ and $\boldsymbol{k}_{\alpha}$ can be block-encoded separately.  
Notably, block-encoding a product requires only the introduction of a $\mathcal{O}(1)$ ancillas and $\mathcal{O}(a)$ Clifford+T gates and can be realized by the following circuit

This construction requires only a single invocation of the block-encoding for each angular momentum component.  
We assume that $\boldsymbol{J}_\alpha$ as well as each $\boldsymbol{k}_{\alpha j}$ are block-encoded using the $\rho$-sparse method, while $\boldsymbol{k}_\alpha$ is implemented as a sum of block-encodings.

The optimal strategy for block-encoding the rotational contribution in Eq.~\ref{eq:KEO_rot} depends on the structure of the matrix $\boldsymbol{\mu}_{\alpha\beta}(\boldsymbol{q})$. For instance, within the rigid-rotor approximation the matrix $\boldsymbol{\mu}_{\alpha\beta}$ is constant, and the operator $\boldsymbol{K}_{Rot}$ acts only on qubits encoding angular momentum basis functions. Furthermore, in the spherical-top basis the matrix associated with $\boldsymbol{K}_{Rot}$ has a particularly simple structure, with exactly three non-zero elements per row~\cite{Bunker1999}. In such cases, there is little advantage in block-encoding the individual contributions of the rotational KEO separately.  

In contrast, for more general forms of $\boldsymbol{\mu}_{\alpha\beta}(\boldsymbol{q})$ it may be preferable to implement each term $\boldsymbol{\mu}_{\alpha\beta}(\boldsymbol{q})\boldsymbol{J}_\alpha\boldsymbol{J}_\beta$ independently. This approach offers two main benefits.  
First, since $\boldsymbol{\mu}_{\alpha\beta}(\boldsymbol{q})$ is independent of the Euler angles, the separate block-encodings $\BE{\boldsymbol{\mu}_{\alpha\beta}}$ and $\BE{\boldsymbol{J}_\alpha}$ act on fewer qubits than a full block-encoding, reducing the overall T-count.  
Second, for each $\alpha$, the unitary $\BE{\boldsymbol{J}_\alpha}$ already appears in the circuit block-encoding the Coriolis operator $\boldsymbol{K}^{cor}$. Because the sum of block-encodings is implemented as a controlled version of the partial block-encoding [see Eq.~\ref{eq:Block_enoding_sum}], it is possible to extend the action of $\BE{\boldsymbol{J}_\alpha}$ to include the rotational contributions simply by adjusting the control structure. Consequently, the only additional components required for the rotational KEO are $\BE{\boldsymbol{\mu}_{\alpha\beta}}$ and $\BE{\boldsymbol{J}_\beta}$.  
Moreover, since $\boldsymbol{\mu}_{\alpha\beta}$ is symmetric, only six independent components require block-encoding.

Assuming separate encodings of the rotational contributions, the circuit implementing the block-encoding of the full Hamiltonian can be written as a product of unitaries:
\begin{equation}\label{eq:BE_second_method_circuit}
    \boldsymbol{U}_H = \boldsymbol{U}_{\boldsymbol{\Theta}}^{(1)} (C_aX_{\tilde{a}_1})\boldsymbol{U}_{\boldsymbol{q}}(C_aX_{\tilde{a}_2}) \boldsymbol{U}_{\boldsymbol{\Theta}}^{(2)},
\end{equation}
where $\boldsymbol{U}_{\boldsymbol{q}}$ acts exclusively on internal degrees of freedom and ancilla registers, while $\boldsymbol{U}_{\boldsymbol{\Theta}}^{(1)}$ and $\boldsymbol{U}_{\boldsymbol{\Theta}}^{(2)}$ act solely on the rotational basis and ancillas. 
\begin{figure}[h]
    \centering
    \includegraphics[width=0.5\linewidth]{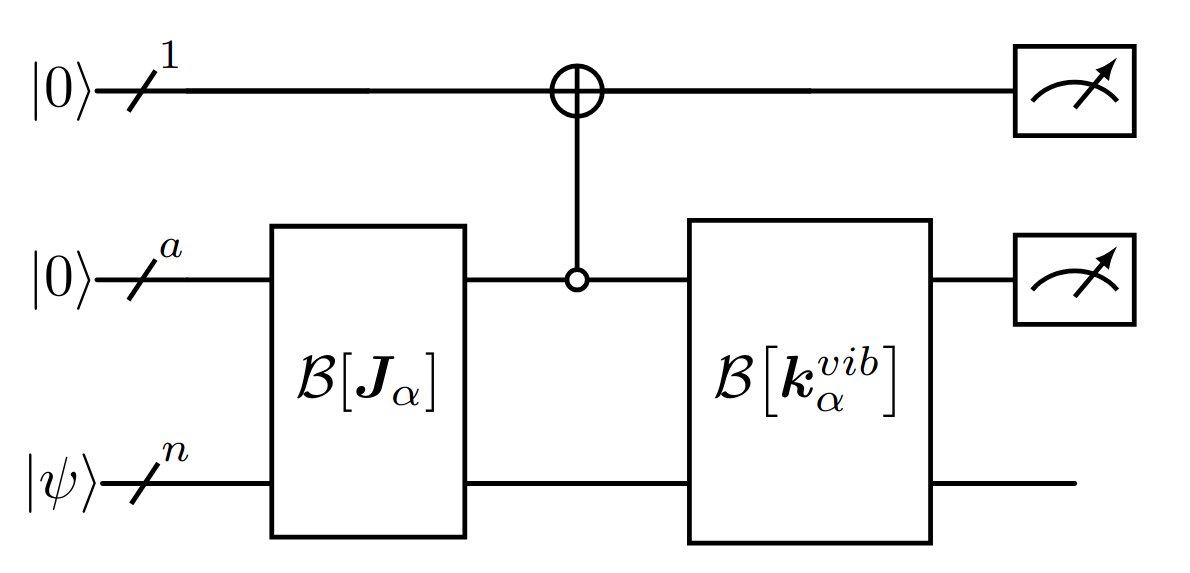}
    \caption{Circuit primitive block-encoding Coriolis terms in the Hamiltonian.}
    \label{eq:prod_BE}
\end{figure}
The $C_aX_{\tilde{a}_i}$ operator is an $X$ gate controlled by the ancilla register $a$ acting on a single additional ancilla $\tilde{a}_i$ required for the product of block-encoding as shown by circuit depicted in Fig.~\ref{eq:prod_BE}. 

These operators are defined as
\begin{equation}\label{eq:U_q}
    \begin{split}
    \boldsymbol{U}_{\boldsymbol{q}} &= \boldsymbol{\Pi}_{00}\otimes \BE{\boldsymbol{V}}
    + \sum_{i\geq j}^D \boldsymbol{\Pi}_{ij} \otimes \BE{\boldsymbol{K}_{ij}^{vib}}
    + \sum_{\alpha = x,y,z}\sum_{i=1}^{D}\boldsymbol{\Pi}_{\alpha i}\otimes \BE{\boldsymbol{k}_{\alpha i}^{cor}} \\ 
    &\quad + \sum_{\alpha} \boldsymbol{\Pi}_{\alpha \alpha}\otimes \BE{\boldsymbol{\mu}_{\alpha \alpha}}
    + \sum_{\alpha > \beta} \left(\boldsymbol{\Pi}_{\alpha \beta}+\boldsymbol{\Pi}_{\beta \alpha}\right)\otimes \BE{\boldsymbol{\mu}_{\alpha \beta}},
    \end{split}
\end{equation}
and
\begin{equation}
    \boldsymbol{U}_{\boldsymbol{\Theta}}^{(i)} = \sum_{\alpha}\left(\boldsymbol{\Pi}^{(i)}_{\alpha}\otimes \BE{\boldsymbol{J}_\alpha} + \big(\id-\boldsymbol{\Pi}^{(i)}_\alpha\big)\otimes\id \right).  
\end{equation}
Here, $\boldsymbol{\Pi}_{ab} = \ket{k_{ab}}\bra{k_{ab}}_{a'}$ are projectors acting on the ancilla register, satisfying $\boldsymbol{\Pi}_{ab}\boldsymbol{\Pi}_{a'b'} = \delta_{aa'}\delta_{bb'}$, with $k_{ab}\in \mathbb{Z}\cap[0,k_{max}]$, $k_{max} = \tfrac{D(D+1)}{2}+3D+9$. The projectors entering $\boldsymbol{U}_{\boldsymbol{\Theta}}^{(i)}$ are
\begin{align}
    \boldsymbol{\Pi}_\alpha^{(1)} &= \sum_{j=1}^D\boldsymbol{\Pi}_{\alpha j}+\sum_{\beta \in \{x,y,z\}}\boldsymbol{\Pi}_{\alpha \beta}, &
    \boldsymbol{\Pi}_\alpha^{(2)} &= \sum_{\beta\in \{x,y,z\}}\boldsymbol{\Pi}_{\beta \alpha}.
\end{align}
If $(k_{max}+1)$ is not a power of two, Eq.~\ref{eq:U_q} must be extended by additional terms to ensure unitarity. These terms, however, play no role in the circuit operation and are omitted here.  
Finally, the partial block-encodings constructed using the $\rho$-sparse method are defined as
\begin{equation}
    (\bra{0}_{a}\otimes\id_\eta)\BE{K^{vib}_{ij}}(\ket{0}_{a}\otimes\id_\eta) 
    = \frac{1}{2\zeta^{vib}_{ij}}\left(\boldsymbol{P}_i^\dagger \boldsymbol{g}_{ij}(\boldsymbol{q})\boldsymbol{P}_j+\boldsymbol{P}_j^\dagger \boldsymbol{g}_{ij}(\boldsymbol{q})\boldsymbol{P}_i\right),
\end{equation}
\begin{equation}
    (\bra{0}_{a}\otimes\id_\eta)\BE{\boldsymbol{k}^{cor}_{\alpha j}}(\ket{0}_{a}\otimes\id_\eta) 
    = \frac{1}{2\zeta^{cor}_{\alpha j}}\left(\boldsymbol{\Gamma}_{\alpha j}(\boldsymbol{q})\boldsymbol{P}_j^\dagger+\boldsymbol{\Gamma}_{\alpha j}(\boldsymbol{q})\boldsymbol{P}_j\right),
\end{equation}
\begin{equation}
    (\bra{0}_{a}\otimes\id_\eta)\BE{\boldsymbol{\mu}_{\alpha \beta}}(\ket{0}_{a}\otimes\id_\eta) 
    = \frac{\boldsymbol{\mu}^{\alpha\beta}(\boldsymbol{q})}{\zeta^\mu_{\alpha\beta}},
\end{equation}
\begin{equation}
    (\bra{0}_{a}\otimes\id_\eta)\BE{\boldsymbol{J}_{\alpha}}(\ket{0}_{a}\otimes\id_\eta) 
    = \frac{\boldsymbol{J_\alpha}}{\zeta^J_{\alpha}},
\end{equation}
\begin{equation}
    (\bra{0}_{a}\otimes\id_\eta)\BE{\boldsymbol{V}}(\ket{0}_{a}\otimes\id_\eta) 
    = \frac{\boldsymbol{V}}{\zeta_{V}}.
\end{equation}
and the whole block-encoding circuit requires $\eta_a = a+a'+\mathcal{O}(1)$ ancilla registers, where $a = \eta+2$ and $a' = \lceil \log_2{k_{max}} \rceil = \mathcal{O}(\log{D})$.
Further cost reductions are possible. 
In the DVR representation, the vibrational metric tensor elements $\boldsymbol{G}_{ij}(\mathbf{q})$ can often be factorized as a product of functions,
\begin{equation}
\boldsymbol{G}_{ij}(\mathbf{q}) = \boldsymbol{G}_{ij}^{(1)}(\mathbf{q}_1) \boldsymbol{G}_{ij}^{(2)}(\mathbf{q}_2),
\end{equation}
where $\mathbf{q}_1$ and $\mathbf{q}_2$ denote disjoint subsets of coordinates. Two cases arise depending on the partitioning of $q_i$ and $q_j$.

\begin{enumerate}
    \item Case I: $q_i, q_j \in \mathbf{q}_2$. In this situation, one can apply the product block-encoding method by implementing $\boldsymbol{G}_{ij}^{(1)}(\mathbf{q}_1)$ separately from  $\frac{1}{2}\left(\boldsymbol{P}_i^\dag \boldsymbol{G}_{ij}^{(2)}(\boldsymbol{q})\boldsymbol{P}_j+\boldsymbol{P}_j^\dag \boldsymbol{G}_{ij}^{(2)}(\boldsymbol{q})\boldsymbol{P}_i\right)$. Moreover, in certain coordinate systems the factor $\boldsymbol{G}_{ij}^{(1)}(\mathbf{q}_1)$ has the same form for multiple indices. For example, in polyspherical coordinates, several angular terms are multiplied by factors of $1/R_i^2$. In such cases, $\boldsymbol{G}_{ij}^{(1)}(\mathbf{q}_1)$ need only be implemented once, with its action extended to the relevant indices by modifying the controls in Eq.~\ref{eq:Block_enoding_sum}.
    \item Case II: $q_i \in \mathbf{q}_1$, $q_j \in \mathbf{q}_2$. Here, the operators $\boldsymbol{P}_i^\dagger \boldsymbol{G}_{ij}^{(1)}(\mathbf{q}_1)$ and $\boldsymbol{G}_{ij}^{(2)}(\mathbf{q}_2) \boldsymbol{P}_j$ act on disjoint sets of qubits, allowing them to be block-encoded independently and combined as a product block-encoding.
\end{enumerate}

\section{Construction of $O_F$ oracle for block-encoding diagonal matrices}
\label{sec:appendix-OF}
For the purposes of this work, it is useful to consider a construction of $O_F$ for a special case of block-diagonal matrices, that is matrices with elements expressed as:
\begin{equation}\label{eq:T_free_OF}
    M_{kk',mm'} = M^{(m)}_{kk'}\delta_{mm'},
\end{equation}
where $k,k'\in Z \cap \left[0,2^{n_1}\right)$ and $m,m'\in  \Z \cap\left[0,2^{n_2}\right)$ for some integers $n_1,n_2$. The state denoting the $j$-th row of this matrix can be expressed as
\begin{equation}
    \ket{j }_\eta=\ket{ k+2^{n_1}m}_\eta = \ket{k}_{\eta_1}\ket{m}_{\eta_2}.
\end{equation} 
Since the number of non-zero elements in each row is exactly $2^{\eta_1}$, the second input state of $O_F$ can be decomposed as $\ket{l}_\eta=\ket{l}_{\eta_1}\ket{0}_{\eta_2}$. The action of the oracle is then given by
\begin{equation}
    O_F\ket{j}_\eta\ket{l}_\eta = O_F\ket{k}_{n_1}\ket{m}_{\eta_2}\ket{l}_{\eta_1}\ket{0}_{\eta_2} = \ket{k}_{\eta_1}\ket{m}_{\eta_2}\ket{l}_{\eta_1
    }\ket{m}_{\eta_2}
\end{equation}
Therefore, $O_F$ can be implemented using only $\eta_2$ CNOT gates.

\section{Utilizing permutation symmetry}
A reduction in the computational cost is possible when the Hamiltonian exhibits symmetry under permutation of a pair of coordinates $(q_i, q_j)$. In this case, the Hamiltonian can be decomposed as
\begin{equation}
\boldsymbol{H} = \boldsymbol{H}_i + \boldsymbol{H}_j + \boldsymbol{H}',
\end{equation}
with
\begin{equation}
\boldsymbol{H}_i = \sum_{k\neq j}^{3A-3} \left(\boldsymbol{P}_i^\dag \boldsymbol{G}_{ik}\boldsymbol{P}_k +\boldsymbol{P}_k^\dag \boldsymbol{G}_{ik}\boldsymbol{P}_i\right)+ \boldsymbol{P}_i^\dag\boldsymbol{G}_{ij}\boldsymbol{P}_j \end{equation} 
\begin{equation} \boldsymbol{H}_j = \sum_{k\neq i}^{3A-3} \left(\boldsymbol{P}_j^\dag \boldsymbol{G}_{jk}\boldsymbol{P}_k +\boldsymbol{P}_k^\dag \boldsymbol{G}_{jk}\boldsymbol{P}_j\right)+ \boldsymbol{P}_j^\dag \boldsymbol{G}_{ij}\boldsymbol{P}_i 
\end{equation}
and $\boldsymbol{H}' = \boldsymbol{H} - \boldsymbol{H}_i - \boldsymbol{H}_j$.

The block-encoding of the full Hamiltonian can be then expressed as a sum of two effective operators,
\begin{equation} \BE{\boldsymbol{H}} = (HAD\otimes \id_{a+\eta})\left(\ket{0}\bra{0}\otimes\BE{\boldsymbol{H}_{eff}^{(i)}}+\ket{1}\bra{1}\otimes\BE{\boldsymbol{H}_{eff}^{(j)}}\right)(HAD\otimes \id_{a+\eta}) \end{equation}
with
\begin{align}
\boldsymbol{H}_{\mathrm{eff}}^{(i)} &= \boldsymbol{H}_i + \tfrac{1}{2}\boldsymbol{H}', &
\boldsymbol{H}_{\mathrm{eff}}^{(j)} &= \boldsymbol{H}_j + \tfrac{1}{2}\boldsymbol{H}' .
\end{align}
and $HAD$ is the Hadamard gate.
From the symmetry property it follows that

\begin{equation} 
\BE{\boldsymbol{H}_{eff}^{(i)}} = SWAP_{ij} \BE{\boldsymbol{H}_{eff}^{(j)}} SWAP_{ij} 
\end{equation}
where $SWAP_{ij}$ exchanges the states of the $i$-th and $j$-th registers. This gate can be realized as a tensor product of single-qubit swap operations acting on the relevant qubits. Accordingly, the full block-encoding becomes
\begin{eqnarray} 
\BE{\boldsymbol{H}} = (HAD\otimes \id)(CSWAP)_{ij}\BE{\boldsymbol{H}_{eff}^{(i)}}(CSWAP)_{ij}(HAD\otimes \id)
\end{eqnarray}
where $(CSWAP)_{ij}$ denotes a controlled-$\mathrm{SWAP}_{ij}$ gate conditioned on a single ancilla qubit. Consequently, the portion of the circuit responsible for block-encoding the terms in $\boldsymbol{H}_j$ can be replaced by single-qubit SWAP gates. This substitution introduces only $\mathcal{O}(\log_2 n_i)$ additional Clifford+T gates and requires one extra ancilla qubit, where $n_i$ is the number of basis functions associated with the coordinate $q_i$.
\section{PES Data}
For grid calculations, we used Hermite and Legendre grid. The Hermite grid is defined be two parameters: equilibrium bond length $R_0$ and characteristic length $l=\sqrt{\frac{\hbar}{\mu\omega}}$, such that $R_i = R_0+x_il$ where $x_i$ is $i$-th zero of the appropriate Hermite polynomial. Similarly, for Legendre grid we have $\theta_i = \theta_0-\frac{\theta_{max}}{\pi}\arccos{y_i}$, where $y_i$ corresponds to $i$-th zero of a appropriate Legendre polynomial. Table~\ref{tab:grids_parameters} shows the parameters used for the molecules analyzed in \ref{sec:block_encoding_pes}.

\begin{table}[h]
    \centering
    \begin{tabular}{c|c}
         Molecule &Parameters\\
         \hline
         H2O& $r_0= 0.957$, $l=0.0234$ \\
         ETSH&$\theta_{max}=\pi$ \\
         DMS& $\theta_{max}=\pi$ \\
         PH3& $l=0.06$, $\theta_{max}^{(1)}=\frac{\pi}{6}$, $\theta_{max}^{(2)}=\frac{\pi}{2}$ \\
         CH4& $l=0.05$, $\theta_{max}=\frac{\pi}{15}$ \\
         PSB3& $r_0^{(1)}=0.0912$, $r_0^{(2)}=0.025$, $l=0.0067$ $\theta_{max}=\pi$ \\
    \end{tabular}
    \caption{Parameters used to define the appropriate grids. Values of $r_0$ and $l$ are given in Angstrom.}
    \label{tab:grids_parameters}
\end{table}
\begin{table}[]
\begin{tabular}{|c|c|c|c|c|c|c|c|c|c|}
\hline
$n$   & $13$ & $15$ & $16$ & $18$ & $19$ & $21$ & $22$ & $24$ & $25$ \\ \hline
$n_1$ & $4$  & $5$  & $5$  & $6$  & $6$  & $7$  & $7$  & $8$  & $8$ \\ \hline
$n_2$ & $4$  & $5$  & $5$  & $6$  & $6$  & $7$  & $7$  & $8$  & $8$ \\ \hline
$n_3$ & $5$  & $5$  & $6$  & $6$  & $7$  & $7$  & $8$  & $8$  & $9$ \\ \hline
\end{tabular}
\caption{Basis set sizes used for the fit described in Table~\ref{tab:PES-polyaotmic} for H2O.}
\end{table}

\begin{table}[]
\begin{tabular}{|c|c|c|c|c|c|c|c|c|c|}
\hline
$n$  & $10$ & $12$ & $14$ & $16$ & $18$ & $20$ & $22$ & $24$ & $26$ \\ \hline
$n_1$ & $5$  & $6$  & $7$  & $8$  & $9$  & $10$ & $11$ & $12$ & $13$ \\ \hline
$n_2$ & $5$  & $6$  & $7$  & $8$  & $9$  & $10$ & $11$ & $12$ & $13$ \\ \hline
\end{tabular}
\caption{Basis set sizes used for the fit described in Table~\ref{tab:PES-polyaotmic} for ETSH.}
\end{table}

\begin{table}[]
\begin{tabular}{|c|c|c|c|c|c|c|c|c|c|c|}
\hline
$n$  & $10$ & $12$ & $14$ & $16$ & $18$ & $20$ & $22$ & $24$ & $26$ & $28$ \\ \hline
$n_1$ & $5$  & $6$  & $7$  & $8$  & $9$  & $10$ & $11$ & $12$ & $13$ & $14$ \\ \hline
$n_2$ & $5$  & $6$  & $7$  & $8$  & $9$  & $10$ & $11$ & $12$ & $13$ & $14$ \\ \hline
\end{tabular}
\caption{Basis set sizes used for the fit described in Table~\ref{tab:PES-polyaotmic} for DMS.}

\end{table}
\begin{table}[]
\begin{tabular}{|c|c|c|c|c|c|c|c|c|c|}
\hline
$n$ & $15$ & $18$ & $19$ & $20$ & $21$ & $24$ & $25$ & $26$ & $27$ \\ \hline
$n_1$ & $3$ & $3$ & $4$ & $3$ & $4$ & $4$ & $5$ & $4$ & $5$ \\ \hline
$n_2$ & $2$ & $3$ & $3$ & $3$ & $3$ & $4$ & $4$ & $4$ & $4$ \\ \hline
$n_3$ & $2$ & $3$ & $3$ & $3$ & $3$ & $4$ & $4$ & $4$ & $4$ \\ \hline
$n_4$ & $2$ & $3$ & $3$ & $3$ & $3$ & $4$ & $4$ & $4$ & $4$ \\ \hline
$n_5$ & $3$ & $3$ & $3$ & $4$ & $4$ & $4$ & $4$ & $5$ & $5$ \\ \hline
$n_6$ & $3$ & $3$ & $3$ & $4$ & $4$ & $4$ & $4$ & $5$ & $5$ \\ \hline
\end{tabular}
\caption{Basis set sizes used for the fit described in Table~\ref{tab:PES-polyaotmic} for PH3.}
\end{table}

\begin{table}[]
\begin{tabular}{|c|c|c|c|c|c|}
\hline
$n$  & $27$ & $24$ & $21$ & $18$ & $15$ \\ \hline
$n_1$ & $3$  & $3$  & $3$  & $2$  & $2$  \\ \hline
$n_2$ & $3$  & $3$  & $3$  & $2$  & $2$  \\ \hline
$n_3$ & $3$  & $2$  & $3$  & $2$  & $1$  \\ \hline
$n_4$ & $3$  & $2$  & $2$  & $2$  & $1$  \\ \hline
$n_5$ & $3$  & $2$  & $2$  & $2$  & $1$  \\ \hline
$n_6$ & $3$  & $3$  & $2$  & $2$  & $2$  \\ \hline
$n_7$ & $3$  & $3$  & $2$  & $2$  & $2$  \\ \hline
$n_8$ & $3$  & $3$  & $2$  & $2$  & $2$  \\ \hline
$n_9$ & $3$  & $3$  & $2$  & $2$  & $2$  \\ \hline
\end{tabular}
\caption{Basis set sizes used for the fit described in Table~\ref{tab:PES-polyaotmic} for CH4.}
\end{table}

\begin{table}[]
\begin{tabular}{|c|c|c|c|c|c|c|c|c|c|}
\hline
$n$  & $14$ & $15$ & $17$ & $18$ & $20$ & $21$ & $23$ & $24$ & $26$ \\ \hline
$n_1$ & $4$  & $5$  & $5$  & $6$  & $6$  & $7$  & $7$  & $8$  & $8$  \\ \hline
$n_2$ & $5$  & $5$  & $6$  & $6$  & $7$  & $7$  & $8$  & $8$  & $9$  \\ \hline
$n_3$ & $5$  & $5$  & $6$  & $6$  & $7$  & $7$  & $8$  & $8$  & $9$  \\ \hline
\end{tabular}
\caption{Basis set sizes used for the fit described in Table~\ref{tab:PES-polyaotmic} for PSB3.}
\end{table}

\section{Estimating norms for polyspherical orthogonal coordinates}
\label{app:polyspherical}
For the polyspherical coordinates, the momentum operators are expressed as $\hat{P}_i = -i\frac{\partial}{\partial q_i}$ for $q_i \in\{R_k, u_l, \phi_m\}$ with $u_l = \cos{\theta_l}$. The inner product is given by
\begin{equation}
    \bra{\phi}\psi\rangle = \int d\tau \overline{\phi}(x)\psi(x) 
\end{equation}
with 
\begin{equation}
    d\tau = d\alpha d\beta\sin\beta d\gamma \left(\prod_{k=1}^{N-1}dR_k \right)\left(\prod_{l=1}^{N-2}du_l\right)\left(\prod_{m}^{N-3}d\phi_m\right).
\end{equation}
Matrix elements of the momentum operator corresponding to $R_i$ in the harmonic oscillator basis are given by
\begin{equation}
    \bra{m}\hat{P}_{R_i}\ket{n} = i\sqrt{\frac{m\hbar\omega}{2}}(\sqrt{n+1}\delta_{m n+1} - \sqrt{n}\delta_{m,n-1})
\end{equation}
Therefore, the norm associated with the radial momentum operator can be written as: $\lambda_{P_R} = \sqrt{\frac{m\hbar\omega}{2}(N_R+1)}$. For the angular coordinates, we consider the Gauss-Legendre DVR. Legendre polynomials satisfy the recurrence relation \cite{Riley2006}:
\begin{equation}
    (2n+1)P_n(x) = \frac{d}{dx}(P_{n+1}(x)-P_{n-1}(x))
\end{equation}
which leads to
\begin{equation}
    \frac{d}{dx}P_{n+1} = \sum_{k=0}^{n}(2(n-2k)+1)P_{n-2k}(x)
\end{equation}
From orthogonality condition
\begin{equation}
    \int_{-1}^{1}dx P_n(x)P_m(x) = \frac{2}{2n+1}\delta_{nm}
\end{equation}
one finds
\begin{equation}
    \int_{-1}^1dx P_n(x)\frac{d}{dx}P_m(x)= \begin{cases}
        2; \text{ if } (n<m) \land (n+m=1 \text{ mod } 2)\\
        0; \text{ otherwise}\\
    \end{cases}
\end{equation}
For $L_2$ normalized states $\ket{n}$ defined by $\bra{\theta}n\rangle = \sqrt{\frac{2n+1}{2}}P_n(\cos{\theta})$ one finds
\begin{equation}
    \bra{n}\hat{P}_u\ket{m}  = \begin{cases}
        \sqrt{(2n+1)(2m+1)}; \text{ if } (n<m) \land (n+m=1 \text{ mod } 2)\\
        0;\text{ otherwise}
    \end{cases}
\end{equation}
Therefore $\lambda_{P_u} =\lambda_{P_\phi} = \sqrt{4(N_\theta-1)^2-1}\approx2N_\theta$. In order to estimate the max norm of the metric tensor given in Eq.~(\ref{eq:KEO_PGP}), we first bound each term $G_{ij}$ by an expression:
\begin{equation}
    G_{ij}<MR^{-n_1}\sin^{-n_2}{\theta}
\end{equation}
with $M \in \mathbb{R}_+$ and $n_1,n_2\in\{0,1,2\}$. Then, we evaluate the singular expressions $R^{-n_1}$ and $\sin^{-n_2}{\theta}$ on the corresponding grid. The resulting bounds on the norms are shown in Table~\ref{tab:norm_estimates}.

\begin{figure}[h]
    \centering
    \includegraphics[width=0.4\linewidth]{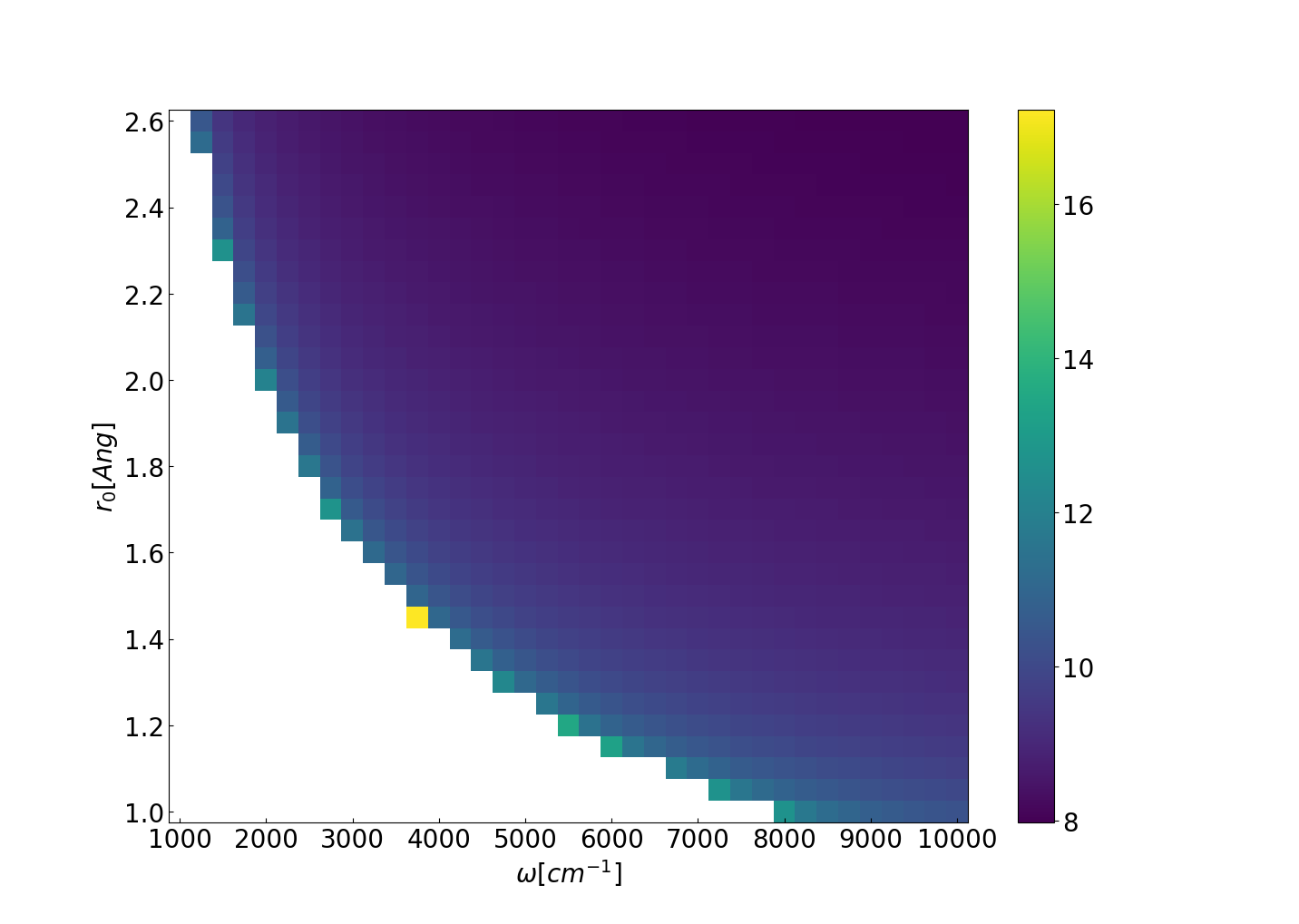}
    \caption{Value of $\log_{10}\lambda_K [\text{cm}^{-1}]$ for different values of parameters $r_0$ and $\omega$ used in the definition of the Gauss-Hermite grid. The remaining parameters are set to: $N_r=N_\theta=N_{\phi}=2^{7}$ and $J=20$, $\theta_{max}=0.9\frac{\pi}{2}$.}
    \label{fig:placeholder}
\end{figure}

\begin{figure}[h]
    \centering
    \includegraphics[width=0.6\linewidth]{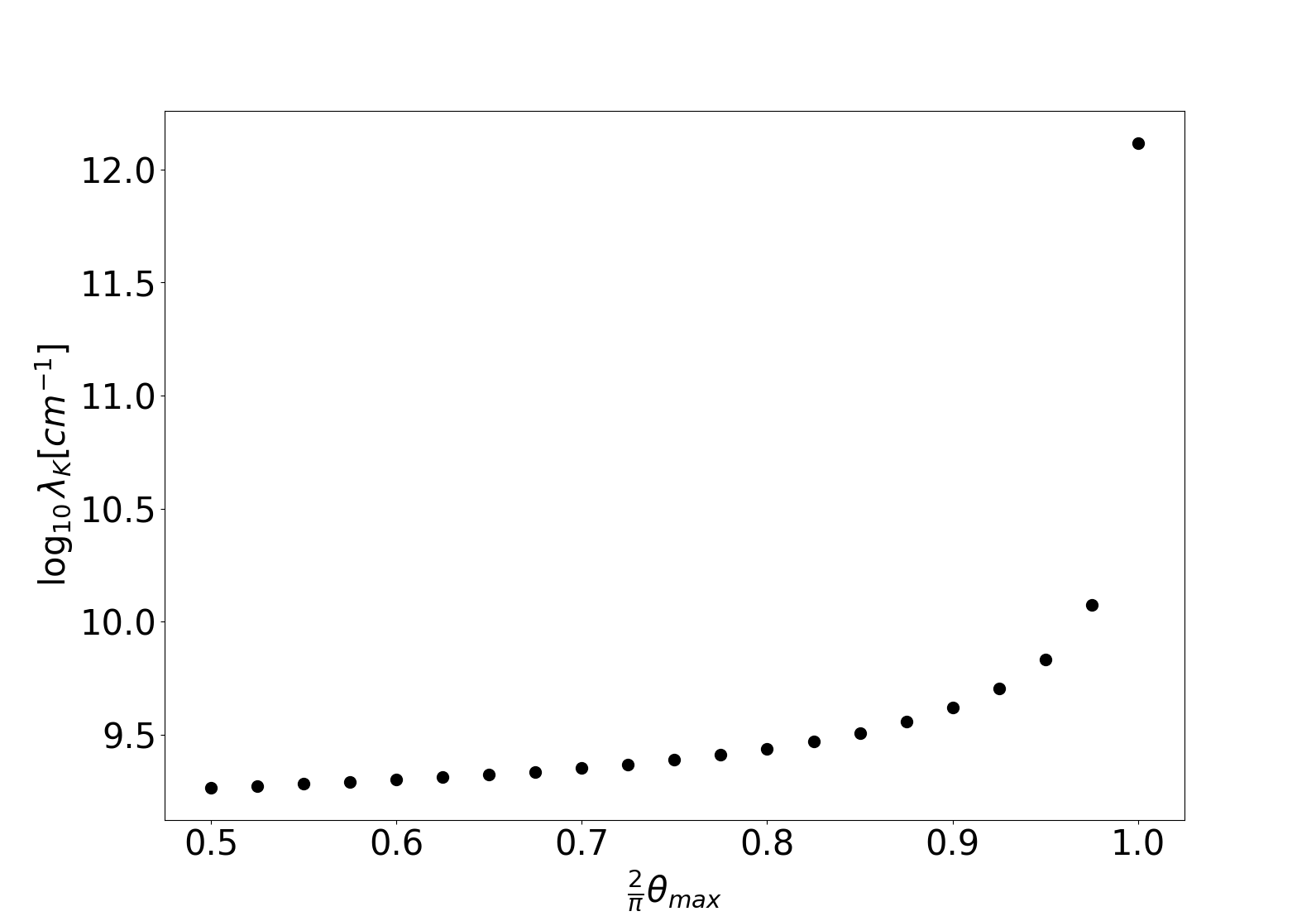}
    \caption{Value of $\log_{10}\lambda_K [\text{cm}^{-1}]$ for polar angles defined of intervals with differing maximal polar coordinates $\theta_{max}$. The remaining parameters are set to: $r_0 = 1.5$, $\omega = 5000$ [cm$^{-1}$], $N_r=N_\theta=N_{\phi}=2^{7}$ and $J=20$.}
    \label{fig:placeholder2}
\end{figure}

\begin{figure}%
    \centering
    \subfloat[]{{\includegraphics[width=6cm]{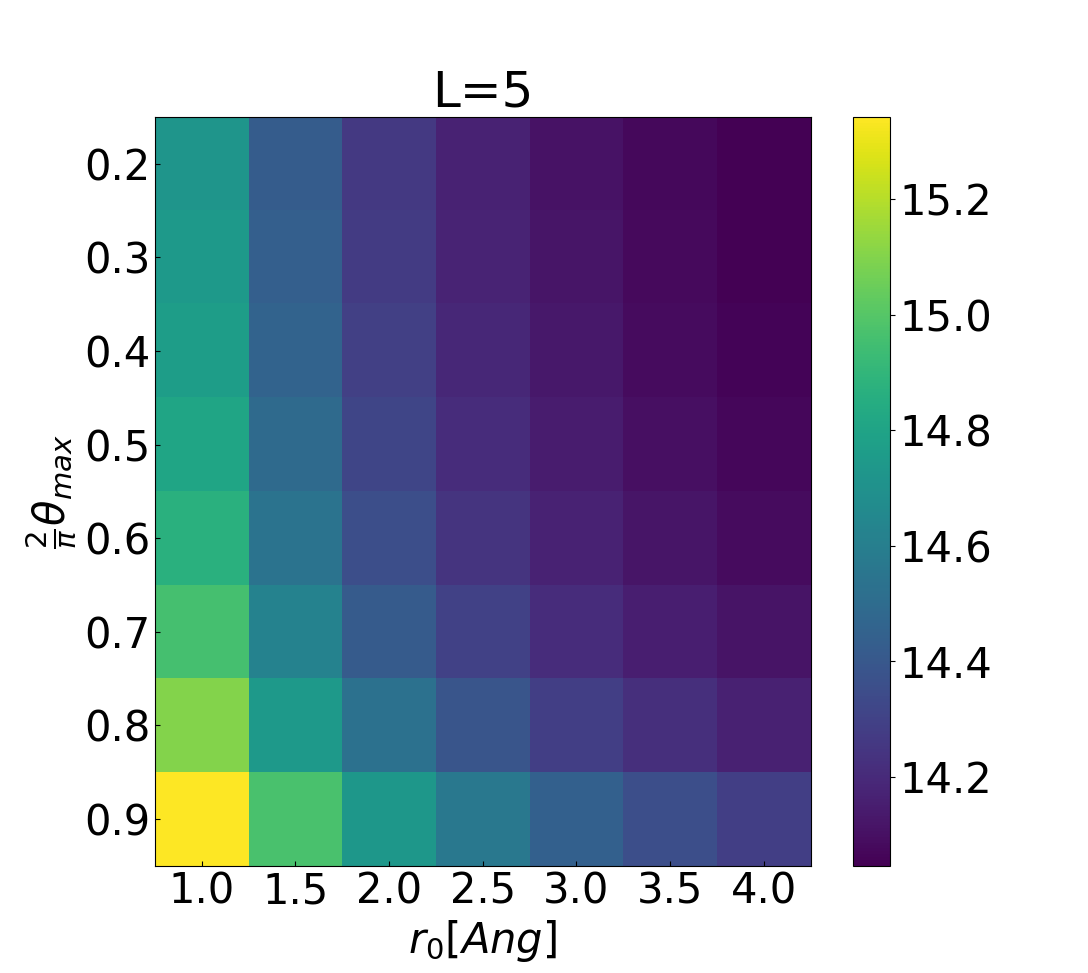} }}%
    \subfloat[]{{\includegraphics[width=6cm]{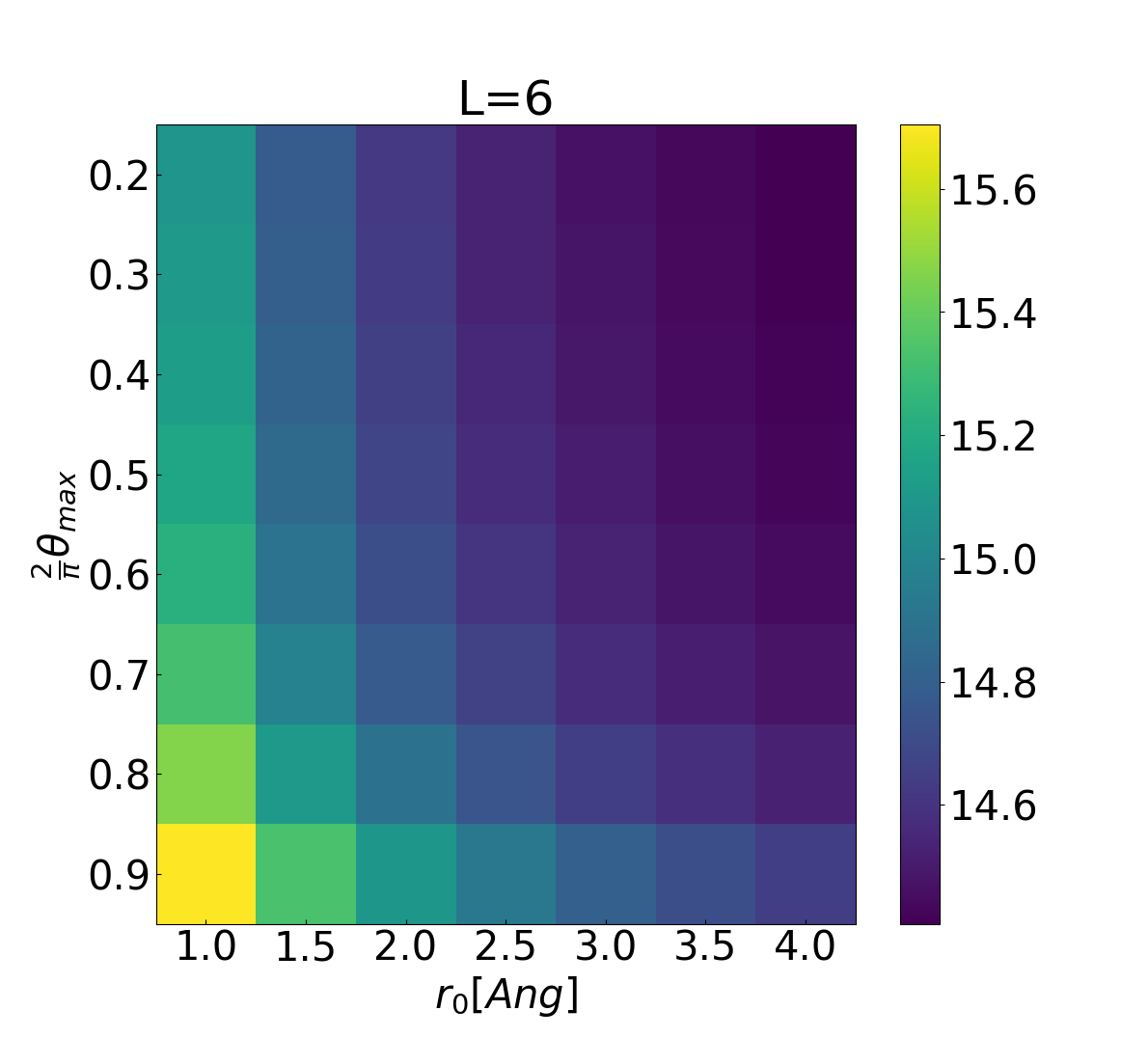} }}%
    \caption{QPE T-count for methane with $N_R = N_\theta = N_\phi=2^4$, $\omega=5000$, $\epsilon = 1$cm$^{-1}$}%
    \label{fig:example}%
\end{figure}
\begin{figure}%
    \centering
    \subfloat[]{{\includegraphics[width=8cm]{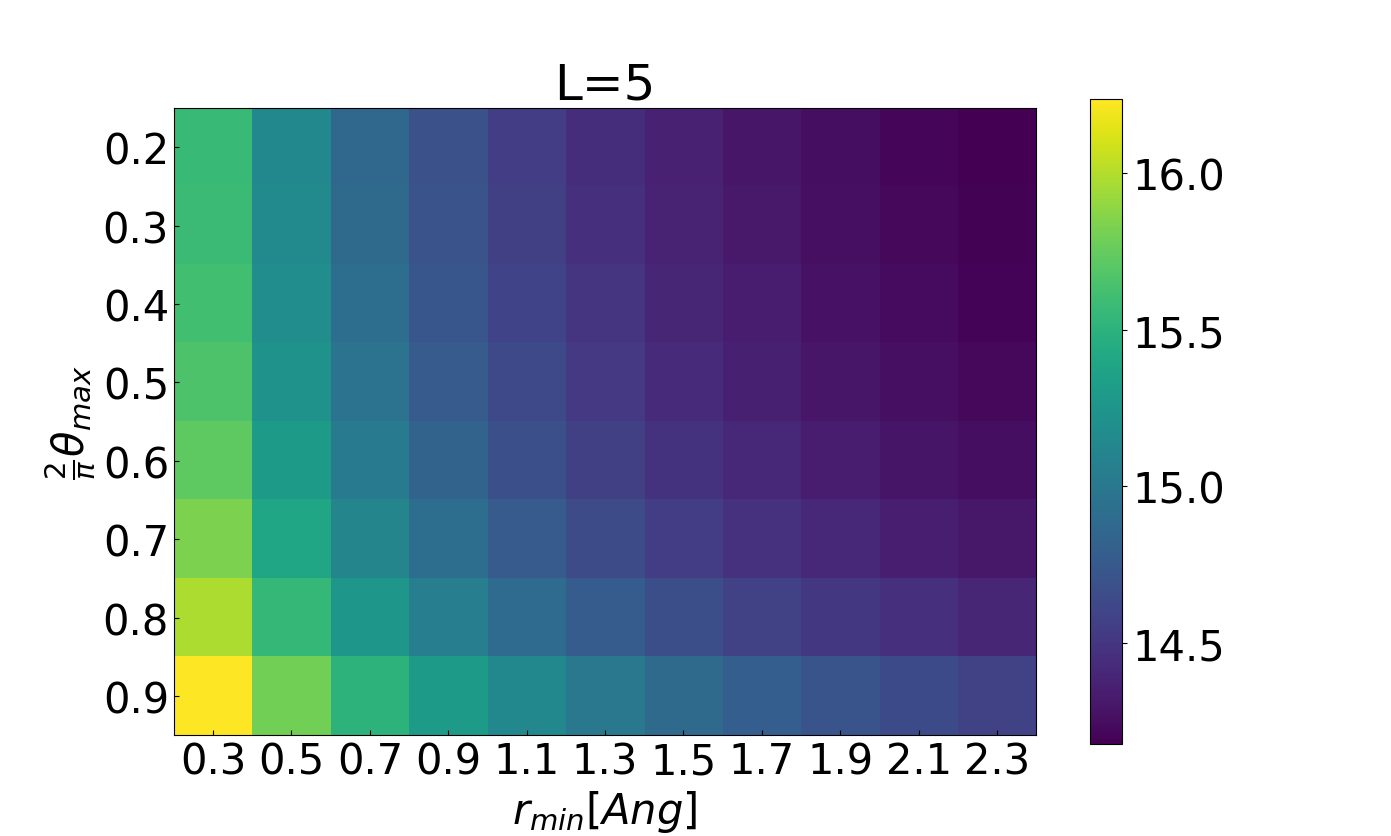} }}%
    \qquad
    \subfloat[]{{\includegraphics[width=8cm]{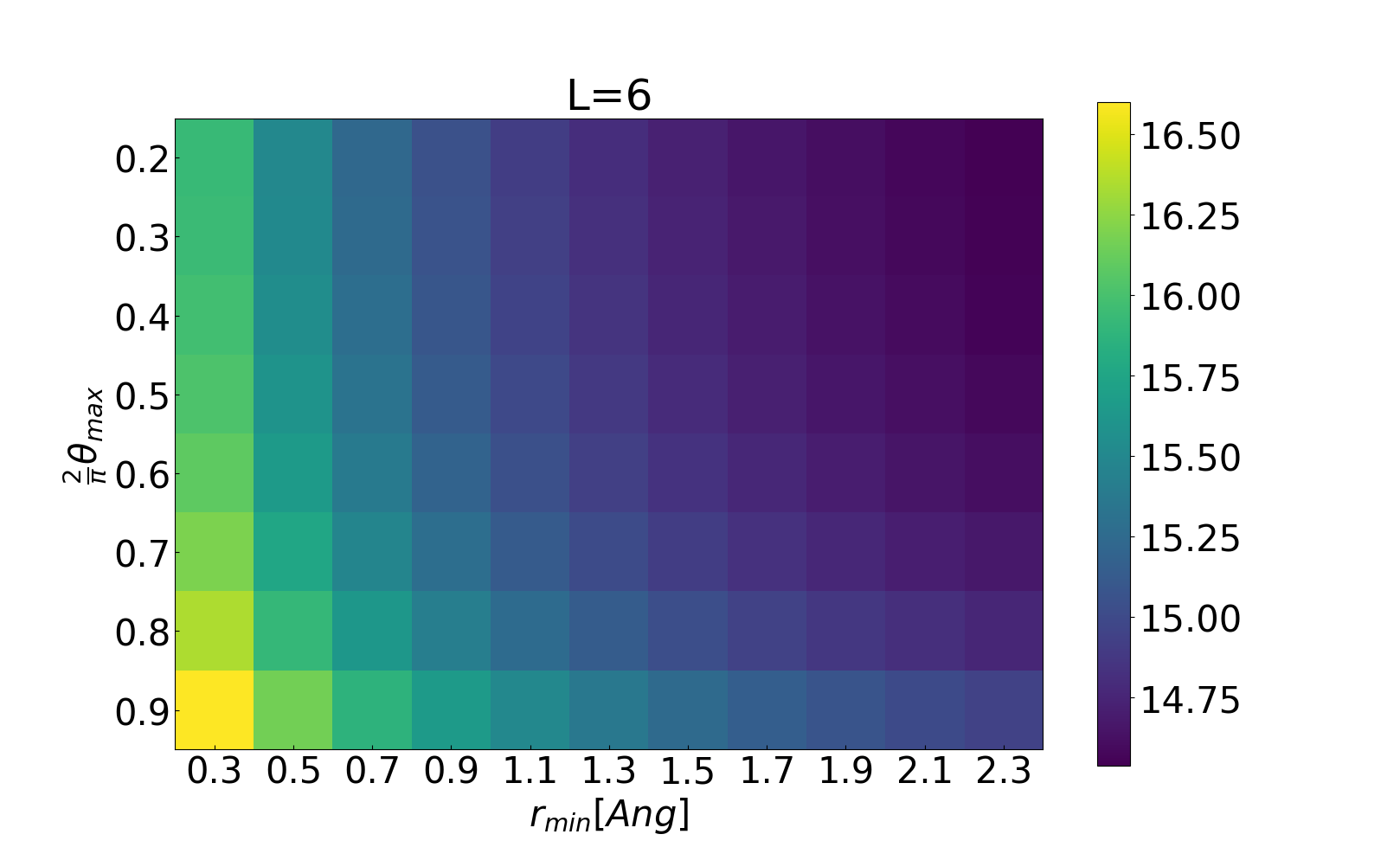} }}%
    \caption{QPE T-count for methane with $N_R = N_\theta = N_\phi=2^4$, $\omega=5000$, $\epsilon = 1$cm$^{-1}$}%
    \label{fig:example2}%
\end{figure}

\begin{figure}%
    \centering
    \subfloat[]{{\includegraphics[width=6cm]{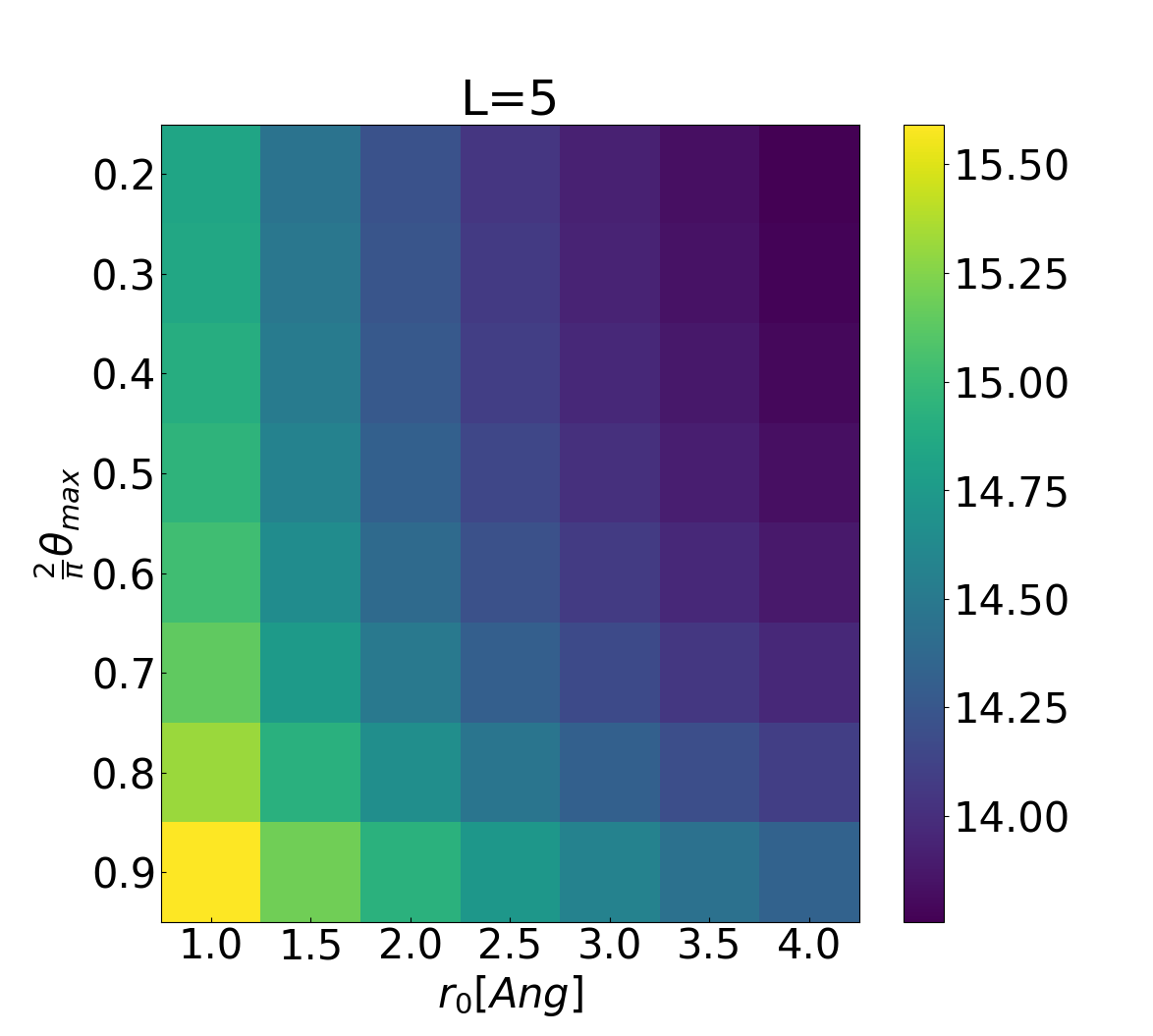} }}%
    \qquad
    \subfloat[]{{\includegraphics[width=6cm]{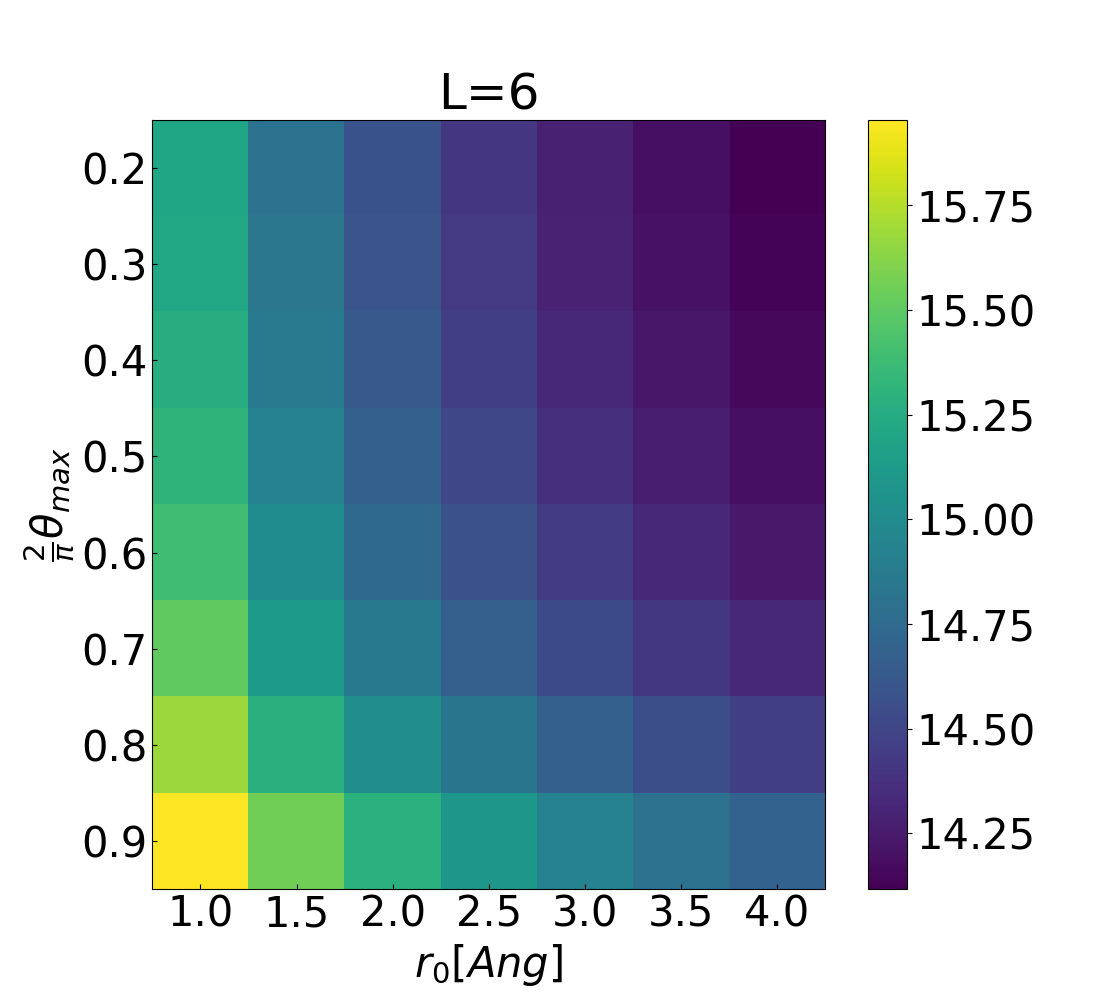} }}%
    \caption{QPE T-count for general $M=39$ molecule with $N_R = N_\theta = N_\phi=2^4$, $\omega=5000$, $\epsilon = 1$cm$^{-1}$}%
    \label{fig:example3}%
\end{figure}
\begin{figure}%
    \centering
    \subfloat[]{{\includegraphics[width=8cm]{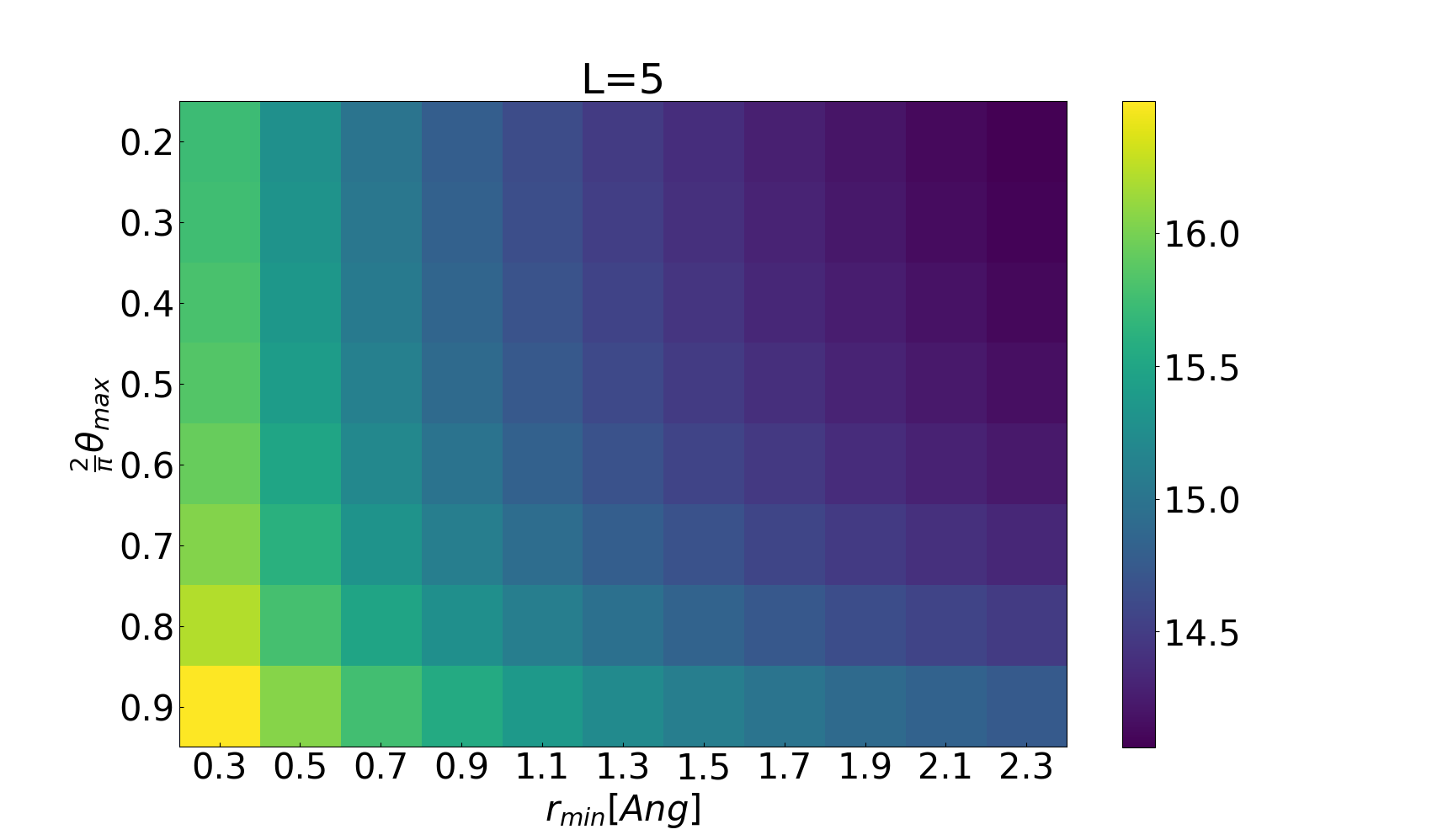} }}%
    \qquad
    \subfloat[]{{\includegraphics[width=8cm]{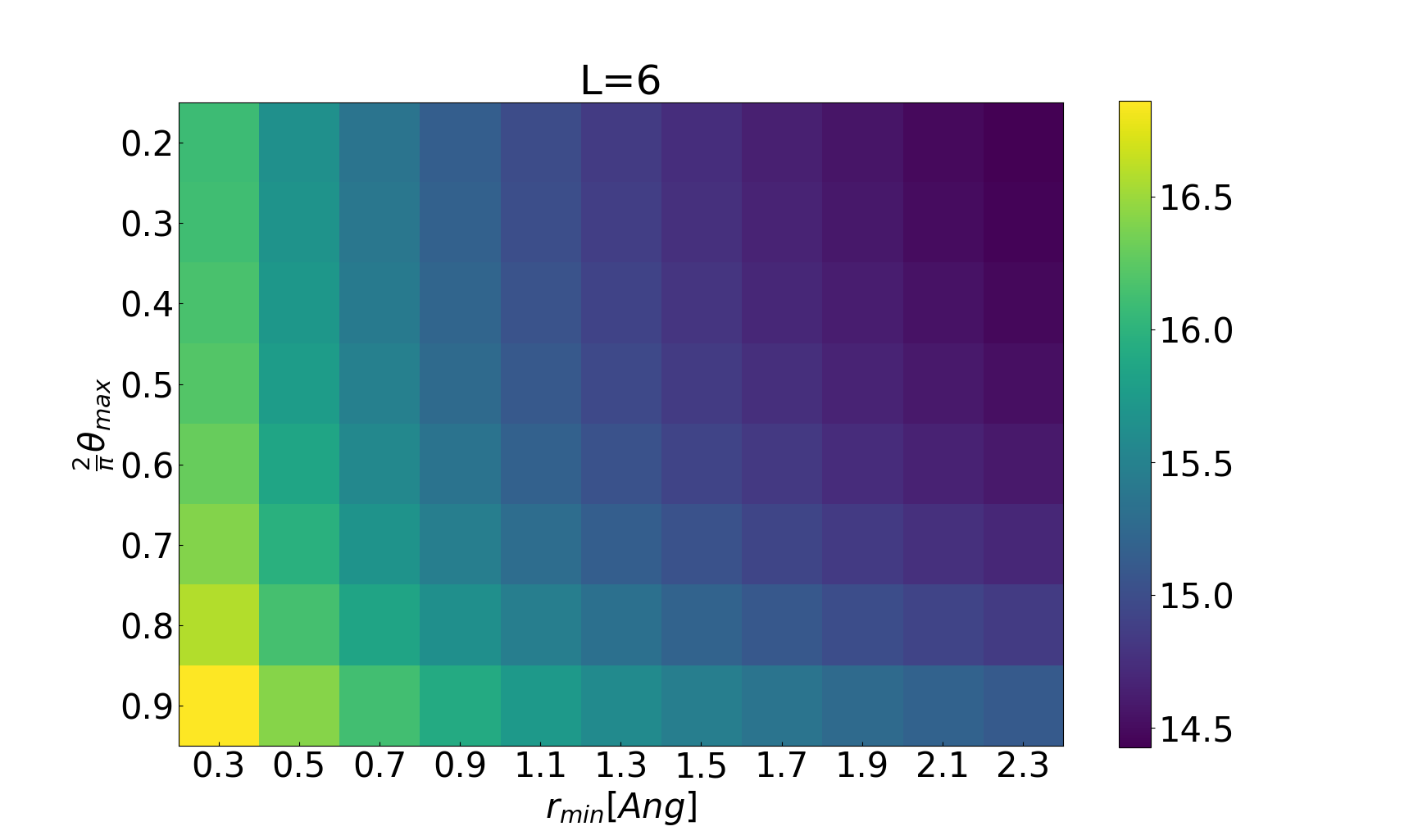} }}%
    \caption{QPE T-count general $D=30$ molecule $N_R = N_\theta = N_\phi=2^4$, $\omega=5000$, $\epsilon = 10$cm$^{-1}$}%
    \label{fig:example4}%
\end{figure}

\begin{table}[H]
    \centering
    \begin{tabular}{c|c|c|c}
Term  &Norm                 & Number of terms & Total\\
    $P_R$ [a.u]&$80$&-&-\\
    $P_U/P_\phi$ [1]&$87$&-&-\\
    $(r_{min}^2\sin^2{\theta_{max}}\mu)^{-1}$ [cm$^{-1}]$&$8.7$ &-&-\\
$ P_RG_{RR}P_R$ [cm$^{-1}$]& $1.1 \times 10^{5}$ & $11$  &  $1.2\times10^6$ \\
$ P_UG_{UU}P_U$ [cm$^{-1}$] & $2.1 \times 10^{4}$ & $121$ &  $2.6\times10^6$ \\
$ P_UG_{U\phi}P_\phi$ [cm$^{-1}$] & $2.8 \times 10^{4}$ & $100$  &  $2.8\times10^6$ \\
$ P_\phi G_{\phi\phi}P_\phi$ [cm$^{-1}$] & $5.5 \times 10^{4}$ & $81$  &  $4.4\times10^6$ \\
    \end{tabular}
    \caption{Norm breakdown for $D=30$ for: $N=2^4$ $r_0 = 2$\AA, $\omega=5000$ cm$^{-1}$ ($r_{min}=1.85 $\AA), $\theta_{max}=0.5\frac{\pi}{2}$, $\mu = 7$ Da}
    \label{tab:placeholder}
\end{table}

\begin{table}[h]
    \centering
    \begin{tabular}{c|c|c}
         Operator& Norm[cm$^{-1}$]&Number of occurences  \\
         \hline
$ G_{RR}$ & $3.2 \times 10^{5}$ & $(L-1)^2$\\
$ G_{UU}$ & $2.8 \times 10^{7}$ & $(L-2)^2$\\
$ G_{UP}$ & $4.0 \times 10^{7}$ & $(L-2)(L-3)$\\
$ G_{PP}$ & $8.4 \times 10^{7}$ & $(L-3)^2$\\
$ G_{ZZ}$ & $1.7 \times 10^{5}$ & $1$\\
$ G_{XX}$ & $1.8 \times 10^{5}$ & $1$\\
$ G_{XY}$ & $1.8 \times 10^{5}$ & $1$\\
$ G_{XZ}$ & $2.5 \times 10^{5}$ & $1$\\
$ G_{XU}$ & $1.6 \times 10^{6}$ & $L-2$\\
$ G_{YU}$ & $1.6 \times 10^{6}$ & $L-2$\\
$ G_{ZU}$ & $1.6 \times 10^{6}$ & $L-2$\\
$ G_{XP}$ & $4.5 \times 10^{6}$ & $L-3$\\
$ G_{YP}$ & $2.3 \times 10^{6}$ & $L-3$\\
$ G_{ZP}$ & $2.2 \times 10^{6}$ & $L-3$\\
    \end{tabular}
    \caption{Norms for $\omega=5000 cm^{-1}$, $\mu=1 Da$, $R_0=1.5 \AA$, $N_\theta=2^{7}$, $N_R = 2^{7}$, $J=20$, $\theta_{max} = 0.5\frac{\pi}{2}$}
    \label{tab:norm_estimates}
\end{table}

\subsection{Comment about norms}\label{sec:comment_about_norms}
For block-encoding an operator $G$ we have the following equality:
\begin{equation}
    U_G\ket{\psi}\ket0 = \frac{1}{\lambda_G}G\ket{\psi}\ket0+\ket{\perp}
\end{equation}
Next, for $U_G$ to be unitary we require
\begin{equation}
    \left|\left|\frac{G}{\lambda_G}\ket{\psi}\right|\right|_2 = \frac{1}{\lambda_G^2}\bra{\psi}G^2\ket{\psi} \leq 1
\end{equation}
for any $\ket{\psi}$. Therefore
\begin{equation}
    \lambda_G \ge \sqrt{\bra{\psi}G^2\ket{\psi}}.
\end{equation}
Let us take $\ket{\psi}$ to be an eigenstate with largest value $g_{max}$. Then
\begin{equation}
    \lambda_G \ge g_{max}
\end{equation}
Therefore, the norm is always greater then the maximal element of a diagonalized matrix. That means, that for any $f(q)$, the norm in the DVR representation is minimized (Note, that in $\rho$-sparse encoding, the norm is given by $\lambda=d|G|_{max}$. While DVR leads to larger $|G|_{max}$, FBR lead to significantly higher number of non-zero elements $d$). Now, the most problematic term in the DVR representation which has a (simplified) form:
\begin{equation}
    P_{\phi_i } G_{\phi_i\phi_j} P_{\phi_j } \approx \frac{P_{\phi_i }  P_{\phi_j }}{R^2\sin{\theta^2}}
\end{equation}
Since momentum operators act on different qubits than the singular term $sin^{-2}\theta$ no method for implementing $P_iG_{ij}P_j$ can obtain a better norm, then the DVR implementation.

For a molecule with 30 fully-coupled modes, we found the norms to be $\lambda\approx1.3\times 10^8 cm^{-1}\approx 600 Ha$, with $N_R=2^7$ and $N_\theta =2^4$. Ref.~\cite{Majland2025} reports $\lambda \approx 10-100$ Hartree for the H$_2$O$_2$ molecule with $6$ vibrational modes and $n=2-10$. Comparing to our method, for $6$ vibrational modes, we obtain $\lambda=5 Ha$ for $N_R=N_{\theta}=8$ and $\omega=5\times10^3 cm^{-1}$, $\mu=1 Da$, $R_0=1.65 \AA$, $N_\theta=2^{7}$, $N_R = 2^{7}$, $\theta_{max} = 0.7\frac{\pi}{2}$.

\section{Resource estimation for the full DVR Hamiltonian}
\label{sec:RE-DVR}
In this section we summarize resource estimation for the case of full DVR Hamiltonian.

In the general case, we consider all elements of the momentum operator matrices $\boldsymbol{P}_i^{\mathrm{DVR}}$ to be nonzero. Let us  consider the cost of implementing $O_F$ for different parts of the circuit. In DVR, momentum operators $\boldsymbol{P}_i$ in general have all elements non-zero, whereas functions  $\boldsymbol{g}_{ij}(\boldsymbol{q}), \boldsymbol{\Gamma}_{\alpha j}(\boldsymbol{q)}$ are diagonal. Therefore all terms in $\boldsymbol{K}^{vib}$ and $\boldsymbol{K}^{cor}$ are in the form given by \ref{eq:T_free_OF} and the oracle $O_F$ only requires $\mathcal{O}(\log_2n_i)$ Clifford gates and no T-gates. The only remaining non-diagonal terms are angular momentum operators $\boldsymbol{J}_x$ and $\boldsymbol{J}_y$, for which $O_F$ can has the same form as in Eq.~(\ref{eq:OF_angular_momentum}) and can also be implemented with the cost of $\mathcal{O}(\log_2{(2J+1)})$  Clifford+T gates. 
Parameters and quantities required to express the costs of partial block-encodings and the total rovibrational Hamiltonian block-encoding in the DVR are summarized in Tables~\ref{tab:T_cost_partial_DVR}.
\begin{table}[]
    \centering
    \begin{tabular}{c|c|c|c|c|c}
                    &$\rho$ &$n$&Number  \\ 
        \hline
         $\BE{\boldsymbol{U}^{vib}_{ii}}$&$n_i$ &$n_{g^{ii}}$& $D$ \\
         $\BE{\boldsymbol{U}^{vib}_{ij}}$&$n_in_j$&$n_{g^{ij}}$& $\frac{D(D-1)}{2}$ \\
         $\BE{\boldsymbol{U}^{cor}_{\alpha j}}$&$n_i$&$n_{\Gamma^{\alpha j}}$& $3D$ \\
         $\BE{\boldsymbol{U}^{\mu}_{\alpha \beta}}$&$1$& $n_{{\mu}^{\alpha\beta}}$&$6$\\
         $\BE{\boldsymbol{U}_{J_z}}$&$1$& $2J+1$&$2$\\
         $\BE{\boldsymbol{U}_{J_{\alpha\neq z}}}$&$2$& $2J+1$&$4$\\
         $\BE{\boldsymbol{U}_V}$&$1$&$\prod_{i=1}^D n_i$&$1$
    \end{tabular}
    \caption{Parameters relevant to the cost of block-encoding the rovibrational Hamiltonian in DVR. $\rho$ denotes the maximal number of non-zero element in a row of a matrix and $n$ is the dimension of a matrix. The last column shows the number of unitaries of a given type required to block-encode the full Hamiltonian.}
    \label{tab:T_cost_partial_DVR}
\end{table}
In summary, the total Clifford+T gate cost for block-encoding the DVR Hamiltonian is given by
\begin{equation}
   C_H(DVR) =   C_{BE} + O(D^2\log_2D) + O\left(D\sqrt{\log_2\frac{1}{\epsilon}}\right) + \mathcal{O}(\log_2{2J+1}) .
\end{equation}
For a clearer intuitive picture of the scaling, let us consider equal number of basis functions for each coordinate $n_i = n$ for all $i$. Additionally, let as assume that all matrix elements of $\boldsymbol{g}_{ij}(\boldsymbol{q})$, $\boldsymbol{\Gamma}_{\alpha j}(\boldsymbol{q})$ and $\boldsymbol{\mu}_{\alpha \beta}(\boldsymbol{q})$ are functions of at most $m$ coordinates, with $m>2$. That leads to $n_{g_{ij}}=n_{\Gamma_{\alpha j}}=n_{\mu_{\alpha \beta}} = n^{m}$. Finally, let each matrix have at most $\rho=\rho_i$ non-zero elements.

\begin{table}[]
    \begin{tabular}{c|c}
                    &Clifford+T-count  \\ 
        \hline
         $\boldsymbol{K}^{vib}$& $2MC_D(n^{^{m+1}})+M(M-1)C_D(n^{m+2})$ \\
         $\boldsymbol{K}^{cor}$& $6MC_D(n^{m+1})+4C_D(4J+2)+C_D(2J+1)$\\
         $\boldsymbol{K}^{rot}$& $C_D(n^m)+4C_D(4J+2)+C_D(2J+1)$\\
         $\boldsymbol{V}$& $C_D(n^{D})$\\
    \end{tabular}
    \caption{Summary of Clifford+T costs of partial block-encodings corresponding to different parts of the Hamiltonian using the DVR approach. For simplicity, we assume equal number of basis functions $n_i = n$ and equal sizes of $\boldsymbol{g}_{ij}(\boldsymbol{q}),\boldsymbol{\Gamma}_{ij}(\boldsymbol{q})$ and $\boldsymbol{\mu}_{ij}(\boldsymbol{q})$ matrices: $n_{g_{ij}}=n_{\Gamma_{\alpha j}} =  n_{\mu_{\alpha\beta}} = n^m$. Here, $m$ is the number of coordinates that the matrices $\boldsymbol{g}_{ij}(\boldsymbol{q)}$, $\boldsymbol{\Gamma}_{\alpha j}(\boldsymbol{q})$, $\boldsymbol{\mu}_{\alpha\beta}(\boldsymbol{q)}$ depend on.}
    \label{tab:T_cost_Ham}
\end{table}
The total cost of block-encoding the DVR Hamiltonian is given  by 
\begin{equation}
    C_{BE} =8DC_{D}(n^{m+1})+D(D-1)C_{D}(n^{m+2})+6C_D(n^m)+2C_D(2J+1)+8C_D(4J+2)+C_D(n^D).
\end{equation}
and a breakdown of costs associated with individual terms is summarized in \ref{tab:T_cost_Ham}.

\paragraph{Resource estimation for DVR Hamiltonian in polyspherical coordinates}
Tables~\ref{tab:T_cost_partial_spherical},~\ref{tab:poly_spher_scaling} present resource estimation for block-encoding the full DVR Hamiltonians. We note that for $A<5$, the scaling with $n$ improves relative to the values reported in Table~\ref{tab:poly_spher_scaling}, since the most costly terms, such as $\boldsymbol{U}^{vib}(\phi_i,\phi_j)$, are absent. For instance, in the case of a triatomic molecule the cost of implementing the KEO scales as $O(n^{\frac{3}{2}})$.
\begin{table}[]
    \centering
    \begin{tabular}{c|c|c|c|c|c|c}
                    &$\rho$ &$n$&Number &Clifford+T ($n=n_i$)&  Reduced Clifford+T\\ 
        \hline
         $\boldsymbol{U}^{vib}({R_i,R_i})$&$n_R$ &$n_R$& $A-1$ &$2C_D(2)$ &$2C_D(n^2)$\\
         $\boldsymbol{U}^{vib}({u_i,u_i})$&$n_\theta$ &$n_R^2n_\theta$& $A-2$& $2C_D(n^4)$& $4C_D(n^2)+4C_D(n)$\\
         $\boldsymbol{U}^{vib}({\phi_i,\phi_i})$&$n_\phi$ &$n_R^3n_\theta^2 n_\phi$& $A-3$& $2C_D(n^7)$ &$12C_D(n^2)+30C_D(n)$\\
         $\boldsymbol{U}^{vib}({u_i,u_j})$&$n_\theta^2$ &$n_Rn_\theta^2n_\phi^2$& $\frac{(A-2)(A-3)}{2}$& $2C_D(n^7)$ & $4C_D(n^2)+6C_D(n)$\\
         $\boldsymbol{U}^{vib}({\phi_i,\phi_j})$&$n_\phi^2$ &$n_R^2n_\theta^3n_\phi^2 $& $\frac{(A-3)(A-4)}{2}$& $2C_D(n^9)$&$24C_D(n^2)+30C_D(n)$\\
         $\boldsymbol{U}^{vib}({u_i,\phi_i})$&$n_\theta n_\phi$ &$n_Rn_\theta^2n_\phi$& $A-3$& $2C_D(n^6)$&$4C_D(n^2)+2C_D(n)$\\
        $\boldsymbol{U}^{vib}({u_i,\phi_j})$&$n_\theta n_\phi$ &$n_Rn_\theta^3 n_\phi^2$& $(A-3)^2$& $2C_D(n^8)$&$12C_D(n^2)+12C_D(n)$\\
        \hline
        $\boldsymbol{U}^{cor}(\alpha,u_j)_{\alpha=x,y}$&$n_\theta$ &$n_R n_\theta n_\phi$& $2(A-2)$& $2C_D(n^4)$&$2C_D(n^2)+2C_D(n)$\\
        $\boldsymbol{U}^{cor}(z,u_j)$&$n_\theta$ &$n_R n_\theta^2 n_\phi$& $A-2$& $2C_D(n^5)$&$2C_D(n^2)+3C_D(n)$\\
        $\boldsymbol{U}^{cor}(x,\phi_j)$&$n_\phi$ &$n_R n_\theta^2 n_\phi$& $A-3$& $2C_D(n^5)$&$4C_D(n^2)+6C_D(n)$\\
        $\boldsymbol{U}^{cor}(y,\phi_j)$&$n_\phi$ &$n_R n_\theta n_\phi$& $A-3$& $2C_D(n^4)$& $2C_D(n^2)+2C_D(n)$\\
        $\boldsymbol{U}^{cor}(z,\phi_j)$&$n_\phi$ &$n_R^2 n_\theta^2 n_\phi$& $A-3$& $2C_D(n^6)$&$6C_D(n^2)+12C_D(n)$\\
        \hline
        $\boldsymbol{U}_{\mu}(z,z)$&$1$ &$n_R^2n_\theta$& 1 & $C_D(n^3)$& $6C_D(n)$\\
        $\boldsymbol{U}_{\mu}(\alpha,\alpha)_{\alpha=x,y}$&$1$ &$n_R$& 2 & $C_D(n)$&$C_D(n)$\\
        $\boldsymbol{U}_{\mu}(x,z)$&$1$ &$n_Rn_\theta$& 1 & $C_D(n^2)$&$2C_D(n)$\\
        \hline
        $\boldsymbol{U}_{J_z}$&$1$& $2J+1$&$2$& $C_D(2J+1)$&$C_D(2J+1)$\\
        $\boldsymbol{U}_{J_{\alpha= x,y}}$&$2$& $2J+1$&$4$&$2C_D(4J+2)$&$2C_D(4J+2)$\\
        \hline
         $\boldsymbol{U}_V$&$1$&$n_R^{A-1}n_\theta^{A-2}n_\phi^{A-3}$&$1$&$C_D(n^{(3A-6)})$&$C_D(n^{(3A-6)})$
    \end{tabular}
    \caption{Summary of parameters for calculating the cost of block-encoding various parts of the rovibrational Hamiltonian and the T-count for the polyspherical internal coordinates using DVR method. $\rho$ denotes the maximal number of non-zero element in a row of a matrix, while $n$ denotes the dimension of a matrix. The rightmost column shows the T-count assuming $n_R=n_\theta=n_\phi = n$. The last column shows the improved scaling when utilizing of the sum of products form of the $\boldsymbol{G}_{ij}$ as described in \ref{sec:block_encoding_reducing_costs}. For clarity, the lower indices were replaced by arguments in unitaries, e.g. $\boldsymbol{U}_{ab}\rightarrow \boldsymbol{U}(a,b)$.}.
    \label{tab:T_cost_partial_spherical}
\end{table}
\begin{table}[]
    \centering
    \begin{tabular}{c|c|c|c|c|c}
                 KEO term   &$\boldsymbol{K}^{vib}$&$\boldsymbol{K}^{cor}$&$\boldsymbol{K}^{rot}$& $\boldsymbol{V}_{(SEL-SWAP)}$ &  $\boldsymbol{V}_{(WH)}$\\ 
        \hline
        T-cost  &$O\left(A^2n^{\frac{9}{2}}\right)$&$O\left(An^{3}\right)+O(\sqrt{2J+1)}$&$O\left(2N^{\frac{3}{2}}\right)+O(\sqrt{2J+1)}$& $O\left(n^{\frac{1}{2}{(3A-6)}}\right)$ & $O\left(n^{\alpha{(3A-6)}}\right)$\\
        T-cost(Reduced)  &$O\left(A^2n \right)$&$O\left(An\right)+O(\sqrt{2J+1)}$&$O\left(2n \right)+O(\sqrt{2J+1})$& $O\left(n^{\frac{1}{2}{(3A-6)}}\right)$ & $O\left(n^{\alpha{(3A-6)}}\right)$\\
    \end{tabular}
    \caption{Asymptotic T-count scaling for different parts of the Hamiltonian expressed in polyspherical coordinates using the DVR method. The first 4 columns assume $C_D(n) = \mathcal{O}(\sqrt{n})$, achieved by SELECT-SWAP QROM for diagonal unitary synthesis, while the last column assume WH-QROM. Here, $\alpha$ is a potential dependent constant. The second row shows the improved scaling when utilizing of the sum of products form of the $\boldsymbol{G}_{ij}$ as described in \ref{sec:block_encoding_reducing_costs}.}
    \label{tab:poly_spher_scaling}
\end{table}

\section{Resource calculation procedure for SELECT-SWAP QROM}
\label{sec:appendix-SELECT-SWAP}
For the SELECT-SWAP method, we employ a procedure that yields either the exact complexity or a rigorous lower bound, summaries in the three points given below:
\begin{enumerate}
	\item The number of qubits is $2 \eta + \lambda d$, where $\lambda \geqslant 1$ is an integer. This expression is exact. 
	\item The number of CNOT gates equals the number of nonzero digits in the corresponding function. Specifically, if $f : \F_2^\eta \rightarrow \F_2^d$ is the function for the QROM, then  
	\begin{equation}  
		\mbox{CNOT count} = \sum\limits_{x \in \F_2^\eta} h \( f(x) \),  
	\end{equation}  
	where $h(f(x))$ counts the Hamming weight of $f(x)$. This assumes that XORing a nonzero digit in $f$ requires at least one CNOT gate.  

\item The Toffoli counts and depths are given by ref.~\cite{low_trading_2024} ({Table~2}), with the convention that each Toffoli costs $4T$-gates, and that the Toffoli depth coincides with the $T$ depth. This yields  
	\begin{equation}  
		\( \mbox{Toffoli count}, \mbox{Toffoli depth} \) = \( \left\lceil \tfrac{2^\eta}{\lambda} \right\rceil + 2 d \lambda, \left\lceil \tfrac{2^\eta}{\lambda} + \log_2 \( \lambda \) \right\rceil \).  
	\end{equation}  

\end{enumerate}
We treat $\lambda$ as a hyperparameter and, for each PES dataset, we determine the optimal value $\lambda = \lambda_{\min}$ that minimizes the Toffoli count of the SELECT-SWAP QROM. 

\section{Estimating multiple energy levels}
\label{sec:appendix-landscape}
When multiple energy levels are required, either QPE must be executed independently multiple times with different trial states, or an amplitude amplification algorithm can be used jointly with QPE. The total cost of QPE consists of the trial state preparation cost, $C_{\text{trial}}$, and the cost of the QPE core circuit, which scales as $C_{QPE}=\mathcal{\widetilde O}(C_{H}\zeta/\varepsilon) ))$. $C_H$ denotes the cost of block-encoding the Hamiltonian, as given for example, by eq.~\ref{eq:FBR-DVR-Ham-resources}.
The parameter $\zeta$ represents the total block-encoding scaling constant, as listed e.g. in Table~\ref{tab:HO2_cost}. The $\widetilde{\mathcal{O}}$ notation omits logarithmic factors. When $N_{\text{eval}}$ eigenvalues are required, the total $T$-gate cost scales as
\begin{equation}
\widetilde{\mathcal{O}}\left(N N_{\text{eval}}\left(N + C_H \frac{\zeta}{\varepsilon}\right)\right),
\label{eq:multiple-eigenvalues-QPE}
\end{equation}
where the factor $N N_{\text{eval}}$ accounts for the QPE execution overhead related to the average overlap of a random trial state with an eigenstate of the Hamiltonian and the number of eigenvalues requested.
Physics-informed trial states can reduce this overhead, albeit at the expense of an increased state preparation cost.
Here we assume that a generic trial state preparation cost is $C_{\text{trial}} = \widetilde{\mathcal{O}}(N)$~\cite{low2024}, which adds to the cost of a single QPE execution.

This complexity can be contrasted with the $\mathcal{O}(N \rho M_{\text{Kr}}) = \mathcal{O}(n^D \rho M_{\text{Kr}})$ scaling of iterative classical eigensolvers, where $N$ is the total size of the direct-product basis, $\rho$ is the matrix sparsity, and $M_{\text{Kr}}$ is the dimension of the Krylov subspace. We neglect the Hamiltonian matrix elements calculation cost.
Typically, $M_{\text{Kr}}$ scales linearly with the number of eigenvalues requested and logarithmically with the energy precision of each eigenvalue.
Thus, we are comparing a quantum Clifford+$T$ gate count complexity of $\widetilde{\mathcal{O}}\left(N N_{\text{eval}} C_H \zeta / \varepsilon\right)$ with a classical floating-point operation count of $\mathcal{O}\left(N N_{\text{eval}} \rho \log(1 / \varepsilon)\right)$, where $C_H$ is at least $\widetilde{\mathcal{O}}(N)$ for other methods and $\widetilde{\mathcal{O}}(N^{1/\alpha})$ for our method, with $\alpha \approx 0.25 - 0.8 $. Classical eigensolvers exhibit a more favorable dependence on the eigenvalue precision, scaling as $\mathcal{O}\left(\log \frac{1}{\varepsilon}\right)$, in contrast to the inverse power dependence in quantum computation.

QPE can also be combined with amplitude amplification, as described in Ref.~\cite{GRM25} introducing the quantum landscape scanning method, which requires
\begin{equation}
\widetilde{\mathcal{O}}\left(C_{\text{QPE}} \sqrt{N N_{\text{eval}}}\right)
= \widetilde{\mathcal{O}}\left(\sqrt{N N_{\text{eval}}} C_H \frac{\zeta}{\varepsilon}\right),
\label{eq:landscape}
\end{equation}
T-gates and removes the need to prepare separate trial states for each eigenvalue.
This method also mitigates numerical issues associated with matrix inversion when the right-hand side of the \SE is a non-identity matrix (e.g., due to a non-orthogonal basis or approximate Gram matrix elements, cf. eq.~\ref{eq:HVBR}). In this approach, the trial state can be constructed at $\mathcal{O}(1)$ $T$-gate cost as $\ket{\Psi}=\frac{1}{\sqrt{NK}}\sum_{i=0}^{2^{(n+k)}-1}\ket{\psi_i}\ket{\bar\psi_i} = \frac{1}{\sqrt{NK}}\sum_{i=0}^{2^{(n+k)}-1}\ket{i}\ket{i} $,
where $\ket{\psi_i} = \ket{\gamma_i}_k \otimes \ket{\phi_i}_n$ denotes the $i$th eigenstate of $\mathbf{H} - \gamma \mathbf{1}$, and $\ket{ii}$ is the composite state encoding integer indices $i = 1, 2, \ldots, N K$.
Here $K$ denotes the number of grid points $\gamma_i \in {1, \ldots, K}$ at which eigenvalues of $\mathbf{H}$ are scanned, and it scales as $\mathcal{O}(N_{\text{eval}})$.
Using quantum landscape scanning, the performance for finding multiple eigenvalues, including relatively dense spectra, such as those observed in floppy molecules, can be improved quadratically with system size, albeit with an increased cost in eigenvalue precision.

\paragraph{Condition number.}
Rovibrational calculations often lead to a generalized eigenvalue problem. When non-exact quadratures are employed or the basis set is non-orthogonal, the resulting overlap matrix deviates from the identity and may exhibit a large condition number~\cite{GRM25}. In classical FBR calculations using orthogonal basis sets with Gaussian quadratures, the overlap matrix simplifies to the identity, removing the need for matrix inversion. In such cases, one can instead employ quantum landscape scanning as proposed in ref.~\cite{GRM25}, which inherently avoids issues associated with ill-conditioned matrices.

By contrast, quantum algorithms that rely on the Harrow–Hassidim–Lloyd (HHL) procedure~\cite{Harrow2009} for matrix inversion, combined with quantum phase estimation, face significant practical limitations. Although the HHL algorithm offers asymptotically better scaling than exact classical inversion-$\mathcal{O}(\log(N) d^2 \kappa^2 / \epsilon)$ versus $\mathcal{O}(N^3)$, where $d$ denotes matrix sparsity, and it performs poorly for systems with large condition numbers. Classical iterative eigensolvers such as the Arnoldi or Lanczos methods exhibit a more favorable dependence, scaling as $\mathcal{O}(\kappa^{1/2})$.

\end{document}